\documentclass[prd,aps,floats,floatfix,superscriptaddress,preprintnumbers,showpacs,eqsecnum,nofootinbib,twocolumn]{revtex4}
\usepackage{amsfonts,amsmath,epsfig,amssymb,latexsym,color,graphicx,array,subfigure,bm,float,hyperref,nameref }

\definecolor{red}{rgb}{1,0,0}
\definecolor{blue }{rgb}{0,0,1}
\definecolor{green}{rgb}{0,1,0}

\usepackage{amsfonts}
\usepackage{dcolumn}
\usepackage {diagbox}
\allowdisplaybreaks[1]
\usepackage{stackengine}
\usepackage{appendix}

\newcommand{\bea}{\begin{eqnarray}}
\newcommand{\ena}{\end{eqnarray}}
\newcommand{\beann}{\begin{eqnarray*}}
\newcommand{\enann}{\end{eqnarray*}}

\newcommand{\dsl}{\pa \kern-0.5em /}

\newcommand{\pa}{\partial}

\newcommand{\nn}{\nonumber\\}

\newcommand{\vect}[1]{\!\!\mbox{ \boldmath $#1$}}
\newcommand{\gsim}{\, \mbox{\raisebox{-1.ex}
{$\stackrel{\textstyle>}{\textstyle\sim}$}}\,}

\allowdisplaybreaks
\begin{document}

\date{\today}

\title{Dynamics of Binary System  around a Supermassive Black Hole :\\ Binary Scattering and Eccentric vZLK Oscillations}

\author{Kei-ichi Maeda}
%\email{maeda"at"waseda.jp}
\affiliation{
Department of Physics, Waseda University, 3-4-1 Okubo, Shinjuku, Tokyo 169-8555, Japan}
\affiliation{Center for Gravitational Physics and Quantum Information, Yukawa Institute for Theoretical Physics, Kyoto University, 606-8502, Kyoto, Japan}

\author{Hirotada Okawa}
%\email{maeda"at"waseda.jp}
\affiliation{Faculty of Software and Information Technology, Aomori University, Seishincho, Edogawa, Tokyo 134-0087, Japan
}

\begin{abstract}

We study the dynamics of a binary orbiting a supermassive black hole (SMBH), focusing on both binary scattering in unbound orbits and eccentric von Zeipel–Lidov–Kozai (vZLK) oscillations in bound orbits. The motion is described in a local inertial frame in Kerr spacetime, where tidal effects are encoded in the Riemann curvature.

For unbound (parabolic and hyperbolic) orbits, we identify four scattering regimes—adiabatic, tidally affected, chaotic, and disruptive—depending on the binary semi-major axis. As the binary becomes softer, tidal interactions near periapsis lead to strong eccentricity excitation, large changes in the orbital parameters, and eventually chaotic behavior or tidal disruption, with a sensitive dependence on the argument of periapsis.

For eccentric bound (elliptic) orbits, the vZLK mechanism differs qualitatively from the standard one, although the $z$-component of the angular momentum in the local inertial frame remains approximately conserved. The evolution proceeds on a dynamical timescale and exhibits step-like changes driven by repeated periapsis passages, which can be interpreted as a sequence of scattering events. We refer to this behavior as scattering-type vZLK oscillations.

The rotation of the SMBH also modifies the oscillation profiles, although its effect is less significant than the dependence on the initial orbital parameters.

These results suggest a unified picture of periapsis-driven tidal dynamics in galactic nuclei.

\end{abstract}

\maketitle

%\tableofcontents

%\setcounter{page}{1}

%\newpage

%%%%%%%%%%%%%%%%%%%%%%%%%%%%%%%%%%%%%%%
%%%%%%%%%%%%%%%%%%%%%%%%%%%%%%%%%%%%%%%
\section{Introduction}
\label{Introduction}
%%%%%%%%%%%%%%%%%%%%%%%%%%%%%%%%%%%%%%%%%%%%%%%%%%%%%%%%%%%%%%%%%%%%%%%%%%%%%%

The discovery of gravitational waves (GWs) by the LIGO–Virgo–KAGRA (LVK) Collaboration~\cite{GW150914,Abbott_2020,Abbott_2021}  has opened a new window in astronomy and physics.
We now have a unique opportunity to test gravity in the strong-field regime, for example by measuring the propagation speed of GWs~\cite{GW170817}.
The origin of unexpectedly massive stellar-mass black holes (BHs) remains mysterious~\cite{GW150914}.
Understanding the redshift distribution of black holes (BHs) and their environments holds significant promise~\cite{test1,test2,test3,test4,test5}.
Realizing this potential requires accurate and comprehensive modeling of GW signals across a wide range of source parameters.

While current detections by the LVK Collaboration have primarily involved isolated binary systems, hierarchical triple systems may also play an important role~\cite{Martinez_2020,Gerosa_2021}.
The dynamics of hierarchical triple systems has long been a central topic in celestial mechanics. In such systems, the inner binary can undergo large-amplitude oscillations of eccentricity and inclination due to the perturbation from a distant third body~\cite{vonZeipel10,Lidov62,Kozai62}. In this paper, we refer to this phenomenon as the von Zeipel–Lidov–Kozai (vZLK) mechanism, following our previous works \cite{Maeda:2023tao, Maeda:2023uyx, Maeda:2025row}.

The vZLK mechanism has been extensively studied in a wide range of astrophysical contexts, including planetary systems, stellar triples, and compact-object binaries~\cite{naoz13b,Naoz12,Naoz2020,tey13,Li15,Will14a,Will14b,Liu:2019tqr,Liu:2021uam,Lim2020,Fang_2019a,Fang_2019b}. (see also \cite{Naoz2016} for a review). In many applications, it has been recognized as an efficient channel for producing highly eccentric binaries and enhancing merger rates through gravitational radiation. Most of these studies are based on the secular approximation, in which the orbital motion is averaged over short timescales and the evolution proceeds on timescales much longer than the orbital period. Extensions of this framework to relativistic systems, particularly in the vicinity of a supermassive black hole, have also been investigated, where additional effects such as relativistic precession can significantly modify the secular evolution \cite{AntoniniPerets2012,LiuMunozLai2015,KatzDongSari2011}.
Recent studies have further explored three-body systems and their GW emission~\cite{Amaro-Seoane2010, Antonini2012,hoang18,Antonini2016,Meiron2017,Robson2018,Lisa2018,Lisa2019,Hoang2019,Loeb2019,Gupta_2020,kuntz2022transverse,Chandramouli_2022}. Notably, in the presence of a massive tertiary, the vZLK timescale can be reduced to as short as a few years, making recurrent GW signals driven by vZLK oscillations potentially detectable~\cite{Hoang2019, Gupta_2020}. Indirect signatures of triple systems in GW observations have also been investigated, for example through periastron shifts in binary pulsars undergoing vZLK oscillations~\cite{Haruka2019,Suzuki:2020zbg}.

In dense galactic nuclei surrounding supermassive black holes (SMBHs), binaries can naturally evolve into hierarchical triple systems~\cite{Heggie1975,Hut1993,Samsing2014,Riddle2015,Fabio2016,stephan2019}. Recent results from the LVK Collaboration suggest that hierarchical mergers may constitute an important formation channel for binary BH coalescences~\cite{sym13091678,Gayathri_2020,Gerosa_2021}.
Binaries orbiting a SMBH may enter a regime in which the assumptions underlying the secular approximation are not fully satisfied. When the center-of-mass orbit is sufficiently compact or highly eccentric, the tidal interaction from the central object becomes strongly localized near periapsis. In such situations, the secular approximation may not provide an accurate description, and the evolution can exhibit features associated with dynamical-timescale interactions.

In this study, we treat a binary system orbiting a SMBH as a perturbation in the background spacetime of the SMBH. While a single object behaves as a test particle in the SMBH gravitational field, the dynamics of a binary system are more complex due to strong mutual interactions.
To describe the binary’s motion, we construct a local inertial frame and employ Newtonian gravitational dynamics, assuming moderate internal interactions. The local inertial frame is defined using Fermi normal coordinates or Fermi–Walker transport~\cite{1963JMP.....4..735M,Nesterov_1999,Delva:2011abw}. This approach has been widely applied to studies of tidal effects near SMBHs as well as to binary dynamics~\cite{Banerjee_2019,PhysRevD.71.044017,Cheng_2013,Kuntz_2021,Gorbatsievich_Bobrik,Chen_Zhang,camilloni2023tidal,
Maeda:2023tao,Maeda:2023uyx,Zhang_2024,Camilloni_2024,Cocco:2025adu,Cocco:2025udb,Cocco:2026lkr}.

In our previous works\cite{Maeda:2023tao, Maeda:2023uyx, Maeda:2025row}, 
assuming that the center-of-mass orbit is either circular or spherical, we investigated the vZLK mechanism in the strong-field regime around a SMBH.
For a circular center-of-mass orbit, we found that vZLK oscillations can persist even near the innermost stable circular orbit (ISCO) for sufficiently inclined compact binaries. While highly compact binaries exhibit regular oscillations, softer binaries display chaotic vZLK oscillations with irregular periods and amplitudes. For binaries with high initial inclination, we also observed orbital flips of the relative inclination between the inner and outer orbits. Notably, we found that these results were largely independent of the black hole spin, as the Riemann curvature on the equatorial plane in Kerr spacetime coincides with that in Schwarzschild spacetime.
Since binaries around SMBHs are not necessarily confined to the equatorial plane, particularly if they form via dynamical capture, we also studied the case of a spherical center-of-mass orbit. We found that the libration in the latitudinal direction significantly affects the vZLK oscillations: as the libration amplitude increases, the oscillation period decreases, while the maximum eccentricity grows, especially in the chaotic regime.
Notably, when the binary is sufficiently soft yet remains stable, the oscillation period is reduced to the dynamical timescale rather than the secular one.
A natural question, then,  is how these results are modified when the center-of-mass orbit is eccentric, which is the focus of the present paper.

A closely related problem is the scattering of binaries by a supermassive black hole. In unbound encounters, tidal interactions during a single periapsis passage can significantly modify the orbital parameters of the binary, leading to eccentricity excitation, energy exchange, or tidal disruption. These processes have been studied in the context of stellar dynamics in galactic nuclei and are known to exhibit strong dependence on the initial conditions\cite{Hills1988,Alexander2005,AntoniniPerets2012}. Despite the similarity of the underlying physics, binary scattering and vZLK oscillations have largely been treated separately in the literature.

In this paper, we investigate both binary scattering and eccentric vZLK oscillations for a binary system orbiting a rotating black hole described by the Kerr spacetime. We consider binaries moving on both bound (elliptic) and unbound (parabolic and hyperbolic) equatorial orbits around the SMBH. The binary motion is described in a local inertial frame, where the tidal effects of the SMBH are incorporated through the Riemann curvature.

Our main goals are as follows. First, we classify the outcomes of binary scattering by an SMBH, identifying different regimes such as adiabatic, tidally affected, chaotic, and disruptive scattering. Second, we explore the properties of vZLK oscillations when the center-of-mass orbit is eccentric, focusing on the emergence of step-like evolution and scattering-like behavior. 

This paper is organized as follows. In Sec. II, we formulate the equations of motion of a binary in a local inertial frame in Kerr spacetime. In Sec. III, we analyze binary scattering for unbound orbits. In Sec. IV, we investigate eccentric vZLK oscillations for bound orbits and discuss their characteristic features. Section V is devoted to a summary and discussion.

Appendices A and B provide additional details: Appendix A presents analytic solutions for test-particle motion in Kerr spacetime, while Appendix B summarizes a general framework for binary dynamics in an arbitrary background spacetime.

\textit{Notation}: Greek indices range from 0 to 3, Roman indices from 1 to 3. Hatted indices denote tetrad components in an observer’s proper reference frame, while barred symbols correspond to static tetrad-frame quantities. We set $G = c = 1$ unless otherwise specified.

 %%%%%%%%%%%%%%%%%%%%%%%%%%
%%%%%%%%%%%%%%%%%%%%%%%%%%
\section{A Binary in Equatorial Orbit}
%%%%%%%%%%%%%%%%%%%%%%%%%%
%%%%%%%%%%%%%%%%%%%%%%%%%

 In this paper, we discuss the dynamics of a binary on an eccentric orbit in the equatorial plane. Eccentric orbits are classified into three types: elliptic, parabolic, and hyperbolic. The orbits in Kerr spacetime can be expressed analytically in terms of elliptic functions, as summarized in Appendix \ref{test_particle_Kerr}.

When a binary follows an unbound (parabolic or hyperbolic) orbit, it represents a binary scattering by an SMBH, since the binary approaches the SMBH only once. The binary may be disrupted by tidal forces or scattered away into an excited state with different orbital parameters. In the case of a bound eccentric orbit, the binary periodically approaches the SMBH, which may alter the behavior of the vZLK mechanism. We shall discuss both cases in this paper.

%%%%%%%%%%%%%%%%%%%%%%%%%%
%%%%%%%%%%%%%%%%%%%%%%%%%%
\subsection{Basic equations for  a binary  in the local inertial frame}
%%%%%%%%%%%%%%%%%%%%%%%%%%
%%%%%%%%%%%%%%%%%%%%%%%%%
Suppose that a binary system 
consists of two point particles with the masses $m_1$ and $m_2$, 
which are much smaller than the SMBH mass $M$. 
Although a binary system can be treated as perturbations in Kerr spacetime backgound, 
each component of a binary system does not follow the geodesic
 in Kerr spacetime because the mutual attractive force between two components
 is much larger than the gravitational force by the SMBH.
In order to discuss a binary motion, as shown in \cite{Maeda:2023tao,Maeda:2023uyx}, 
we first construct a local inertial frame, and then put a self-gravitating binary system.
The tidal force by the Kerr black hole is evaluated by the Riemann curvature in this frame.
In Appendix \ref{binary_in_curved_ST}, 
we summarize how to find the basic equations for  a binary system orbiting around the SMBH.

In this subsection, we write down the equations of motion (EOM) of a binary system in the non-rotating 
inertal frame.
A binary consists of two point particles, which positions are 
given by $\vect{\mathsf{x}}_1$ and $\vect{\mathsf{x}}_2$ in the  non-rotating
local inertial frame.
 Introducing the center-of-mass coordinates and the relative coordinates by
\beann
\vect{\mathsf{R}}&=&{m_1\vect{\mathsf{x}}_1+m_2\vect{\mathsf{x}}_2\over m_1+m_2},
\\
\vect{\mathsf{r}}&=& \vect{\mathsf{x}}_2-\vect{\mathsf{x}}_1
\,,
\enann
we find the Lagrangian up to 0.5PN order in terms of  $\vect{\mathsf{R}}$  and $\vect{\mathsf{r}}$.
As shown in Appendix \ref{binary_in_curved_ST}, we choose $\vect{\mathsf{R}}=0$ 
by introducing small acceleration of the observer in 0.5PN order,
which orbit is slightly deviated from the geodesics. The deviation is given by Eq. (\ref{eq_CM}).
 
 The EOM for the relative coordinates is
  given by the Lagrangian up to 0.5PN in the non-rotating  inertial frame 
 \beann
{\cal L}_{\rm rel}
%\left(\vect{r}, {d\vect{r}\over d\tau}\right)
&=&
{1\over 2}\mu \left({d\vect{\mathsf{r}}\over d\tau}\right)^2+  {G m_1m_2\over \mathsf{r}}
+{\cal L}_{{\rm rel}\mathchar`- \bar{\cal R}}
+{\cal L}_{1/2\mathchar`-{\rm rel}}\,,
\enann
where 
\beann
{\cal L}_{{\rm rel}\mathchar`-\bar{\cal R}}
&=&
-{1\over 2}\mu 
\bar{\cal R}_{\tilde 0 \tilde k \tilde 0 \tilde \ell}\mathsf{r}^{\tilde k}\mathsf{r}^{\tilde \ell}
\\
{\cal L}_{1/2\mathchar`-{\rm rel}}&=&- {2\over 3} \mu{(m_1-m_2)\over (m_1+m_2)}R_{\tilde 0\tilde k \tilde j \tilde \ell}\mathsf{r}^{\tilde k}\mathsf{r}^{\tilde \ell}{d\mathsf{r}^{\tilde j}\over d\tau}
\,,
\enann
and $\mu = m_1 m_2/(m_1+m_2)$ is the reduced mass.

To complete the EOM, 
we have to write down the components of the Riemann tensor of Kerr spacetime in the non-rotating inertial frame.
Kerr metric is given  in Boyer-Lindquist (BL) coordinates as
\bea
d\bar s^2
&=&
-\frac{\Delta}{\Sigma}\left[dt-a(1-\zeta^2)d\phi\right]^2
\nonumber \\
&&
+\frac{1-\zeta^2}{\Sigma}\left[
(\mathfrak{r}^2+a^2)d\phi-adt\right]^2
\nonumber \\
&&
+\frac{\Sigma}{\Delta}d\mathfrak{r}^2+\frac{\Sigma}{1-\zeta^2} d\zeta^2
\,,
\label{Kerr_metric}
\ena
%\end{}
where
\beann
\zeta&=&\cos\theta
\\
\Sigma&=& \mathfrak{r}^2+a^2\zeta^2
\\
\Delta&=& \mathfrak{r}^2-2M\mathfrak{r}+a^2
\,.
\enann
 $M$ is a gravitational mass of a SMBH
 and $a$ is its proper angular momentum.
 
In order to evaluate the curvature components in the EOM of a binary, we approximate the binary orbit by  
 the geodesic in  a rotating SMBH. We can ignore the contributions by the deviations from the geodesic 
 because they are higher-order terms.
  Since the geodesic in Kerr spacetime  is described by the analytic functions of Mino time  
$\mathsf{t}$,
it may be better to discuss a binary motion with respect to the Mino time\cite{Mino:2003yg}.
The action $S$ is given by
\beann
S=\int d\tau {\cal L}=\int d\mathsf{t} \tilde  {\cal L}
\enann
where
\beann
\tilde  {\cal L}&=& \tilde  {\cal L}_{\rm N}+\tilde  {\cal L}_{1/2}
\enann
with
\beann
\tilde  {\cal L}_{\rm N}&\equiv&\Sigma {\cal L}_{\rm N}={\mu\over 2\Sigma} \dot{\vect{\mathsf{r}}}^2+  {G\Sigma m_1m_2\over \mathsf{r}}
-{\mu\Sigma\over 2} 
\bar{\cal R}_{\tilde 0\tilde k \tilde 0 \tilde \ell}\mathsf{r}^{\tilde k}\mathsf{r}^{\tilde\ell}
\\
\tilde  {\cal L}_{1/2}&\equiv&\Sigma {\cal L}_{1/2}=- {2\over 3} \mu{(m_1-m_2)\over (m_1+m_2)}R_{\tilde 0\tilde k \tilde j \tilde \ell}\mathsf{r}^{\tilde k}\mathsf{r}^{\tilde \ell}\dot{\mathsf{r}}^{\tilde j}
\enann

We then introduce the momentum $\vect{\pi}$ as
\beann
\vect{\pi}\equiv {\pa \tilde {\cal L}\over \pa \dot{\vect{\mathsf{r}}}}=\vect{\pi}_{\rm N}+\vect{\pi}_{1/2}
\enann
where
\beann
\vect{\pi}_{\rm N}&=&{\mu\over \Sigma} \dot{\vect{\mathsf{r}}}
\\
{\pi_{1/2}}_{\tilde j}&=&- {2\over 3} \mu{(m_1-m_2)\over (m_1+m_2)}R_{\tilde 0\tilde k \tilde j \tilde \ell}\mathsf{r}^{\tilde k}\mathsf{r}^{\tilde \ell}
\enann
The Hamiltonian $\tilde {\cal H}$ is defined by
\beann
\tilde {\cal H}&=&\vect{\pi}\cdot \dot{\vect{\mathsf{r}}}-\tilde {\cal L}
\\
&=&\Sigma\left[
{1\over 2\mu}\vect{\pi}_{\rm N}^2-  {Gm_1m_2\over \mathsf{r}}
+{\mu\over 2} 
\bar{\cal R}_{\tilde 0\tilde k \tilde 0 \tilde \ell}\mathsf{r}^{\tilde k}\mathsf{r}^{\tilde\ell}
\right]
\\
&=&\Sigma\left[
{1\over 2\mu}\left(\vect{\pi}-\vect{\pi}_{1/2}\right)^2-  {Gm_1m_2\over \mathsf{r}}
+{\mu\over 2} 
\bar{\cal R}_{\tilde 0\tilde k \tilde 0 \tilde \ell}\mathsf{r}^{\tilde k}\mathsf{r}^{\tilde\ell}
\right]
\,.
\enann
The Hamilton equations are
\beann
\dot{\vect{\mathsf{r}}}&=&{\pa \tilde{\cal H}\over \pa \vect{\pi}}
={\Sigma\over \mu}
\left(\vect{\pi}-\vect{\pi}_{1/2}\right)={\Sigma\over \mu}
\vect{\pi}_{\rm N}
\\
\dot{\vect{\pi}}&=&-{\pa \tilde{\cal H}\over \pa \vect{\mathsf{r}}}
\enann
which are explicitly written as
\beann
\dot{\mathsf{r}}_{\tilde j}&=&{\Sigma\over \mu}
\left(\pi_{\tilde j}+ {2\over 3} \mu{(m_1-m_2)\over (m_1+m_2)}R_{\tilde 0\tilde k \tilde j \tilde \ell}\mathsf{r}^{\tilde k}\mathsf{r}^{\tilde \ell}\right)
\\
\dot{\pi}_{\tilde j}
%&=&-\mu\Sigma\left[{G(m_1+m_2)\mathsf{r}_{\tilde j}\over \mathsf{r}^3}+\bar{\cal R}_{\tilde 0\tilde k \tilde 0 \tilde j}\mathsf{r}^{\tilde k}\right]+\dot{\vect{\mathsf{r}}}\cdot {\pa \vect{\pi}_{1/2}\over \pa r^{\tilde j}}
%\\
&=&-\mu\Sigma\left[{G(m_1+m_2)\mathsf{r}_{\tilde j}\over \mathsf{r}^3}+\bar{\cal R}_{\tilde 0\tilde k \tilde 0 \tilde j}\mathsf{r}^{\tilde k}\right]
\\
&&
- {2\over 3} \mu{(m_1-m_2)\over (m_1+m_2)}\left(R_{\tilde 0\tilde k \tilde \ell \tilde j}+R_{\tilde 0\tilde j \tilde \ell \tilde k}\right)
\mathsf{r}^{\tilde k}\dot{\mathsf{r}}^{\tilde \ell}
\enann
For an equal mass binary $(m_1=m_2)$, we obtain
the simple equations of motion as
\beann
\dot{\mathsf{r}}_{\tilde j}&=&{\Sigma\over \mu}
\pi_{\tilde j}
\label{EOM1}
\\
\dot{\pi}_{\tilde j}&=&-\mu\Sigma\left[{G(m_1+m_2)\mathsf{r}_{\tilde j}\over \mathsf{r}^3}+\bar{\cal R}_{\tilde 0\tilde k \tilde 0 \tilde j}\mathsf{r}^{\tilde k}\right]
\label{EOM2}
\enann
In what follows, we assume that $m_1=m_2$ for simplicity.

 %%%%%%%%%%%%%%%%%%%%%%%%%%
%%%%%%%%%%%%%%%%%%%%%%%%%%
\subsection{The local inertial frame and the curvature components in the equatorial plane}
%%%%%%%%%%%%%%%%%%%%%%%%%%
%%%%%%%%%%%%%%%%%%%%%%%%%

A local inertial reference frame in Kerr spacetime, in which a binary system is located, 
is given by the tetrad $e_{\tilde \alpha}^{~\mu}$.
Here we consider only a local inertial reference frame in the equatorial plane ($\theta=\pi/2$).
The one temporal tetrad vector $\equiv e_{\tilde 0}^{~\mu}$ 
is chosen to be the 4-velocity $u^\mu\equiv e_{\tilde 0}^{~\mu}$, which is given by
\beann
e_{\tilde 0\mu}=u_\mu=\left(-E\,,{\dot{\mathfrak{r}}\over \Delta}\,,0\,, L\right)
\,,
\enann
for a given proper energy $E$ and  $z$ component of 
proper angular momentum $L$.  
The three spatial tetrad vectors $e_{\tilde i}^{~\mu}~(i=1, 2, 3)$ are
parallelly transported along the  4-velocity $u^\mu$.
One spatial unit vector parallelly transported  is chosen as
\beann
e_{\tilde 3\, \mu}=
\sigma
\left(
0\,,
0\,,
-\mathfrak{r} \,,
0
\right)\,,
\enann
where 
$\sigma=+1$ and $-1$ correspond to the prograde and retrograde orbits, respectively.

The other two spatial unit tetrad vectors ($e_{\tilde 1\, \mu}\,, e_{\tilde 2\, \mu}$) 
are given by 
the tetrad vectors of rotating inertial frame ($e_{\hat 1\, \mu}\,, e_{\hat 2\, \mu}$) 
such that\cite{Marck1983,vanDeMeent2020} 
\beann
e_{\tilde 1\, \mu}&=&e_{\hat 1\, \mu}\cos \Psi-e_{\hat 2\, \mu}\sin\Psi
\\
e_{\tilde 2\, \mu}&=&e_{\hat1\, \mu}\sin\Psi+e_{\hat  2\, \mu}\cos\Psi
\,,
\enann
where
\beann
e_{\hat 1\, \mu}&\equiv & 
{1\over \sqrt{\ell^2+\mathfrak{r}^2}}
\left(
- \dfrac{\dot{\mathfrak{r}}}{\mathfrak{r}}\,,
\dfrac{\mathfrak{r}}{\Delta}\left(E\mathfrak{r}^2-a\ell\right)\,,
0\,,
a\dfrac{\dot{\mathfrak{r}}}{\mathfrak{r} }
\right)
\\
e_{\hat 2\, \mu}&\equiv & {\sigma\over \sqrt{\ell^2+\mathfrak{r}^2}}
\left(
-(\ell E+a)\,,
\dfrac{\ell}{\Delta} \dot{\mathfrak{r}} \,,
0\,,
\ell(\ell+aE)+a^2+\mathfrak{r}^2
\right)
\,,
\enann
where $\ell=L-aE$. 
%with
%\beann
%\Xi\equiv \sqrt{\ell^2\over \ell^2+\mathfrak{r}^2}.
%\enann
The rotation angle $\Psi$ satisfies the following evolution equation:\cite{Marck1983,vanDeMeent2020}
\beann
\dot \Psi=\sigma {(\ell E+a)  \mathfrak{r}^2\over \ell^2+\mathfrak{r}^2}\,.
\enann

The transformation $\Lambda_{\hat\lambda}^{~\bar \alpha}$ 
from the Carter's tetrad ($e_{\bar \alpha}^{~\mu}$) to 
the above non-rotating inertial frame tetrad 
($e_{\tilde\alpha}^{~ \mu}$)
gives the curvature components in the local inertial frame, and then we find the basic Lagrangian for a binary system, which center-of-mass follows the geodesic in Kerr spacetime.
Since the components of non-rotating inertial frame are a little complicated, 
we first write down the Lagrangian ${\cal L}_{\rm rel\mathchar`-\bar R}$ in rotating frame, and then 
 transform the coordinates $(x,y)$ in the rotating frame to $(\mathsf{x},\mathsf{y})$ in the non-rotating frame by the transformation 
 \beann
 x&=&\mathsf{x}\cos\Psi+\mathsf{y}\sin\Psi
 \\
 y&=&-\mathsf{x}\sin\Psi+\mathsf{y}\cos\Psi
 \,.
 \enann
The transformation matrix 
$\Lambda_{\hat\lambda}^{~\bar \alpha}\equiv  e_{\hat\lambda \mu}
e^{\bar \alpha\mu}$ 
from the Carter's tetrad  to 
the rotating inertial frame tetrad ($e_{\hat\alpha}^{~ \mu}$)
is given by 
\beann
\Lambda_{\hat 0}^{~\bar \lambda}&=&{1\over  \mathfrak{r}\sqrt{\Delta}}
\left(E\mathfrak{r}^2-a\ell,\dot{\mathfrak{r}}, 0,  \ell\sqrt{\Delta}\right)
\\
\Lambda_{\hat 1}^{~\bar \lambda}
&=& {1\over \sqrt{\Delta(\ell^2+\mathfrak{r}^2)}}\left( \dot{\mathfrak{r}}, 
E\mathfrak{r}^2-a\ell, 0, 
0
\right)
\\
\Lambda_{\hat 2}^{~\bar \lambda}&=&{|\ell|\over \mathfrak{r}\sqrt{\Delta(\ell^2+\mathfrak{r}^2)}}
\left(E\mathfrak{r}^2-a\ell, \dot{\mathfrak{r}}, 0, {(\ell^2+\mathfrak{r}^2)\sqrt{\Delta}\over \ell }\right)
\\
\Lambda_{\hat 3}^{~\bar \lambda}&=&\sigma\left(0, 0, -1,  0\right)
\,.
\enann
We find the curvature components on the equatorial plane 
in the rotating inertial frame are given by
\beann
\bar{\cal R}_{\hat 0\hat 1\hat 0\hat 1}&=&-{2\mathfrak{r}^2+3\ell^2\over \mathfrak{r}^2}{\cal Q}_1
\\
\bar{\cal R}_{\hat 0\hat 2\hat 0\hat 2}&=&{\cal Q}_1
\\
\bar{\cal R}_{\hat 0\hat 3\hat 0\hat 3}&=&{\mathfrak{r}^2+3\ell^2\over \mathfrak{r}^2}{\cal Q}_1
\,,
\enann
where
\beann
{\cal Q}_1={M\over \mathfrak{r}^3}\,.
\enann

We then find the curvature components in the non-rotating inertial frame as
 \beann
 \bar{\cal R}_{\tilde 0\tilde 1\tilde 0\tilde 1} &=&\bar{\cal R}_{\hat 0\hat 1\hat 0\hat 1}\cos^2\Psi +\bar{\cal R}_{\hat 0\hat 2\hat 0\hat 2}\sin^2\Psi
 \\
\bar{\cal R}_{\tilde 0\tilde 2\tilde 0\tilde 2} &=&\bar{\cal R}_{\hat 0\hat 1\hat 0\hat 1}\sin^2\Psi+\bar{\cal R}_{\hat 0\hat 2\hat 0\hat 2}\cos^2\Psi 
 \\
\bar{\cal R}_{\tilde 0\tilde 3\tilde 0\tilde 3} &=&\bar{\cal R}_{\hat 0\hat 3\hat 0\hat 3}
 \\
\bar{\cal R}_{\tilde 0\tilde 1\tilde 0\tilde 2} &=&\left(
\bar{\cal R}_{\hat 0\hat 1\hat 0\hat 1}-\bar{\cal R}_{\hat 0\hat 2\hat 0\hat 2}\right)\cos\Psi\sin\Psi
\,.
 \enann

%%%%%%%%%%%%%%%%%%%%%%%%%%%%%%%%%%%%%%
\subsection{Normalized EOM of a binary system }
%%%%%%%%%%%%%%%%%%%%%%%%%%%%%%%%%%%%%%
%%%%%%%%%%%%%%%%%%%%%%%%%%%%%%%%%%%%%%

For numerical analysis of the binary motion, we shall rewrite
 the basic equations using dimensionless variables.
We introduce dimensionless variables with asterisk $*$  as 
\beann
\tau_* &\equiv&n_0\tau
\\
\mathsf{t}_*&\equiv&n_0\mathfrak{r}_0^2\mathsf{t}
\\
\mathsf{r}^{\tilde j}_*&\equiv&{\mathsf{r}^{\tilde j}\over \mathfrak{a}_0}
\\
\pi_{\tilde{j}}^*&\equiv&{\pi_{\tilde{j}}\over \mathfrak{a}_0 n_0\mu}
\,,
\enann
where $\mathfrak{a}_0$ is the initial ``semi-major axis' of a binary,
which will be defined in the next subsection, 
 and the initial ``mean motion'' $n_0$ is defined by
\beann
n_0\equiv \sqrt{G(m_1+m_2)\over \mathfrak{a}_0^3}
\enann

Suppose that the typical scale length of a binary's center-of-mass  orbit is 
$\mathfrak{r}_0$.
For a bound elliptic orbit, we choose $\mathfrak{r}_0$
 to be the semi-major axis, which is given by 
\beann
\mathfrak{r}_0={1\over 2}(\mathfrak{r}_p+\mathfrak{r}_a)={p_{\rm out}\over 1-e_{\rm out}^2}
\,,
\enann
where $\mathfrak{r}_p$ and $\mathfrak{r}_a$ are the periapsis and apoapsis, respectively.
Here, $p_{\rm out}$ and $e_{\rm out}$ denote the semi-latus rectum and eccentricity of the binary's center-of-mass orbit, defined by
\beann
p_{\rm out}={2\mathfrak{r}_a \mathfrak{r}_p\over \mathfrak{r}_a+\mathfrak{r}_p}
~\,,~~
e_{\rm out}={\mathfrak{r}_a-\mathfrak{r}_p\over \mathfrak{r}_a+\mathfrak{r}_p}
\enann
For unbounded (parabolic or hyperbolic) orbit, since no apoapsis exists, we instead choose $\mathfrak{r}_0=\mathfrak{r}_p$.

We then define the other dimensionless valuables as
\beann
\Sigma_* &\equiv&{ \mathfrak{r}^2\over \mathfrak{r}_0^2}
\\
{\cal R}_{\tilde 0 \tilde{k}\tilde 0\tilde{\ell}}^*&\equiv&\mathfrak{r}_0^2
\bar{\cal R}_{\tilde 0 \tilde{k}\tilde 0\tilde{\ell}}
\,.
\enann

The normalized EOM by use of the dimensionless variables 
are now
\beann
{d\mathsf{x}_*\over d \mathsf{t}_*}&=&\Sigma_*\pi^*_{\mathsf{x}}
\\
{d\mathsf{y}_*\over d \mathsf{t}_*}&=&\Sigma_*\pi^*_{\mathsf{y}}
\\
{d\mathsf{z}_*\over d \mathsf{t}_*}&=&\Sigma_* \pi^*_{\mathsf{z}}
\enann
and 
\beann
{d \pi^*_{\mathsf{x}}\over d \mathsf{t}_*}
&=&-\Sigma_*\left[{\mathsf{x}_*\over \mathsf{r}_*^3}+{1\over n_0^2\mathfrak{r}_0^2}
\left({\cal R}^*_{\tilde 0 \tilde 1\tilde 0\tilde 1}\mathsf{x}_*+
{\cal R}^*_{\tilde 0 \tilde 2\tilde 0\tilde 1}\mathsf{y}_*\right)\right]
\\
{d \pi^*_{\mathsf{y}}\over d \mathsf{t}_*}
&=&-\Sigma_*\left[{\mathsf{y}_*\over \mathsf{r}_*^3}
+{1\over n_0^2\mathfrak{r}_0^2} 
\left({\cal R}^*_{\tilde 0 \tilde 1\tilde 0\tilde 2}\mathsf{x}_*
+{\cal R}^*_{\tilde 0 \tilde 2\tilde 0\tilde 2}\mathsf{y}_*\right)\right]
\\
{d \pi^*_{\mathsf{z}}\over d\mathsf{t}_*}
&=&-\Sigma_*\left[{\mathsf{z}_*\over \mathsf{r}_*^3}+{1\over n_0^2\mathfrak{r}_0^2}
{\cal R}^*_{\tilde 0 \tilde 3\tilde 0\tilde 3} \mathsf{z}_*\right]
\enann

%%%%%%%%%%%%%%%%%%%%%%%%%%%%%%%%
%%%%%%%%%%%%%%%%%%%%%%%%%%%%%%%%
\subsection{Orbital Parameters and Initial Data}
%%%%%%%%%%%%%%%%%%%%%%%%%%%%%%%%
%%%%%%%%%%%%%%%%%%%%%%%%%%%%%%%%
 In order to describe the properties of 
a binary orbit, it is more convenient to use the orbital parameters,
since we expect that
 the binary motion is close to an elliptic orbit, which 
 is described by
\beann
\mathsf{r}={\mathfrak{a}(1-e^2)\over 1+e\cos f}
\,,
\label{elliptic_orbit}
\enann
where $\mathfrak{a}, e$ and $f$ are the semi-major axis
of an elliptic orbit, the eccentricity, and the true anomaly, respectively.
We also introduce 
three angular variables; the argument of periapsis $\omega$, the ascending node $\Omega$ and the inclination angle $I$.

For the elliptic orbit, the relation between the  Cartesian  coordinates $\vect{\mathsf{r}}=(\mathsf{x},\mathsf{y},\mathsf{z})$ of a binary and the orbital parameters $(\omega\,,\Omega\,,  \mathfrak{a}\,, e\,, I\,, f)$ is given by 
\begin{equation}
\resizebox{0.9\linewidth}{!}{$
\begin{array}{rcl}
\begin{pmatrix}
\mathsf{x} \\
\mathsf{y} \\
\mathsf{z}
\end{pmatrix}
&=&
\mathsf{r}
\begin{pmatrix}
\cos \Omega\cos(\omega+f)-\sin\Omega\sin(\omega+f)\cos I \\
\sin \Omega\cos(\omega+f)+\cos\Omega\sin(\omega+f)\cos I\\
\sin(\omega+f)\sin I
\end{pmatrix}
\end{array}
$}
\label{orbital_parameters}
\end{equation}
In order to extract the orbital parameters 
from the orbit given by the Cartesian 
coordinates, one can use the osculating orbit when the orbit is close to an ellipse. 
The 
normalized Laplace-Runge-Lenz vector is defined by
\bea
\vect{e}\equiv \vect{\mathsf{p}}^*\times (\vect{\mathsf{r}}_*
\times \vect{\mathsf{p}}^*)-{\vect{\mathsf{r}}_*\over \mathsf{r}_*}\,,
\label{LRL_vector}
\ena
where $\vect{\mathsf{p}}^*$ denotes the normalized momentum.
Its magnitude $e=|\vect{e}|$
is commonly used for a measure of orbital eccentricity.
The inclination angle $I$ is defined as mutual inclination between angular momenta of the inner and outer  binaries.
\bea
I=\cos^{-1}\left({\mathfrak{l}^*_{\mathsf{z}} \over |\vect{\mathfrak{l}}^*|}\right)
\,,
\label{def_inclination}
\ena
where $\vect{\mathfrak{l}}^*\equiv \vect{\mathsf{r}}_*\times \vect{\mathsf{p}}^*$ is the normalized angular momentum of a binary, which is defined in the non-rotating local inertial frame.

The other two essential angles $\Omega$ and $\omega$ governing the orientation of the orbital plane. 
The longitude of ascending node $\Omega$ is the angle between the reference axis (say $\mathsf{x}$-axis) and node line vector $\vect{N}$, which is defined by $\vect{N} = \hat{\vect{\mathsf{z}}}\times \vect{\mathfrak{l}}^*$, where $\hat{\vect{\mathsf{z}}}$ is normal to the reference plane (the unit vector in the $\mathsf{z}$ direction). 
So $\Omega$ is computed as,
\bea
\Omega = \cos^{-1} (N_{\mathsf{x}}/N) \,.
\label{def_Omega}
\ena
The argument of periapsis $\omega$ is the angle between node line and periapsis measured in the direction of motion. Therefore,
\bea
\omega = \cos^{-1} \bigg(\frac{\vect{N} \cdot \vect{e}}{N\,e}\bigg) \,.
\label{def_omega}
\ena

%%%%%%%%%%%%%%%%%%%%%%%%%%%%%%%%
%%%%%%%%%%%%%%%%%%%%%%%%%%%%%%%%
%\subsection{ Initial Data}
%%%%%%%%%%%%%%%%%%%%%%%%%%%%%%%%
%%%%%%%%%%%%%%%%%%%%%%%%%%%%%%%%

In order to provide the initial data of a binary, i.e., $\mathsf{x}_{*0}\,,\mathsf{y}_{*0}\,,\mathsf{z}_{*0}$ and $\mathsf{p}^*_{\mathsf{x}0}\,,\mathsf{p}^*_{\mathsf{y}0}\,,\mathsf{p}^*_{\mathsf{z}0}$, 
we shall give the initial orbital parameters  $(\omega_0\,,\Omega_0\,,  \mathfrak{a}_0\,, e_0)$.
From (\ref{orbital_parameters}), assuming $f=0$ at $\tau=0$, 
we find
\beann
\mathsf{x}_{*0}
&=&(1-e_0)\left[\cos \Omega_0\cos\omega_0-\sin\Omega_0\sin\omega_0\cos I_0\right]\,,
\\
\mathsf{y}_{*0}
&=&(1-e_0)\left[\sin \Omega_0\cos\omega_0+\cos\Omega_0\sin\omega_0\cos I_0\right]\,,
\\
\mathsf{z}_{*0}
&=&(1-e_0)\sin\omega_0\sin I_0
\,.
\enann
As for the momentum $\mathsf{p}^*_{\mathsf{x}0}\,,\mathsf{p}^*_{\mathsf{y}0}\,,
\mathsf{p}^*_{\mathsf{z}0}$, 
we use the definitions of the orbital parameters 
of the osculating orbit, i.e., Eqs. (\ref{LRL_vector}), (\ref{def_inclination}), 
(\ref{def_Omega}) and (\ref{def_omega}).
The details are found in \cite{Maeda:2023uyx}.

%%%%%%%%%%%%%%%%%%%%%%%%%%%%%%%%
%%%%%%%%%%%%%%%%%%%%%%%%%%%%%%%%
\subsection{Validity and Stability}
\label{validity}
%%%%%%%%%%%%%%%%%%%%%%%%%%%%%%%%
%%%%%%%%%%%%%%%%%%%%%%%%%%%%%%%%
Before showing our numerical results, 
we discuss validity of the present approach and 
the stability conditions. The minimum curvature radius at the radius $\mathfrak{r}_0$ is evaluated as 
\beann
 \ell_{\bar{\cal R}} &\equiv& {\rm min} \left[|\bar{\cal R}_{\hat \mu\hat \nu\hat \rho\hat \sigma}|^{-{1\over 2}}, |\bar{\cal R}_{\hat \mu\hat \nu\hat \rho\hat \sigma ; \hat \alpha}|^{-{1\over 3}}, |\bar{\cal R}_{\hat \mu\hat \nu\hat \rho\hat \sigma ; \hat \alpha;\hat \beta}|^{-{1\over 4}}
\right]
\\
&\sim&
 {\rm min} \left[\left({M\over \mathfrak{r}_0^3}\right)^{-{1\over 2}}\,,\left({M\over \mathfrak{r}_0^4}\right)^{-{1\over 3}}\,,
\left( {M\over \mathfrak{r}_0^5}\right)^{-{1\over 4}}
\right]
\\
&\sim &\mathfrak{r}_0 \left( {\mathfrak{r}_0 \over M}\right)^{1/4}
\,.
\enann
When we put a binary at $\mathfrak{r}=\mathfrak{r}_0$,
 the  binary size $\ell_{\rm binary}$ should satisfy
\beann
\ell_{\rm binary}\ll  \ell_{\bar{\cal R}} 
\enann

The relativistic effect in a binary is not  important when
\beann
\ell_{\rm binary} &\gg& {G(m_1+m_2)\over c^2}
\,,
\enann
which we have to impose because we assume a Newtonian binary in the local inertial frame.

As for stability of a binary, the mutual gravitational interaction  within a binary 
should be much larger than the tidal force by a third body.
The condition is given by
\beann
{Gm_1m_2\over r^2}\gg  {G\mu M  \over \mathfrak{r}_0^3}\,r
\,.
\enann
It gives the constraint on a binary size $\ell_{\rm binary}$ as
\bea
\ell_{\rm binary} 
&\ll &
\left({m_1+m_2\over M}\right)^{1\over 3}\,\mathfrak{r}_0.
\label{stability_tidal}
\ena
Hence, 
for a binary with the size of
\beann
{G(m_1+m_2)\over c^2 }\ll \ell_{\rm binary} 
 \ll 
 \left({m_1+m_2\over M}\right)^{1\over 3}\,\mathfrak{r}_0\,,
 \enann
 we may apply the present Newtonian approach.
 
We also have another criterion for stability
\cite{mardling2001tidal,myllari2018stability}.
In order to avoid a chaotic energy exchange 
instability, we may have to impose 
the condition for the ratio of the circular radius $\mathfrak{r}_0$ to 
the binary size $\ell_{\rm binary} $ 
such that
\beann
{\mathfrak{r}_0\over \ell_{\rm binary} }
\gsim C_{\rm chaotic}\left({M\over m_1+m_2}\right)^p\,,
\label{chaotic_stability}
\enann
when $M\gg m_1, m_2$. 
In the previous paper\cite{Maeda:2023uyx}, we analysed the details in the case of the circular orbit and 
 found that 
$p=1/3$ and $ C_{\rm chaotic}\sim 2-4$.
This condition is described as
 the mutual gravitational force must be one or two order of magnitude larger than 
the tidal force by SMBH.

\begin{figure}[htbp]
\begin{center}
\includegraphics[width=7cm]{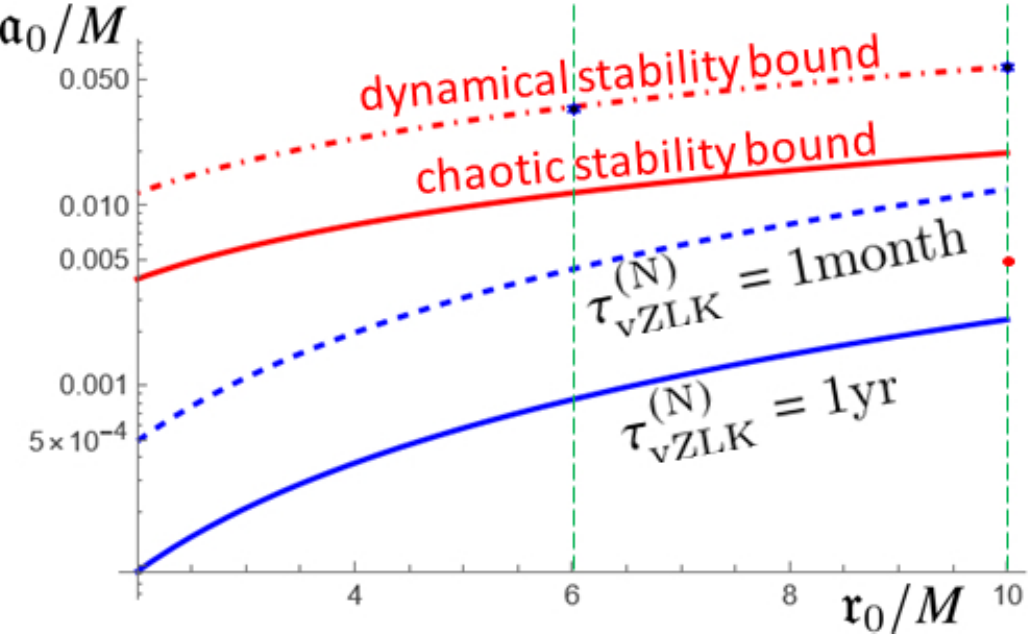}
\caption{The dynamical stability bound (the red dot-dashed curve), the chaotic stability bound (the red solid ),
 and the Newtonian vZLK oscillation time scales 
$\tau_{\rm vZLK}^{\rm (N)}$ of 
1 month (the dashed blue) and 1 year (the solid blue).  
}
\label{fig:parameter_range}
\end{center}
\end{figure}

In Fig.~\ref{fig:parameter_range}, we show the dynamical stability bound (the red dot-dashed curve) and the chaotic stability bound (the red solid curve) for $m_1=m_2=10\,M_\odot$ and $M=10^8\,M_\odot$.
For the vZLK oscillations, we also show two curves corresponding to $\tau^{\rm (N)}_{\rm vZLK}=1$ month (the dashed blue curve) and $1\,\mathrm{yr}$ (the solid blue curve) as references, where $\tau^{\rm (N)}_{\rm vZLK}$ denotes the Newtonian vZLK oscillation period. 
Although this figure provides a rough understanding of the stability conditions, as we will see later, stable systems may exist beyond this bound, particularly in the scattering case.

%%%%%%%%%%%%%%%%%%%%%%%%%%%%%%%%%%%%%%%
\section{Binary Scattering}
\label{binary_scattering}
%%%%%%%%%%%%%%%%%%%%%%%%%%%%%%%%%%%%%%%
We now present our numerical results.
We first consider binary scattering, in which an unbound binary moves approximately on a hyperbolic or parabolic orbit around the SMBH.
Since the gravitational force from the SMBH increases only once as the binary approaches it, we focus on the changes in the binary orbital parameters during the encounter.

Because the binary velocity at infinity is typically nonrelativistic, we mainly present results for parabolic orbits ($E=m$), where $m$ is the rest mass of the binary.
We will also briefly discuss the hyperbolic case for a relativistic binary later.

\subsection{Parabolic Coplanar Binary}
The orbital parameters of a binary system are the  eccentricity $e$, the semi-major axis $a$, the inclination $I$, the argument of periapsis $\omega$, and the longitude of ascending node $\Omega$.
Since  $\Omega$ corresponds on the choice of the reference coordinate system, we choose its initial value as $\Omega_0=0$.
In this subsection, we consider a binary moving on the coplanar orbit, i.e., $I=0$. Although it is the simplest case, the effect of the tidal force of SMBH becomes maximum as we will show it later.

The initial parameters are now the angular momentum of a binary $L_0$
(or the periapsis radius $\mathfrak{r}_p$) for the center-of-mass orbit, and the eccentricity $e_0$, the semi-major axis $\mathfrak{a}_0$, and the argument of periapsis $\omega_0$ 
for a binary.

A binary system is expected to be disrupted by the tidal force of the SMBH when the tidal field is strong ($\mathfrak{r}_p$ is small) and the binary is soft ($\mathfrak{a}_0$ is large).
The critical value of $\mathfrak{a}_0$, known as the Hills radius, is given by the stability condition~(\ref{stability_tidal}), which is also applicable to the scattering case \cite{Hills1988,Bromley2012,HayashiTraniSuto2022,HayashiTraniSuto2025}:
 \beann
 \mathfrak{a}_{\rm H} \approx \left({m_1+m_2\over M}\right)^{1\over 3}\,\mathfrak{r}_p.
 \enann
 In the present paper, we choose $m_1=m_2=10M_\odot$ and $M=10^8 M_\odot$, which gives $ \mathfrak{a}_{\rm H} \approx 0.005848 \mathfrak{r}_p$.
We perform the calculations for the case of $\mathfrak{r}_p=10M$ and 
$6M$, which gives $ \mathfrak{a}_{\rm H} \approx 0.05848M$ and $0.035M$, respectively.
Those values are shown by the blue stars in Fig. \ref{fig:parameter_range}.

We perform our calculations for the ranges $\mathfrak{a}_0=0.005$--$0.5\,M$ with $\mathfrak{r}_p=10M$ and $\mathfrak{a}_0=0.005$--$0.03\,M$ with $\mathfrak{r}_p=6M$.
The Kerr rotation parameter is chosen as $a=0$ (Schwarzschild BH) and $a=M$ (extreme Kerr BH).
Note that there are two cases for the Kerr BH: prograde ($L>0$) and retrograde ($L<0$) orbits.

We first present the results for the Schwarzschild BH case ($a=0$), while the effect of rotation will be discussed later.
We focus on the case of $\mathfrak{r}_p=10M$ because it allows us to include retrograde orbits.
The initial eccentricity is chosen as $e_0=0.01$ (nearly circular binary) or $0.5$ (eccentric binary).

We set the initial data, $\mathfrak{a}_0$ and $\omega_0$, at $\mathfrak{r}=500M$, and analyze the final orbital parameters $e_{\rm fin}$, $\mathfrak{a}_{\rm fin}$, and $\mathfrak{L}_{\rm fin}^2$ at $\mathfrak{r}=500M$, where $\mathfrak{L}$ is the angular momentum of the binary in the local inertial frame.
The initial value is given by $\mathfrak{L}_0^2=\mathfrak{a}_0^2(1-e_0^2)$.
We find several interesting features in the outcomes.

\begin{widetext}

\subsubsection{nearly circular binary $(e_0=0.01)$}
We first present the results for the nearly circular binary case.
The results depend strongly on the argument of periapsis $\omega_0$ for some range of $\mathfrak{a}_0$.
In Fig.~\ref{fig:om-op_Sch_coplanar_e0.01}, we show the final eccentricity, semi-major axis, and angular momentum for the initial semi-major axes $\mathfrak{a}_0=0.01M$, $0.015M$, $0.02M$, and $0.025M$.
The calculations were carried out for angles ranging from $0^\circ$ to $180^\circ$ with a step size of $10^\circ$.
The results for $180^\circ \le \omega_0 < 360^\circ$ are identical to those for $0^\circ \le \omega_0 < 180^\circ$ because of the symmetry of the equal-mass binary ($m_1=m_2$).

In the case of $\mathfrak{a}_0=0.01M$, the final orbital parameters of the binary do not change significantly with the argument of periapsis.
We call this case an adiabatic scattering.
For $\mathfrak{a}_0=0.015M$, although the final values still do not depend strongly on $\omega_0$, the eccentricity is slightly enhanced.
In the case of $\mathfrak{a}_0=0.02M$,  not only the orbital parameters are significantly enhanced by the scattering, but they also depend sensitively on the argument of periapsis.
We call these cases tidally affected scatterings.

For $\mathfrak{a}_0=0.025M$, on the other hand, the binary system is disrupted by the tidal force of the SMBH during the scattering for some range of the argument of periapsis.
We indicate such cases by purple star symbols.
However, for other ranges of the argument of periapsis, the binary remains stable after the scattering.
The final orbital parameters in these stable scattering cases are not systematic, but rather randomly distributed.
We call this case a chaotic scattering.

\begin{figure}[htbp]
\begin{center}
\includegraphics[width=4.8cm]{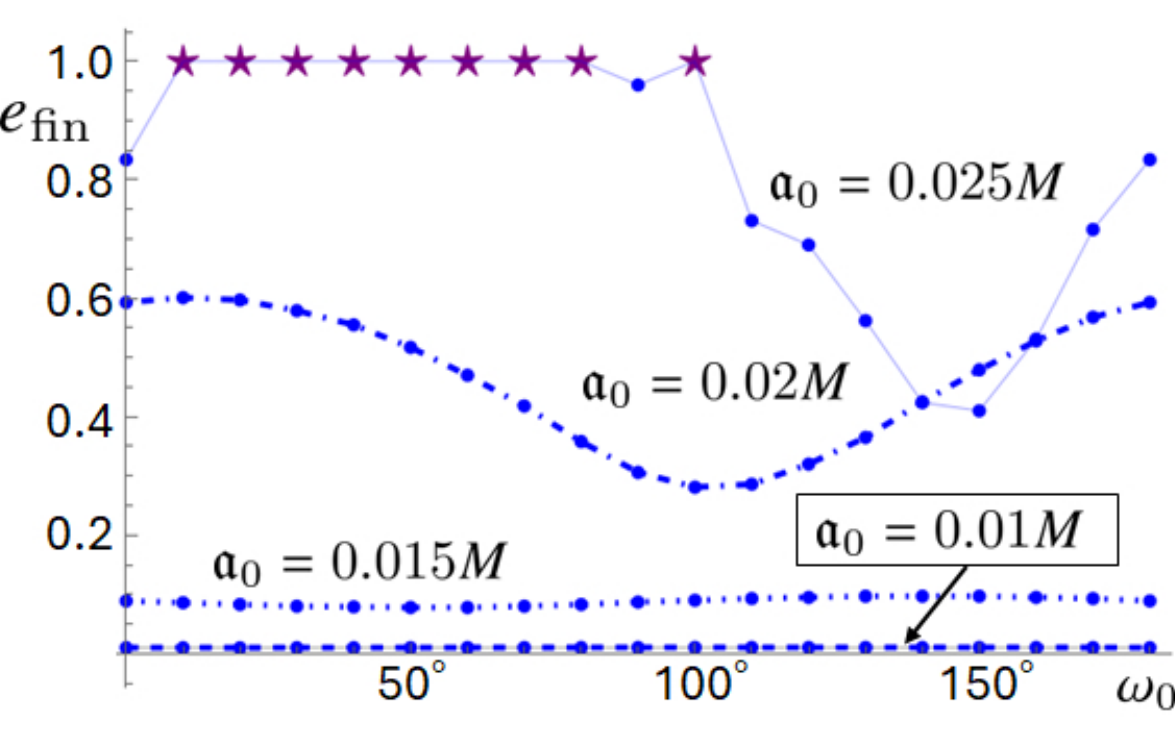}~~
\includegraphics[width=5.cm]{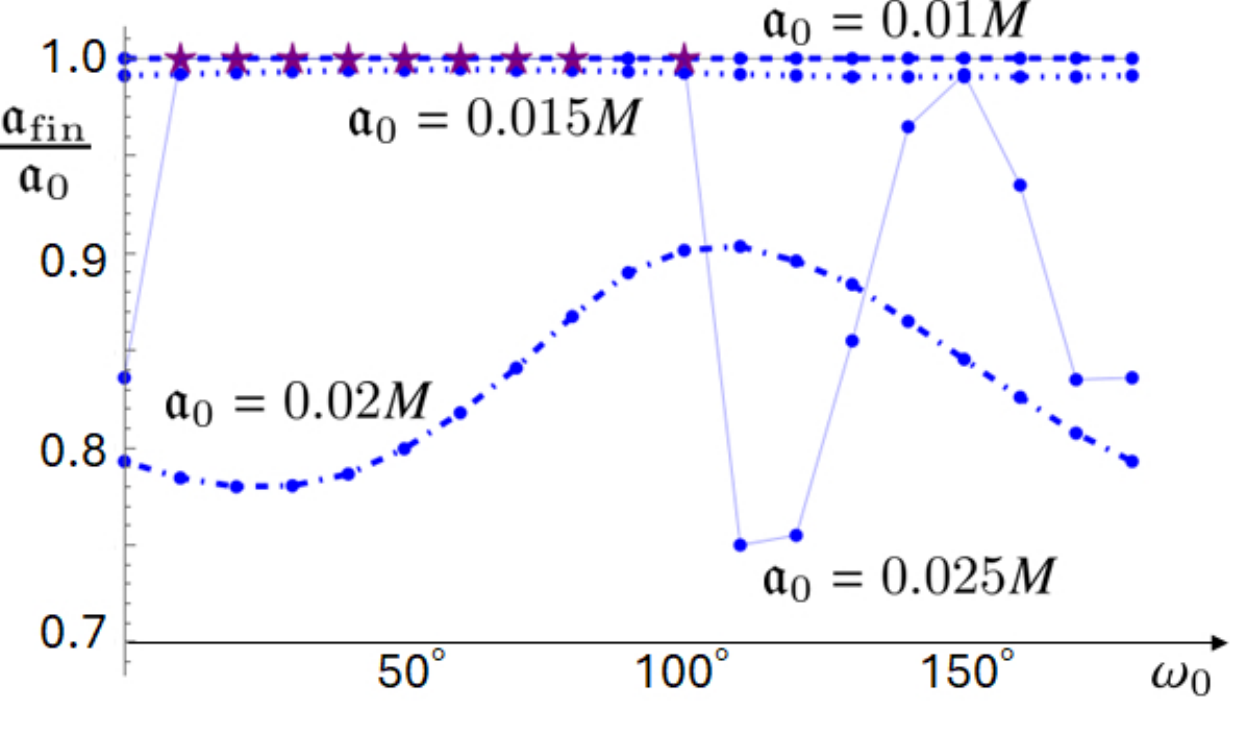}~~
\includegraphics[width=5.cm]{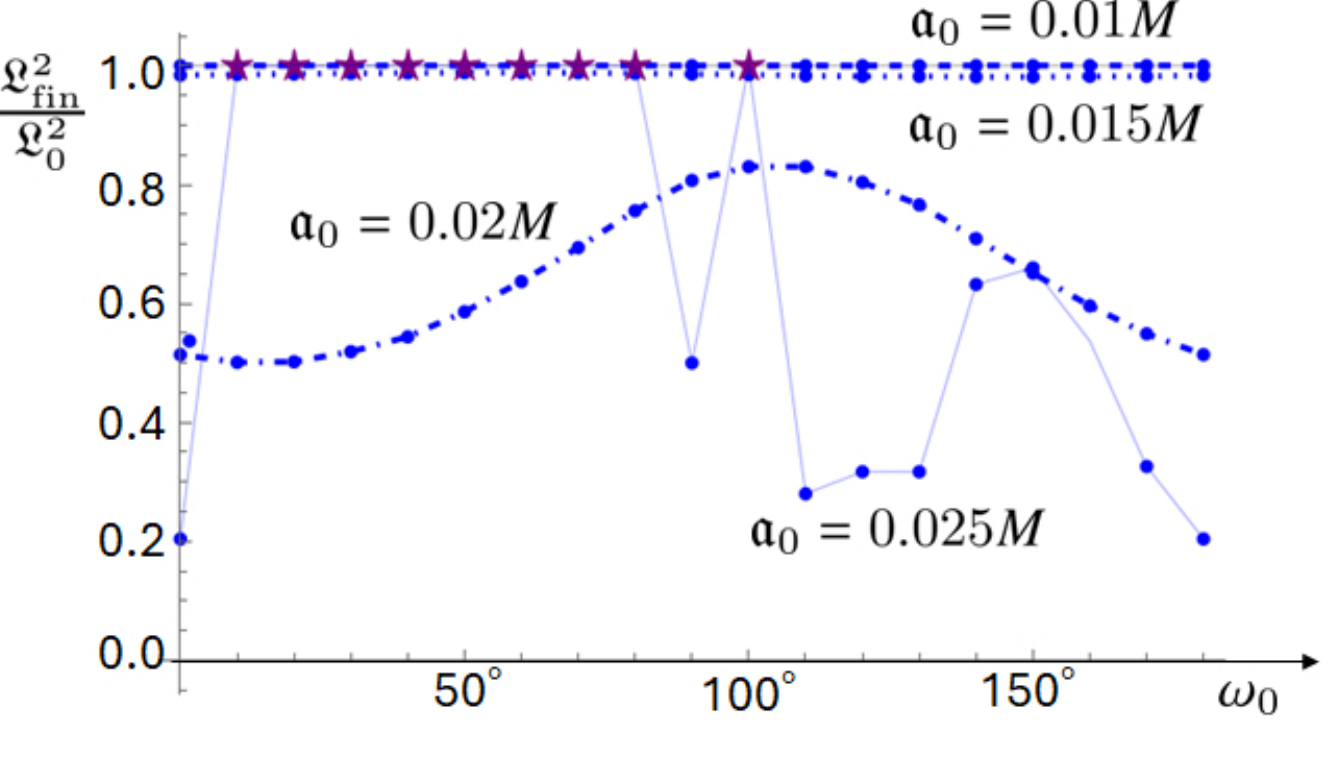}
\\
\hskip 0.5cm (a) final eccentricity \hskip 2cm (b) final semi-major axis \hskip 1.6cm (c) final angular momentum squared
\caption{
Final orbital parameters of the binary after scattering for the initial semi-major axes $\mathfrak{a}_0=0.01M$, $0.015M$, $0.02M$, and $0.025M$.
The left panel (a), middle panel (b), and right panel (c) show the final eccentricity, the ratio of the final semi-major axis to the initial value, and the ratio of the final angular momentum squared to the initial value, respectively.
The final values depend strongly on the argument of periapsis $\omega_0$ for the cases of $\mathfrak{a}_0=0.02M$ and $0.025M$.
The values for $\mathfrak{a}_0=0.02M$ vary smoothly with $\omega_0$, whereas those for $\mathfrak{a}_0=0.025M$ are randomly distributed.
For some ranges of $\omega_0$ ($10^\circ \lesssim \omega_0 \lesssim 80^\circ$ and $\omega_0 \approx 100^\circ$), the binary system is disrupted by the tidal force of the SMBH, which is indicated by purple star symbols.
The final orbital parameters of a binary after scattering for the initial semi-major axis $\mathfrak{a}_0=0.01M, 0.015 M, 0.02M$ and $0.025M$. 
}
\label{fig:om-op_Sch_coplanar_e0.01}
\end{center}
\end{figure}

As a result, binary scatterings can be classified into four cases:
the adiabatic scattering (A), the tidally affected scattering (T), the chaotic scattering (C), and the disruptive scattering (D).
In the A phase, the binary retains nearly the same orbital parameters after the scattering, whereas in the T phase, the binary remains stable but its orbital parameters after the scattering vary smoothly with respect to $\omega_0$.

\begin{figure}[htbp]
\begin{center}
\includegraphics[width=5cm]{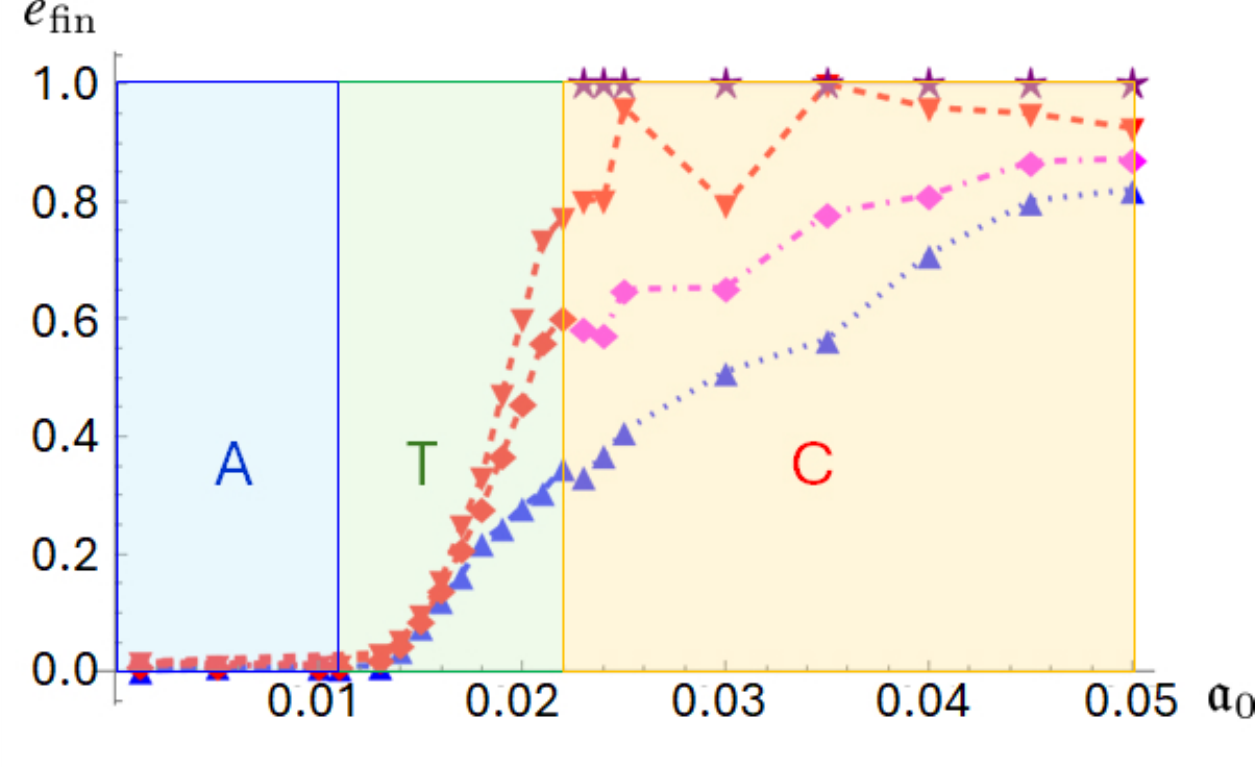}~~
\includegraphics[width=5cm]{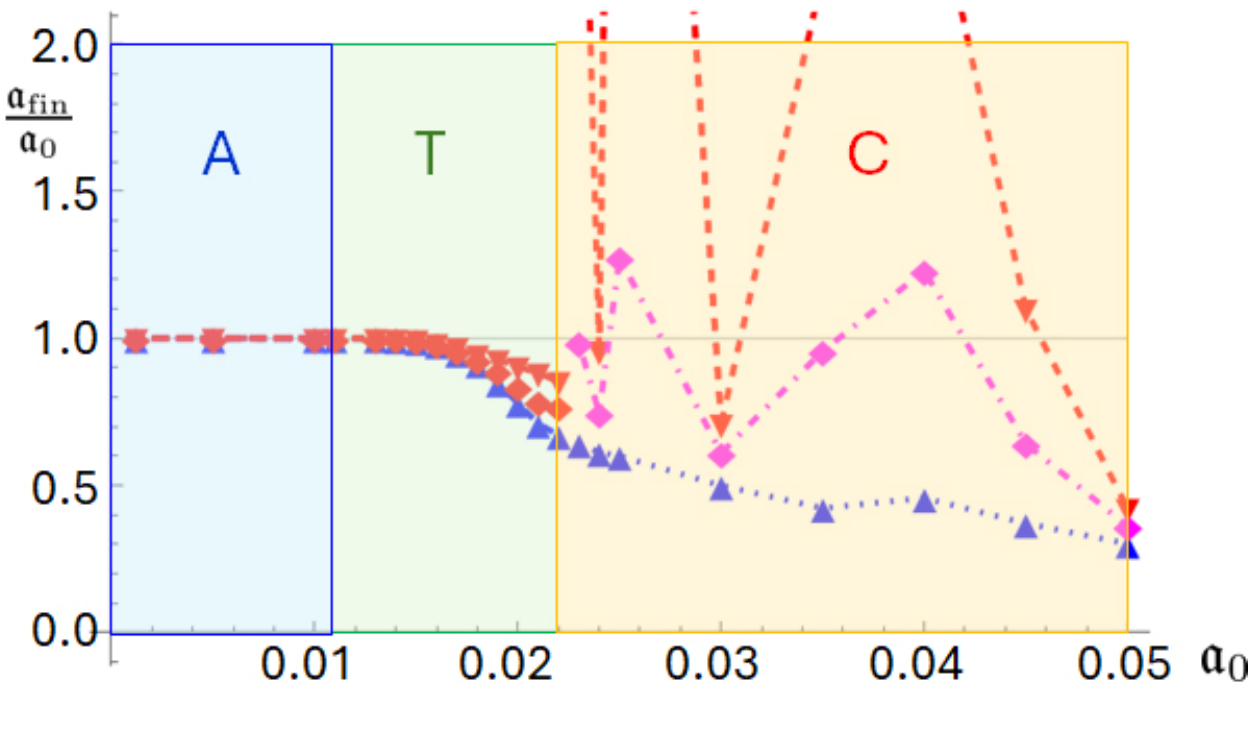}~~
\includegraphics[width=5.2cm]{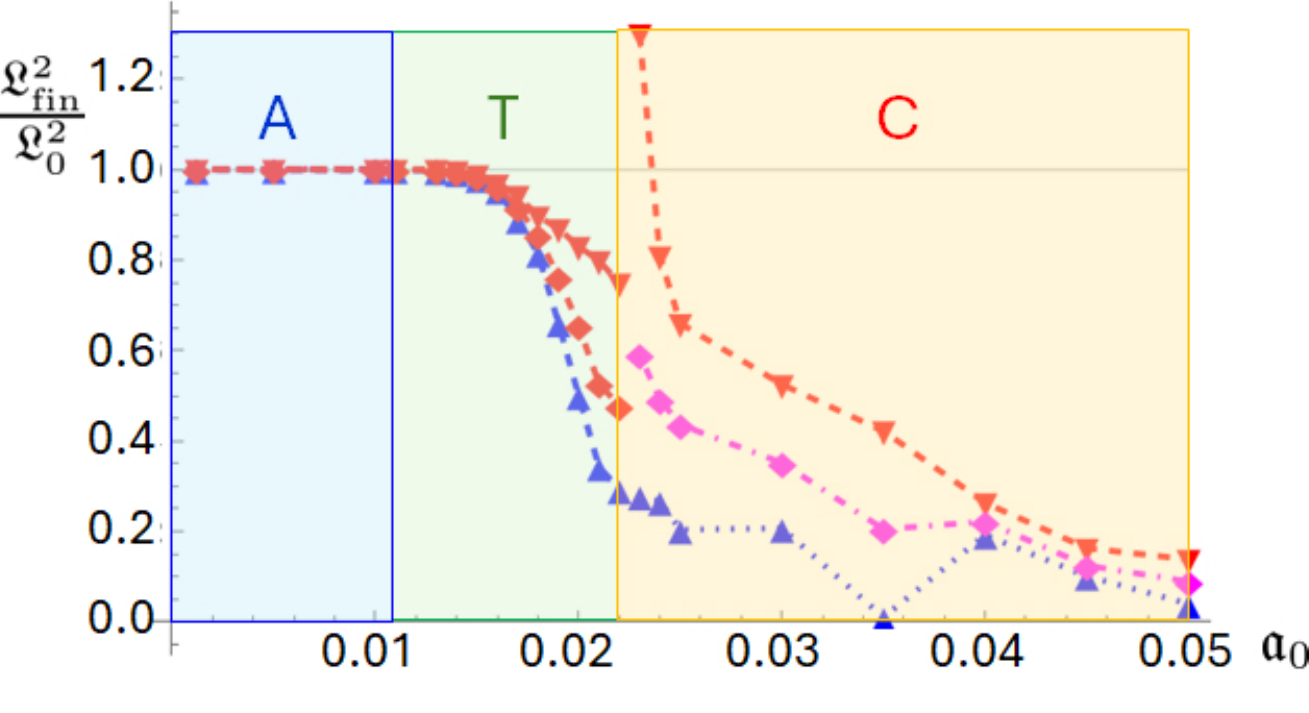}
\\
(a) final eccentricity \hskip 2.2cm (b) final semi-major axis \hskip 1.8cm (c) final angular momentum squared
\caption{The final orbital parameters of a binary after scattering in terms of  the initial semi-major axis $\mathfrak{a}_0=0.005M-0.05M$. 
The left panel figure (a), the middle panel (b) and the right panel (c) denote the final eccentricity, the ratio of the final semi-major axis to the initial value and the ratio of the final angular momentum squared to the initial one, respectively.
The red down triangle and blue up triangle denote the maximum and minimum values of the eccentricity, while the magenta diamond shows
the averaged value.  In the range of the chaotic scattering (C),  a binary system with some values of $\omega_0$ is disrupted by the tidal force of SMBH, which is shown by  a purple star symbol. In this range, the averaged values are taken for the stable binaries.}  
\label{fig:a-op_Sch_coplanar_e0.01}
\end{center}
\end{figure}

In Fig.~\ref{fig:a-op_Sch_coplanar_e0.01}, we summarize the final orbital parameters as functions of the semi-major axis.
We plot three quantities: the maximum values (red downward triangles), minimum values (blue upward triangles), and mean values (magenta diamonds).
The boundary between the A and T phases is defined by a 5\% deviation from the mean value.

The maximum, minimum, and mean values of the final eccentricity in the T phase increase as the initial semi-major axis $\mathfrak{a}_0$ increases, i.e., as the binary becomes softer.
On the other hand, the maximum, minimum, and mean values of the final semi-major axis and angular momentum squared decrease as $\mathfrak{a}_0$ increases, which means that the binary loses energy and angular momentum through the scattering.

In the C phase, some binaries are disrupted by the tidal force.
For the surviving binaries, the above trends remain, but the values no longer change systematically.
In particular, for some binaries, the semi-major axis and angular momentum increase after the scattering.
The C phase extends up to $\mathfrak{a}_0=\mathfrak{a}_{\rm cr}\equiv 0.099M$, for which we find
$e_{\rm fin}=0.997$, $\mathfrak{a}_{\rm fin}=0.0161\mathfrak{a}_0$, and $\mathfrak{L}_{\rm fin}^2\approx 0$.
Note that $\mathfrak{a}_{\rm cr}$ is larger than $\mathfrak{a}_{\rm H}$.
Beyond this semi-major axis, the binary is disrupted for all values of $\omega_0$.

Since our calculations were performed at intervals of $10^\circ$ over the range $0^\circ \le \omega_0 \le 180^\circ$, it is possible that stable scatterings for $\mathfrak{a}_0 \gtrsim 0.1M$ may have been missed by fine-tuning $\omega_0$.

\subsubsection{eccentric binary $(e_0=0.5)$}
 We also analyze the eccentric binary with $e_0=0.5$.
 The $\omega_0$-dependence of the orbital parameters are shown in Fig. \ref{fig:om-op_Sch_coplanar_e0.5}.
 The initial semi-major axis are chosen as $\mathfrak{a}_0=
 0.01M,0.015M$ and $0.02M$.
 $\mathfrak{a}_0=
 0.01M$ gives the adiabatic scattering, while 
  $\mathfrak{a}_0=
 0.015M$ and $0.02M$ are classified into the T phase and the C phase, respectively.

 \begin{figure}[htbp]
\begin{center}
\includegraphics[width=5cm]{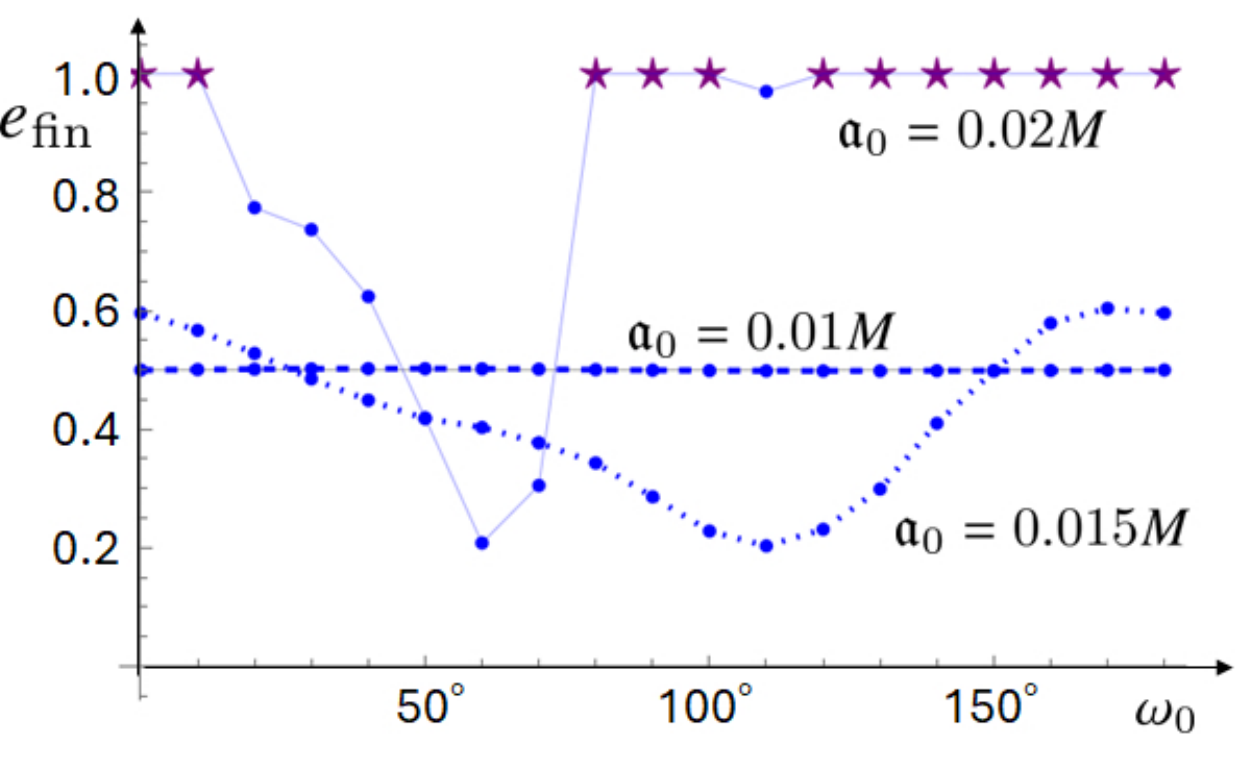}~~
\includegraphics[width=5.cm]{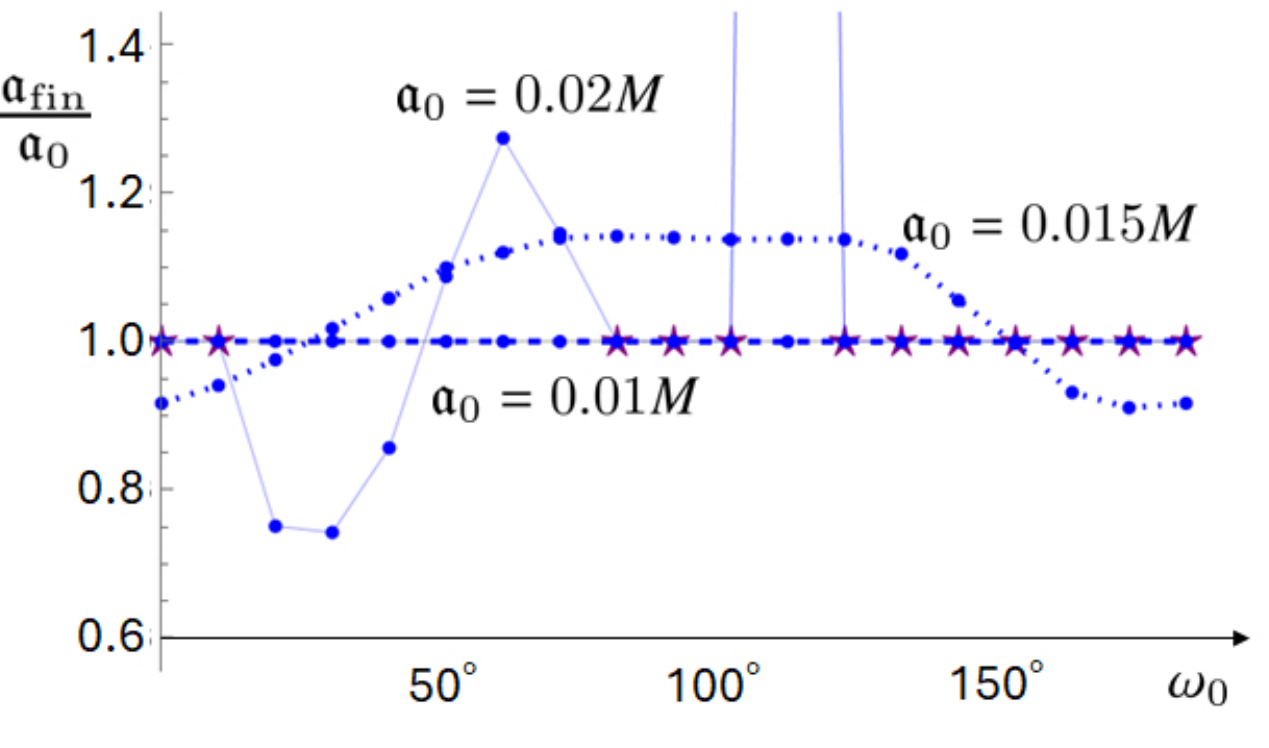}~~
\includegraphics[width=5.cm]{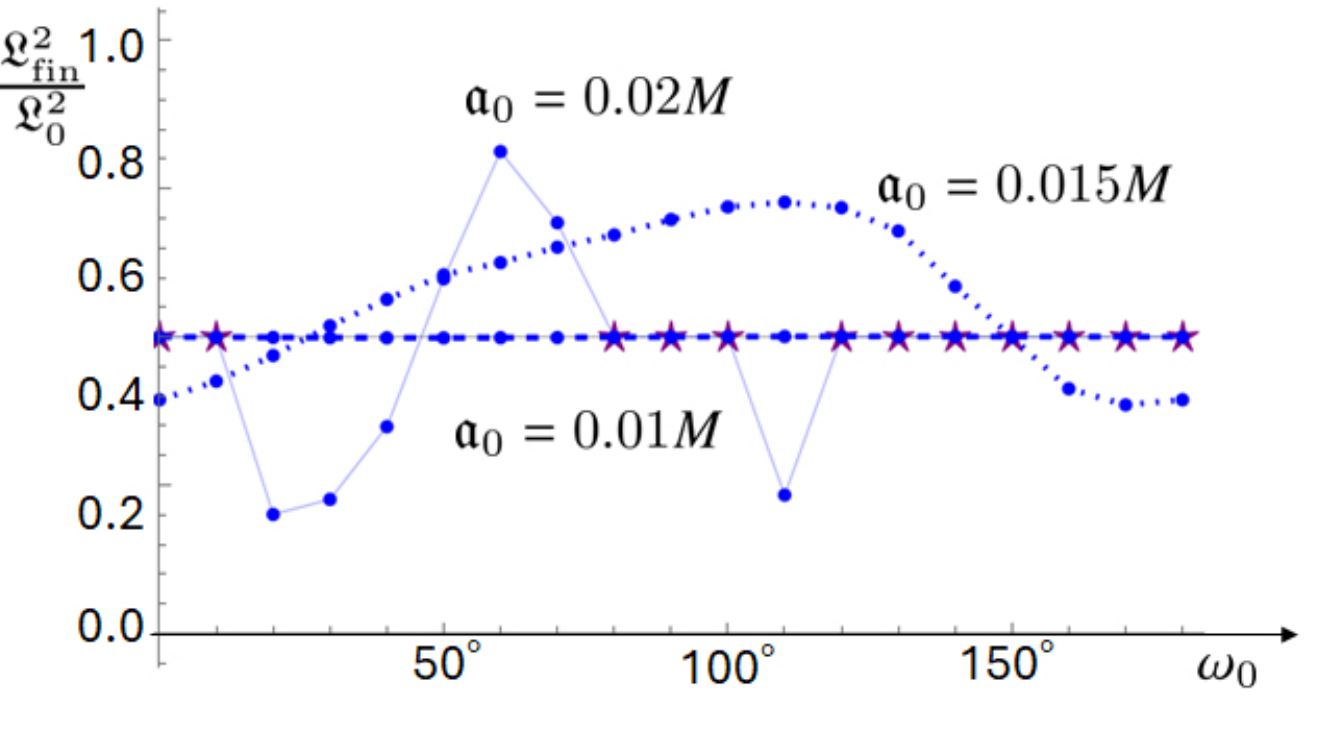}
\\
\hskip 0.5cm (a) final eccentricity \hskip 2cm (b) final semi-major axis \hskip 1.6cm (c) final angular momentum squared
\caption{The same figures as Fig. \ref{fig:om-op_Sch_coplanar_e0.01} for the initial eccentricity $e_0=0.5$.
 }
\label{fig:om-op_Sch_coplanar_e0.5}
\end{center}
\end{figure}

Since the initial eccentricity is not so small, it may increase or 
decrease at the scattering.
The finial semi-major axis and angular momentum are also not
monotonic. They may increase or decrease at the scattering. 
The scattering process is stochastic.

In Fig.~\ref{fig:a-op_Sch_coplanar_e0.5}, we show the dependence of the orbital parameters on $\mathfrak{a}_0$.
Their behavior is similar to that in the case of $e_0=0.01$.
The main difference lies in the ranges of the T and C phases.
The C phase extends over a wider range, whereas the T phase becomes narrower.
This is because a larger value of $e_0$ tends to destabilize the binary for some values of $\omega_0$.
 
\begin{figure}[htbp]
\begin{center}
\includegraphics[width=5.2cm]{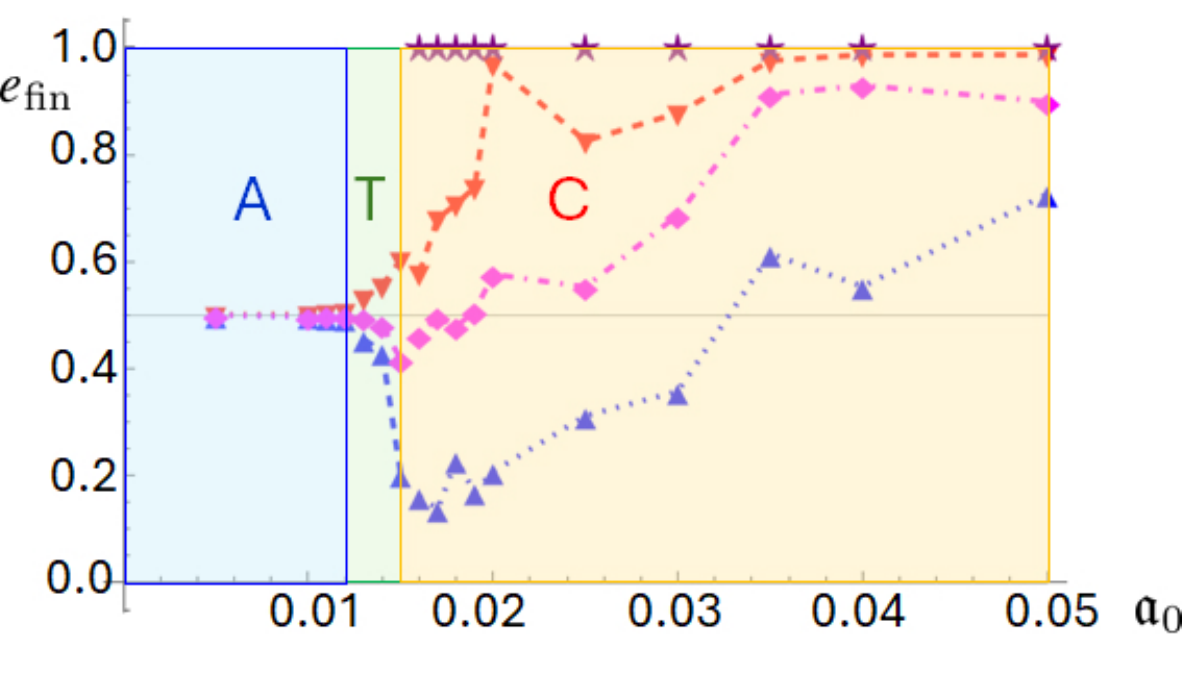}~~
\includegraphics[width=5.2cm]{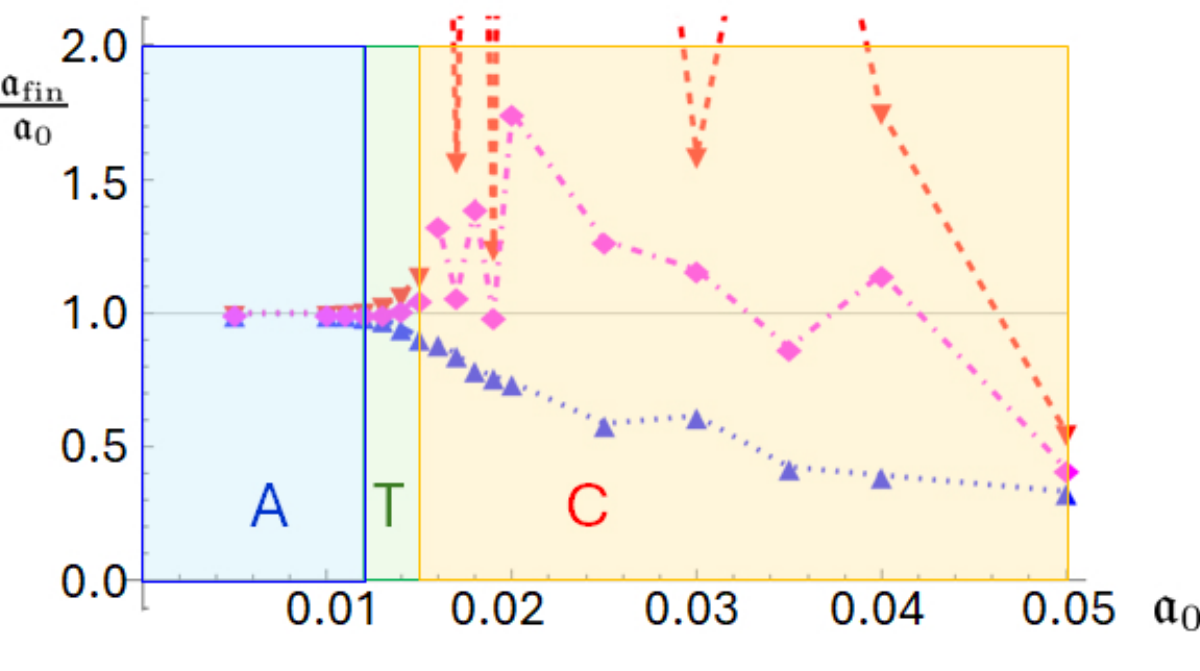}~~
\includegraphics[width=5.2cm]{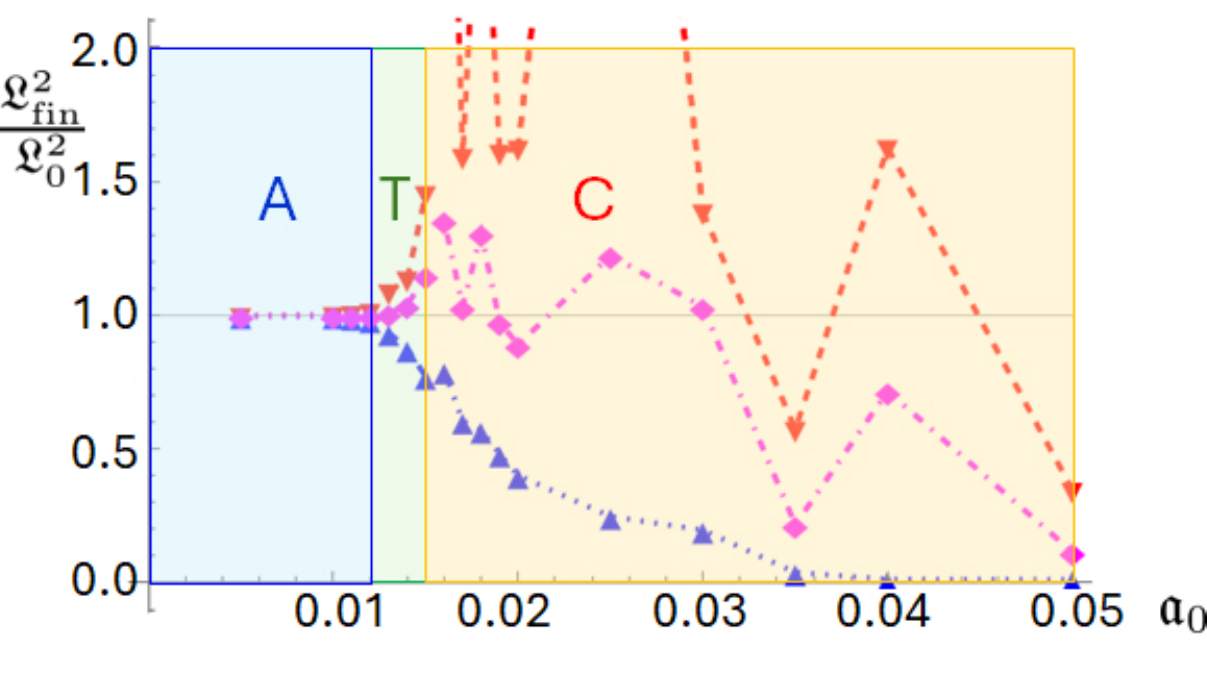}
\\
 \hskip 0.5cm  (a) final eccentricity \hskip 2cm (b) final semi-major axis \hskip 1.8cm (c) final angular momentum squared
\caption{The same figures as Fig. \ref{fig:a-op_Sch_coplanar_e0.01} for the initial eccentricity $e_0=0.5$.
}  
\label{fig:a-op_Sch_coplanar_e0.5}
\end{center}
\end{figure}

\subsubsection{Rotating SMBH $(a=M)$}
\label{rotating_SMBH}
Next, we analyze the rotational effect of the SMBH.
We consider the extreme Kerr black hole ($a = M$).
For the center-of-mass orbital motion, there are two cases: prograde ($L > 0$) and retrograde ($L < 0$) orbits.
To compare the results with the Schwarzschild case ($a = 0$), we analyze orbits with the same periapsis $\mathfrak{r}_p$.
For $\mathfrak{r}_p = 10M$, the angular momentum is
$L = (-1 + 9\sqrt{5})\mu M/4 \approx 4.78115\,\mu M$ for the prograde orbit,
whereas $L = -(1 + 9\sqrt{5})\mu M/4 \approx -5.28115\,\mu M$ for the retrograde orbit.
Note that this differs slightly from the Schwarzschild value $L = \pm 5\mu M$.

In Fig. \ref{fig:om-op_Kerr_coplanar_e0.01}, we show the final orbital parameters in terms of the initial argument of periapsis $\omega_0$
for the initial semi-major axis $\mathfrak{a}_0=0.015M$ and $0.02M$.
We consider a nearly circular binary with an initial eccentricity of $e_0 = 0.01$.
The magenta filled diamond and empty square show the results of the prograde and retrograde orbits, respectively, while the blue circle gives the Schwarzschild case.

\begin{figure}[htbp]
\begin{center}
\includegraphics[width=5cm]{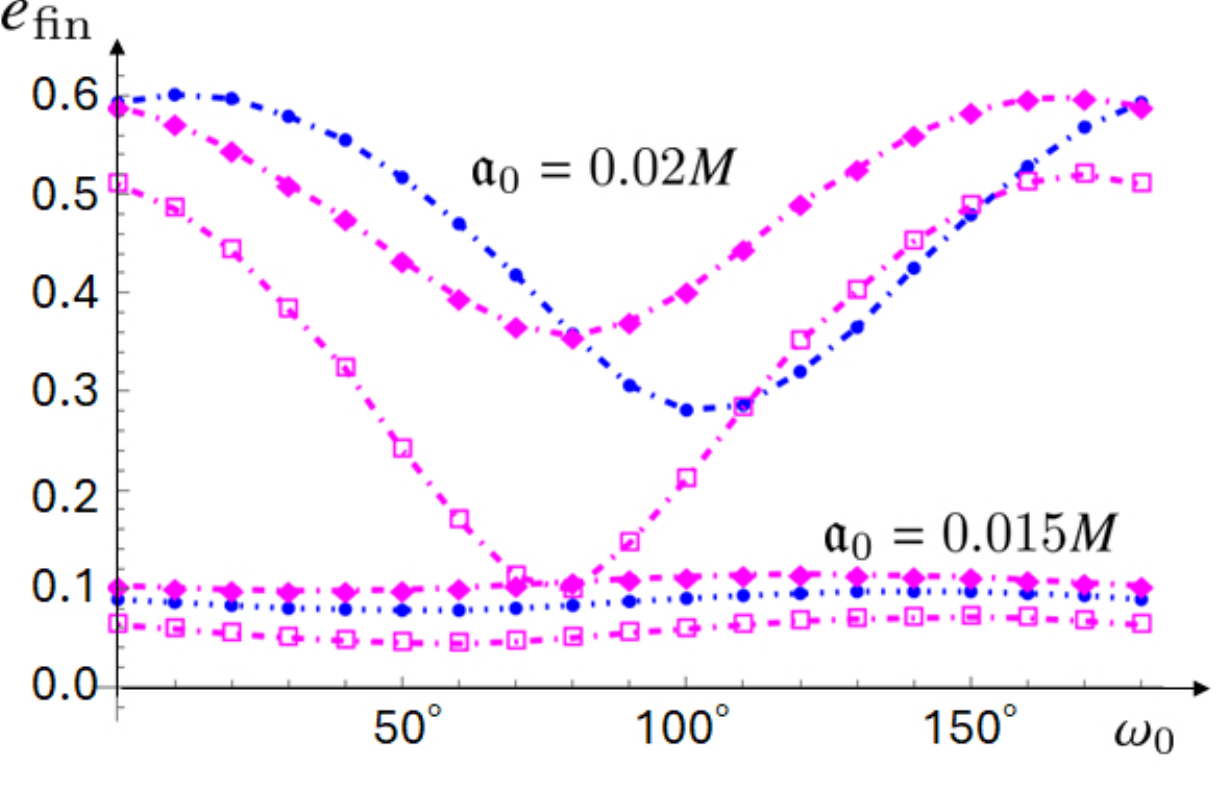}~~
\includegraphics[width=5.cm]{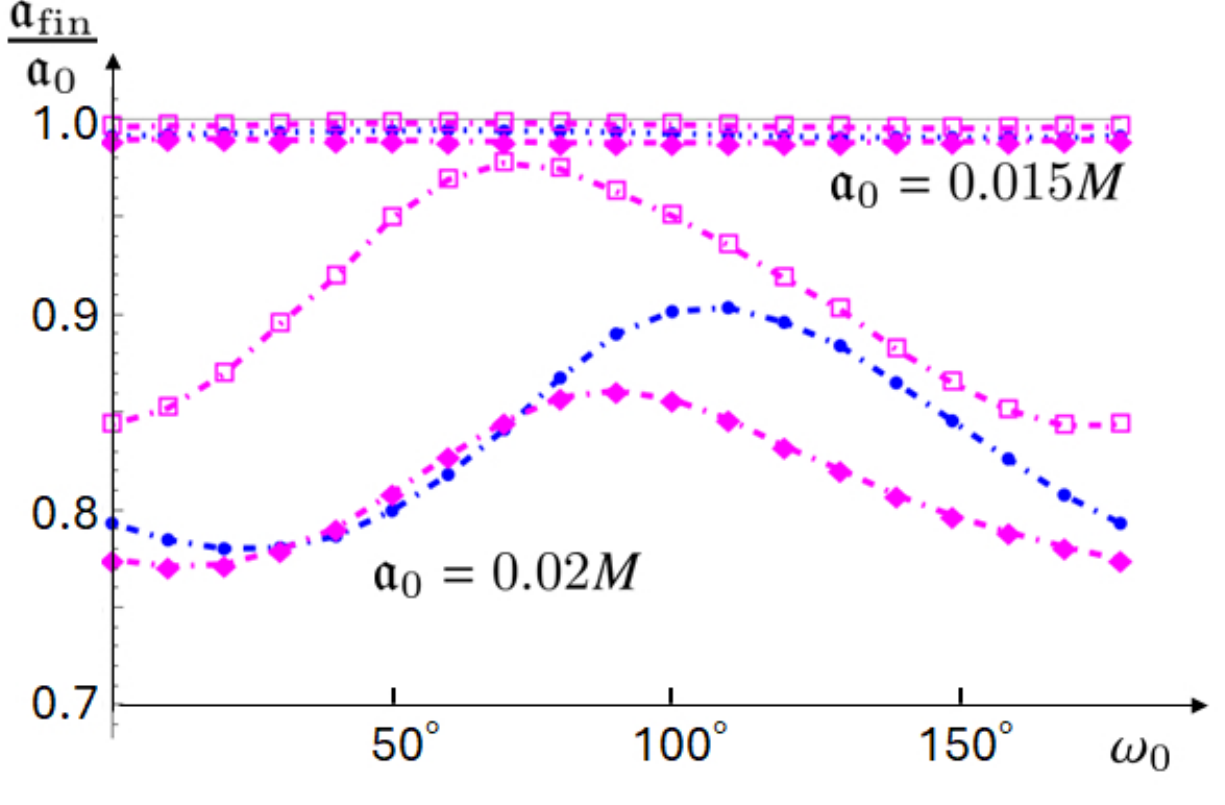}~~
\includegraphics[width=5.cm]{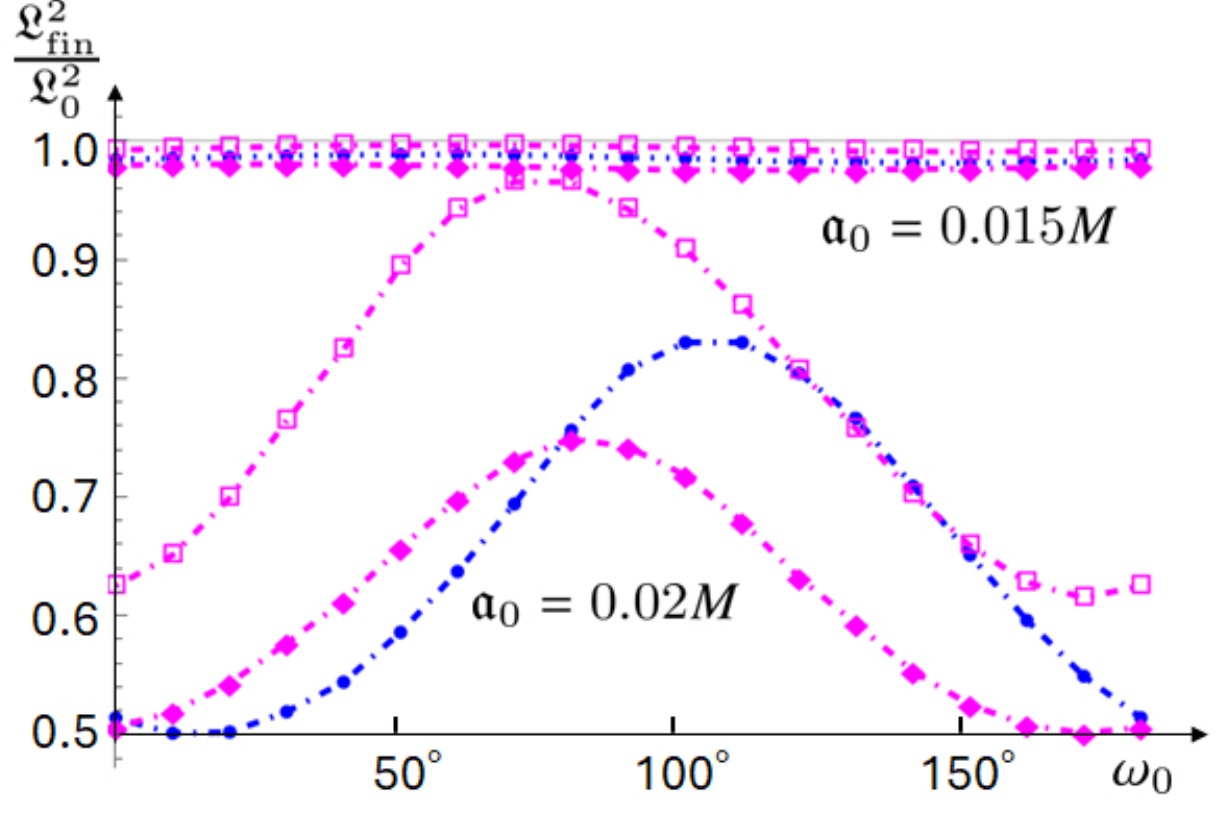}
\\
\hskip 0.5cm (a) final eccentricity \hskip 2cm (b) final semi-major axis \hskip 1.6cm (c) final angular momentum squared
\caption{The comparison of the final orbital parameters between Schwarzschild SMBH ($a=0$) [blue circle]  and the extreme Kerr SMBH ($a=M$) 
[magenta filled diamond (prograde) and magenta empty square (retrograde)]. 
 We show two cases of $\mathfrak{a}_0=0.015M$ and $0.02M$.}
\label{fig:om-op_Kerr_coplanar_e0.01}
\end{center}
\end{figure}

\end{widetext}

From Fig. \ref{fig:om-op_Kerr_coplanar_e0.01}, we find that the eccentricity is more strongly enhanced for prograde orbits, whereas it is suppressed for retrograde orbits. The other orbital parameters show similar behavior. The semi-major axis and the angular momentum decrease more substantially for prograde orbits, whereas the changes are smaller for retrograde orbits.
For eccentric binaries, however, these trends disappear.
For $e_0 = 0.5$, the results show no significant deviation from the Schwarzschild case for either prograde or retrograde orbits in Kerr spacetime.

\begin{figure}[htbp]
\begin{center}
\includegraphics[width=6cm]{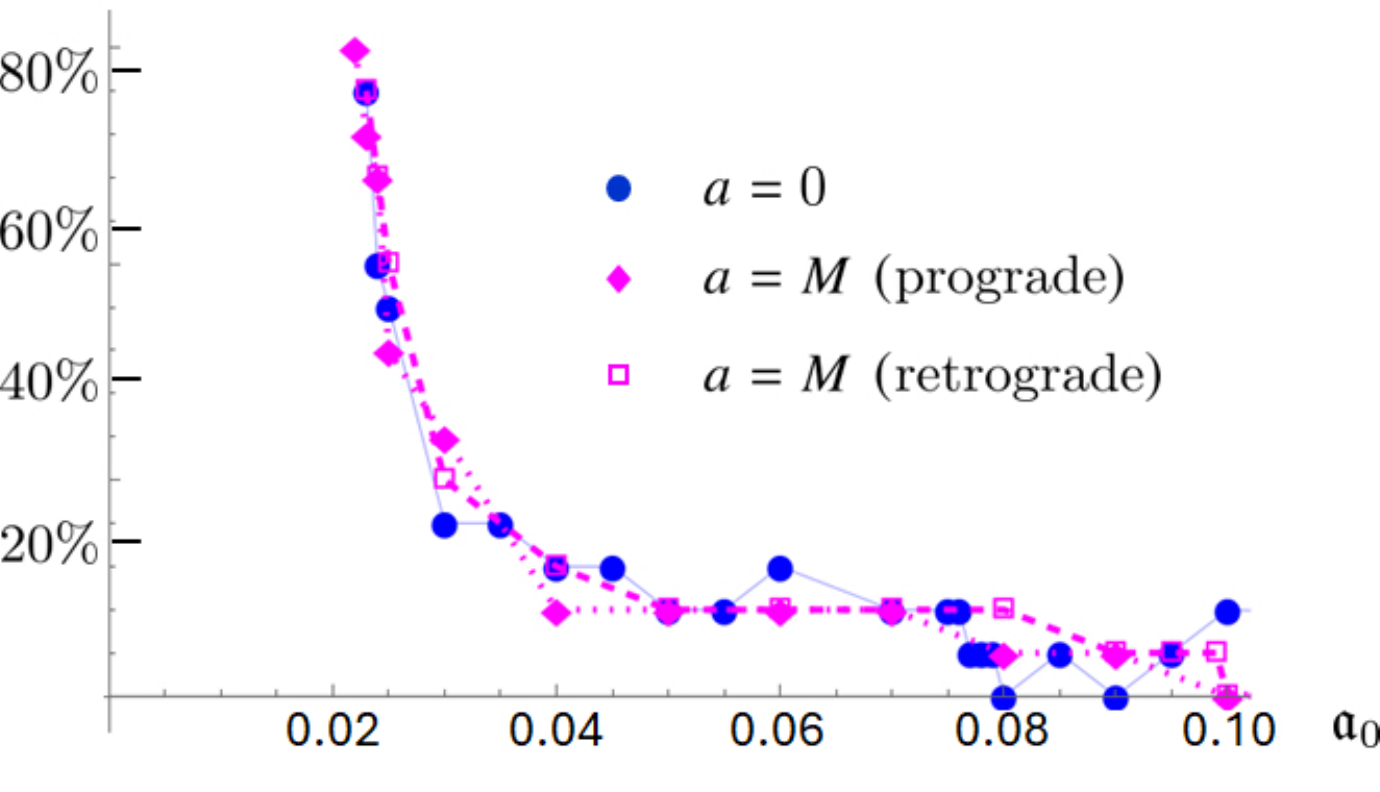}~~
\caption{The  percentage of binaries scattered without disruption
as a function of the semi-major axis $\mathfrak{a}_0$.
The initial eccentricity is $e_0=0.01$.
The blue circles, magenta diamonds and magenta squares denote the case of $a=0$, $a=M$ (prograde) and $a=M$ (retrograde), respectively. }
\label{fig:stability_rate}
\end{center}
\end{figure}

In the chaotic scattering, binaries  with some arguments of periapsis 
are disrupted by the tidal force of SMBH, but the others keep stable.
The proportion of stable binaries depend on the softness of a binary.
In Fig.  \ref{fig:stability_rate}, we show how many percentage of binaries are scattered without disruption
in terms of  the semi-major axis. The proportion decreases rapidly as the semi-major axis increases, 
but it shows a long tail until $\mathfrak{a}_0\approx 0.3$-$0.4 M$.
We also find  the proportions are almost independent of the rotation of SMBH.

\subsection{Parabolic Inclined Binary}

\begin{figure}[htbp]
\begin{center}
\includegraphics[width=6cm]{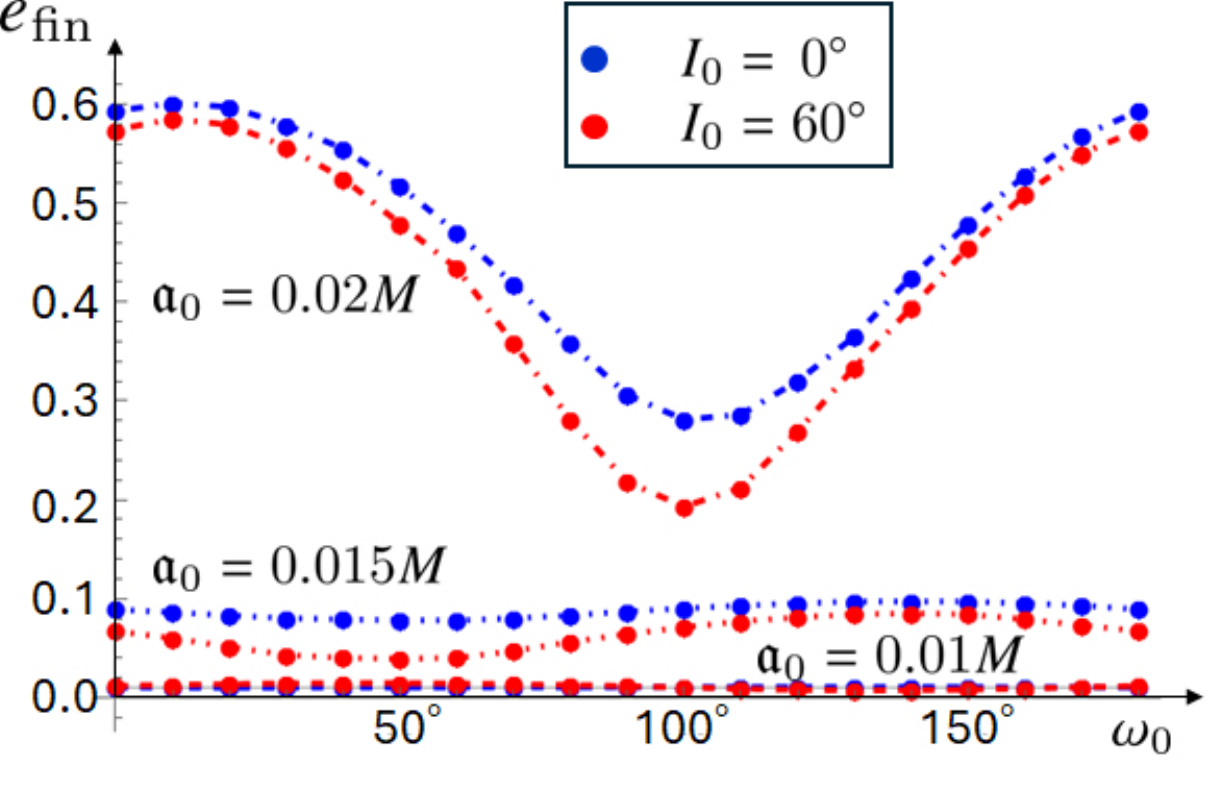}~~
\caption{The final eccentricity for coplanar and inclined binaries in terms of $\omega_0$.
The blue and red circles denote the coplanar and inclined ($I_0=60^\circ$) cases. We choose $a=0$ and $e_0=0.01$, and consider $\mathfrak{a}_0=0.01M, 0.015M$ and $0.02M$.}
\label{fig:om_inclination}
\end{center}
\end{figure}

Fig. \ref{fig:om_inclination}, we compare the final eccentricities in terms of $\omega_0$ 
 in two cases of $I_0=0^\circ$ 
(blue) and $I_0=60^\circ$ (red) with the same other parameters 
($a=0, e_0=0.01$ and $\mathfrak{a}_0=0.01M, 0.015M$ and $0.02M$).
We find the values with $I_0=0^\circ$ are always larger than those with $I_0=60^\circ$
although they still depend highly on $\omega_0$.
This is because the tidal force acts more effectively in the coplanar case than in the inclined binary.

\begin{figure}[htbp]
\begin{center}
\includegraphics[width=6cm]{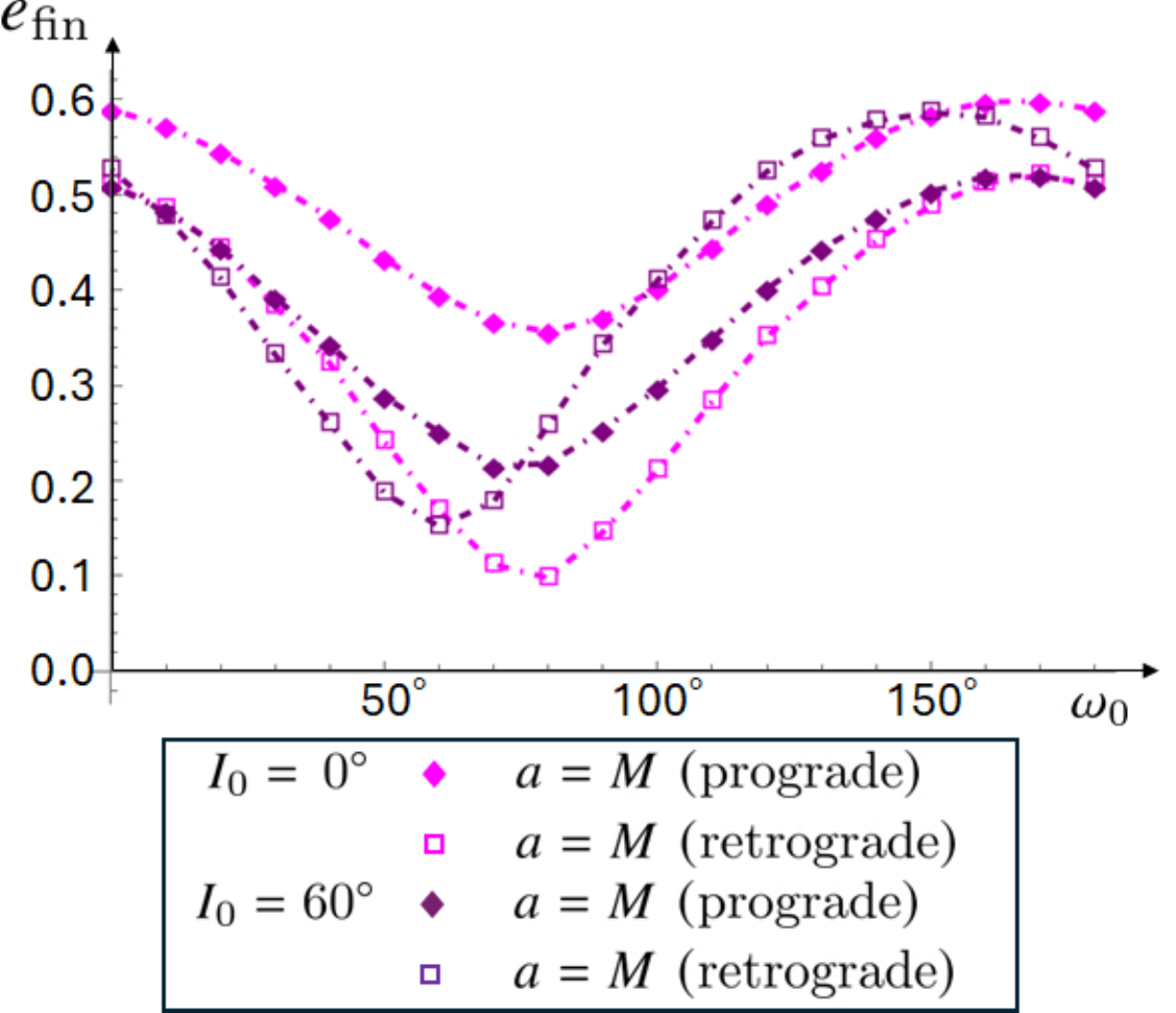}~~
\caption{The same figure for the extreme Kerr black hole ($a = M$) as in the Schwarzschild case shown in Fig.~\ref{fig:om_inclination}.
The diamonds and open squares denote the prograde and retrograde binaries, respectively, while the magenta and purple symbols represent the coplanar and inclined ($I_0 = 60^\circ$) cases, respectively.
We choose $\mathfrak{a}_0 = 0.02M$.}
\label{fig:om_inclination_Kerr}
\end{center}
\end{figure}

It is also true for the rotating SMBH as shown in  Fig. \ref{fig:om_inclination_Kerr}.
Here, we compare the final eccentricities in terms of $\omega_0$ 
 in two cases of $I_0=0^\circ$ 
(magenta) and $I_0=60^\circ$ (purple) with the same other parameters 
($a=M, e_0=0.01$ and $\mathfrak{a}_0=0.02M$).
The filled diamond shows the prograde orbits, while the empty square gives the retrograde orbits.
The enhancement of the eccentricity for the inclined binary is always less effective. 

As discussed in \S\ref{rotating_SMBH},
the eccentricity in the coplanar case is enhanced for prograde orbits,
whereas it is suppressed for inclined binaries.
We therefore compare the following two cases:
a coplanar binary around a Schwarzschild SMBH ($a = 0$)
and an inclined binary around an extreme Kerr SMBH ($a = M$).

\begin{figure}[htbp]
\begin{center}
\includegraphics[width=6cm]{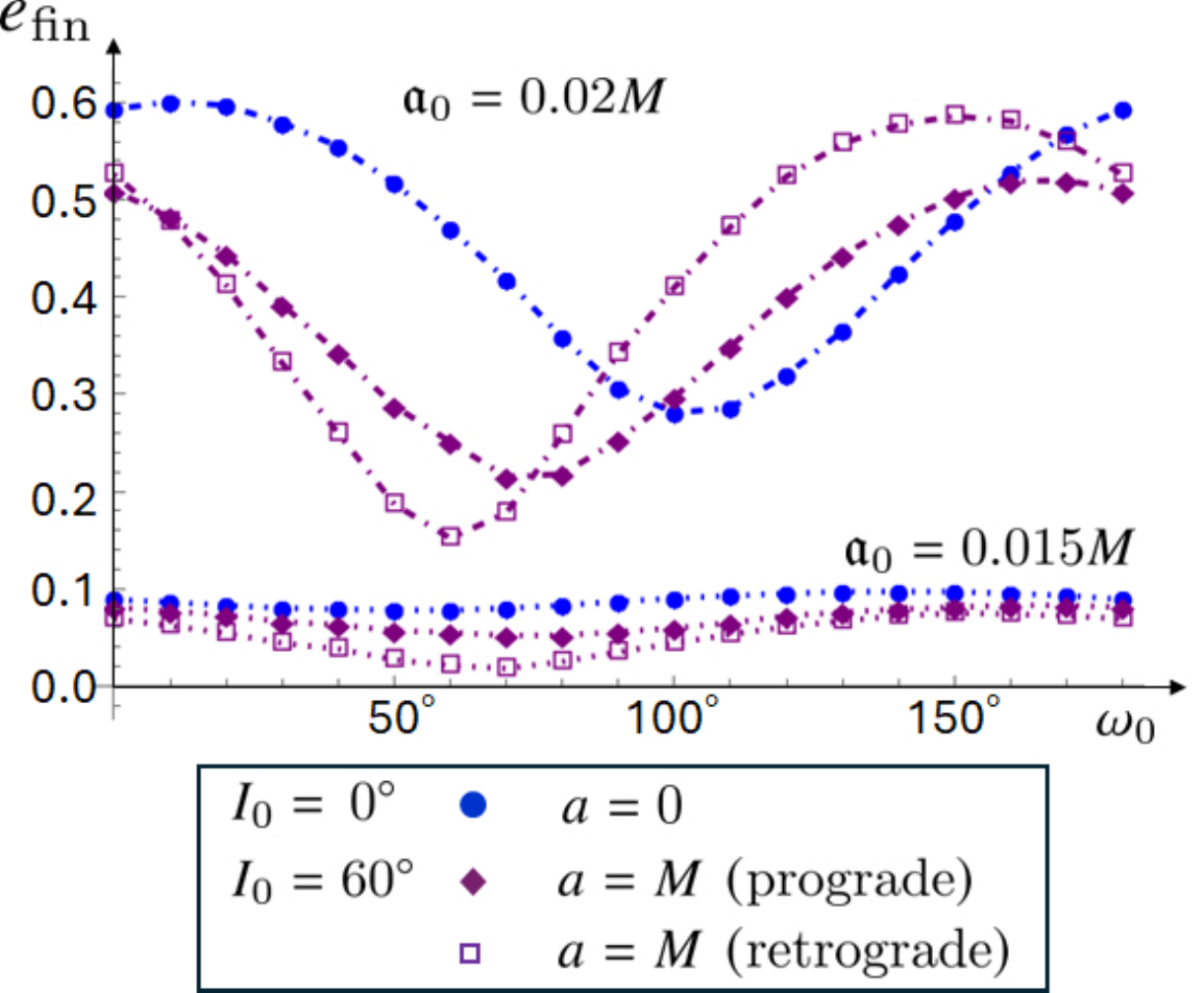}~~
\caption{Comparison between the coplanar binary in a Schwarzschild black hole
($a = 0$) and the inclined prograde and retrograde binaries
($I_0 = 60^\circ$) in an extreme Kerr black hole ($a = M$).
We choose $e_0 = 0.01$ and
$\mathfrak{a}_0 = 0.015M$ and $0.02M$.}
\label{fig:Kerr_inclined}
\end{center}
\end{figure}

In Fig.~\ref{fig:Kerr_inclined},
we adopt the parameters
$a = 0$, $I_0 = 0^\circ$, and
$a = M$, $I_0 = 60^\circ$,
while keeping the other parameters fixed
($e_0 = 0.01$ and $\mathfrak{a}_0 =0.015M,  0.02M$).
From this figure, we find that the rotational effect is not very significant compared with the effect of binary inclination.

\subsection{Hyperbolic Binary}

As mentioned above, the typical velocity of a binary system orbiting an SMBH is usually non-relativistic. Therefore, the energy of the binary system, $E$, is approximately equal to its rest-mass energy:
$E \approx m$.
As a result, the scattering of a binary system on a hyperbolic orbit is nearly identical to that in the parabolic case.

\begin{widetext}

\begin{figure}[htbp]
\begin{center}
\includegraphics[width=5cm]{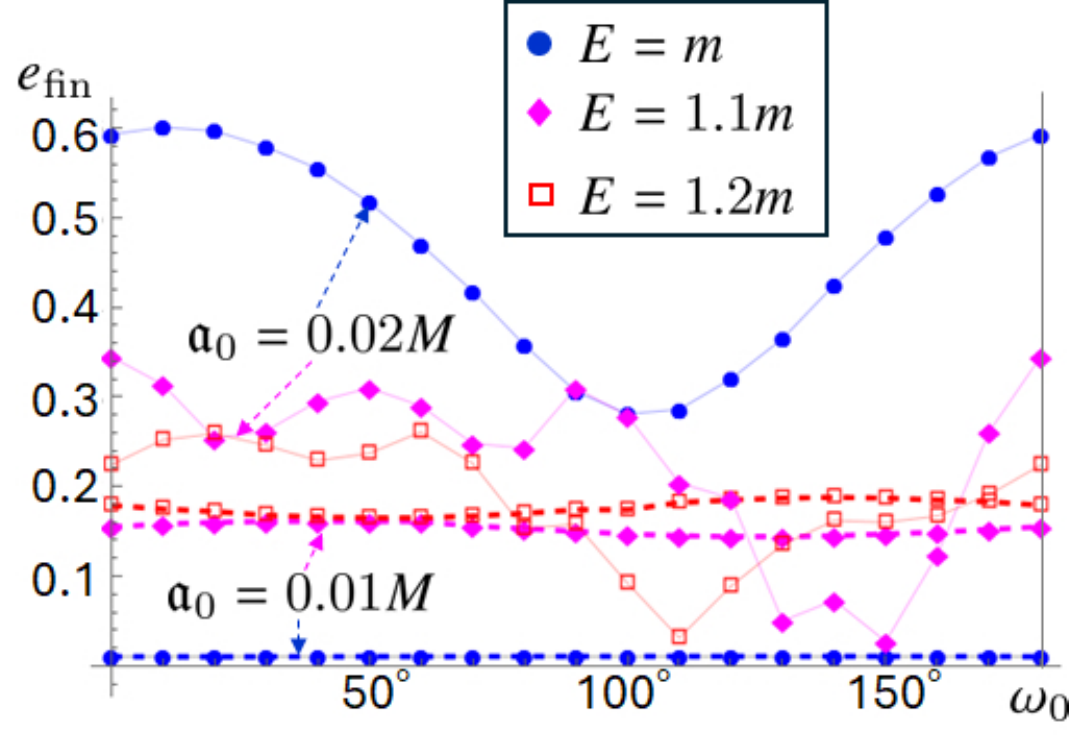}~~
\includegraphics[width=5.cm]{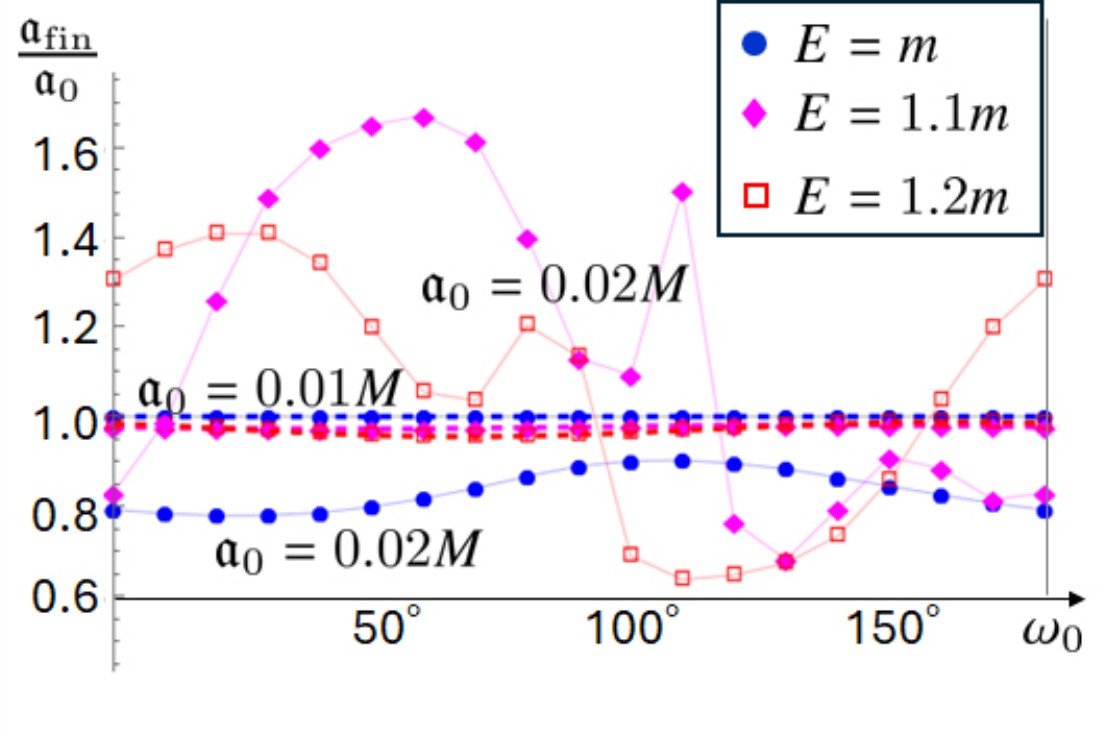}~~
\includegraphics[width=5.cm]{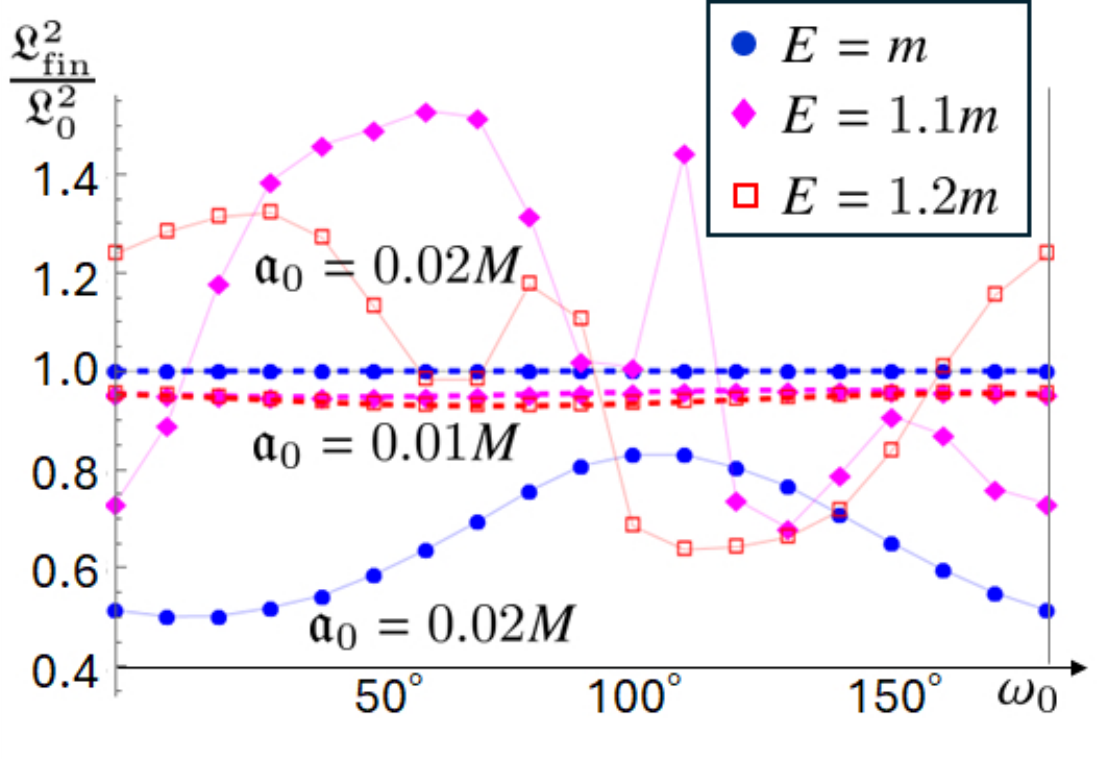}
\\
\hskip 0.5cm (a) final eccentricity \hskip 2cm (b) final semi-major axis \hskip 1.6cm (c) final angular momentum squared
\caption{Comparison between the parabolic ($E = m$) and hyperbolic ($E = 1.1m,\ 1.2m$) binaries for a Schwarzschild black hole ($a = 0$).
We show the final eccentricity (a), semi-major axis (b), and angular momentum squared (c) for $\mathfrak{a}_0 = 0.01M$ and $0.02M$.}
\label{fig:om-op_hyperbolic}
\end{center}
\end{figure}

\end{widetext}

We analyze two cases: $E = 1.1m$ and $E = 1.2m$.
The results are shown in Fig.~\ref{fig:om-op_hyperbolic}.
For a hard binary ($\mathfrak{a}_0 = 0.01M$),
not only the eccentricity but also the semi-major axis and angular momentum
are enhanced in the hyperbolic case, although the difference between
$E = 1.1m$ and $E = 1.2m$ is not very large.
On the other hand, for a soft binary ($\mathfrak{a}_0 = 0.02M$),
the enhancement in the hyperbolic case is not as significant as in the parabolic case.
Instead, the main difference appears in the $\omega_0$ dependence.
In the parabolic case, the dependence is rather smooth, whereas in the hyperbolic case it becomes chaotic.

\section{Eccentric  \textsc{v}ZLK Oscillations}

When a bound binary moves approximately on an elliptical orbit around the SMBH, we find the von Zeipel–Lidov–Kozai (vZLK) mechanism, in which oscillations between the inclination angle $I$ and the eccentricity $e$ of the binary are induced by the tidal force from the tertiary body \cite{vonZeipel10,Lidov62,Kozai62}.
Here we focus on binaries orbiting on eccentric trajectories and discuss several characteristic features that differ from the case moving on a circular orbit \cite{Maeda:2023tao,Maeda:2023uyx}.

\subsection{Schwarzschild SMBH ($a=0$)}

We first present the results for a Schwarzschild SMBH ($a = 0$).

\begin{figure}[h]
\begin{center}
\includegraphics[width=6.5cm]{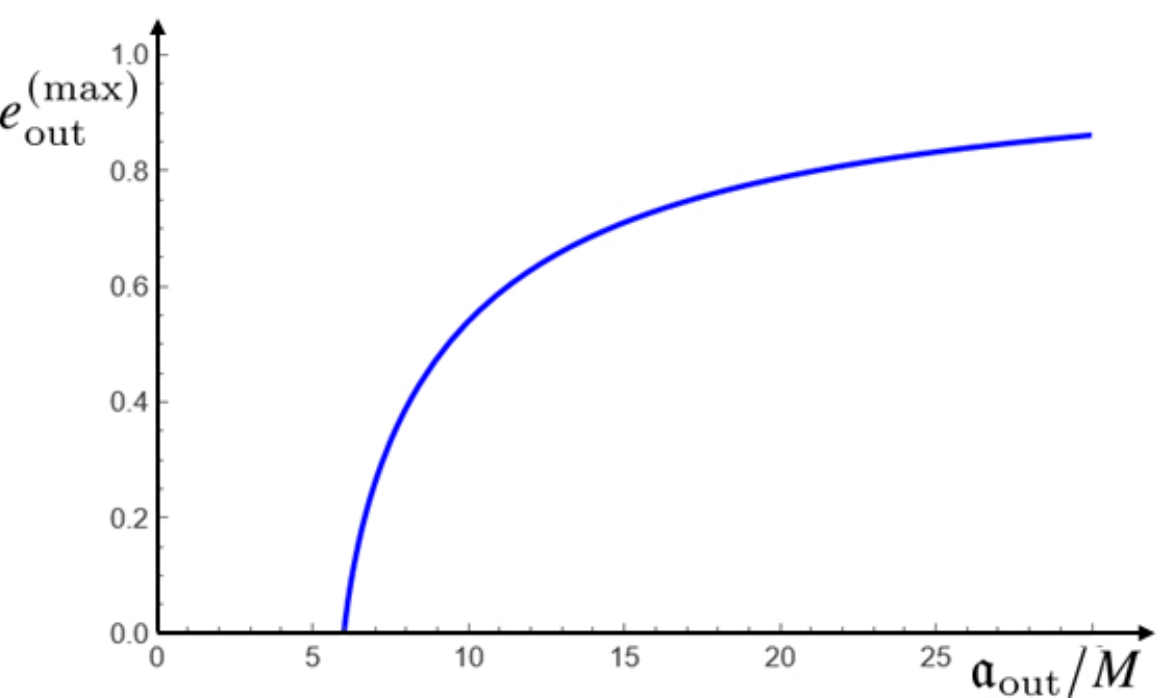}
\caption{Maximum value of the eccentricity, $e_{\rm out}^{\rm (max)}$, for fixed $\mathfrak{a}_{\rm out}$ in the Schwarzschild black hole case.}
\label{fig:emax_r0}
\end{center}
\end{figure}

The center-of-mass orbit is characterized by the semi-major axis $\mathfrak{a}_{\rm out}$ and the eccentricity $e_{\rm out}$.
For a given $\mathfrak{a}_{\rm out}$,
there exists a maximum value of the eccentricity,
$e_{\rm out}^{\rm (max)}$, given by
\beann
e_{\rm out}^{\rm (max)}=
{-1 + \sqrt{1 - 6\mathfrak{a}_{\rm out} + \mathfrak{a}_{\rm out}^2}\over \mathfrak{a}_{\rm out}}
\,,
\enann
which is shown in Fig.~\ref{fig:emax_r0}.
The value of $e_{\rm out}^{\rm (max)}$ vanishes at
$\mathfrak{a}_{\rm out} = 6M$,
which corresponds to the ISCO radius in Schwarzschild spacetime.

\begin{figure}[htbp]
\begin{center}
\includegraphics[width=5.5cm]{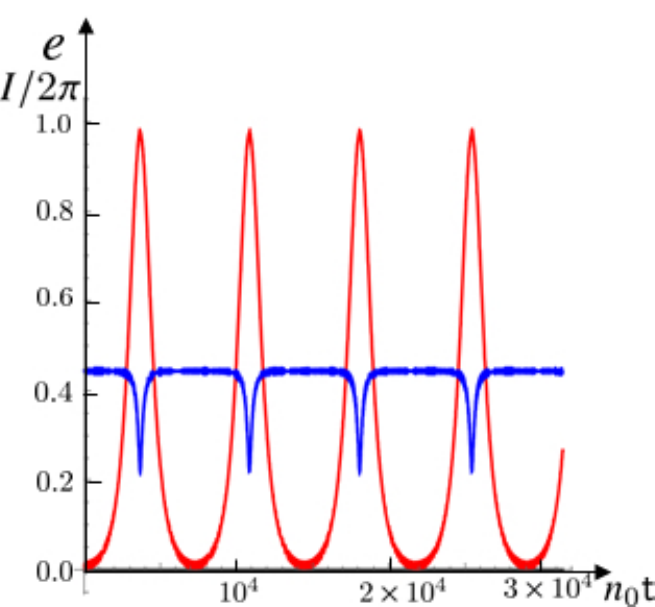}\\
(a) $I_0=80^\circ$
\\
\includegraphics[width=5.5cm]{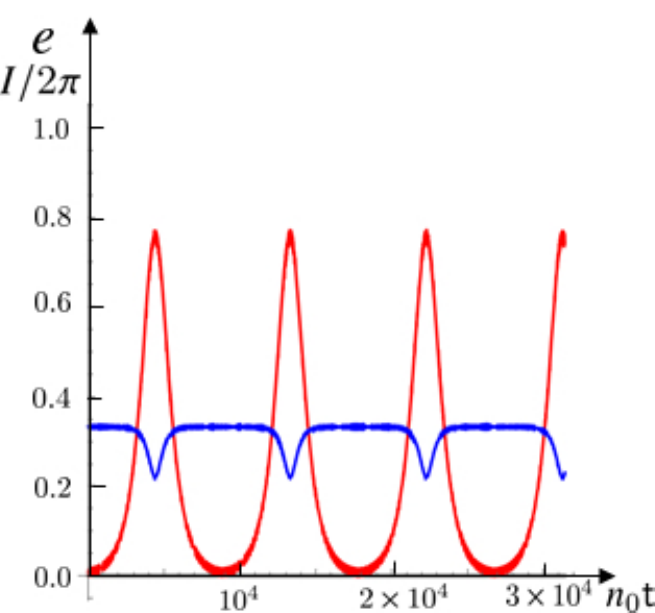}\\
(b) $I_0=60^\circ$
\\
\includegraphics[width=5.5cm]{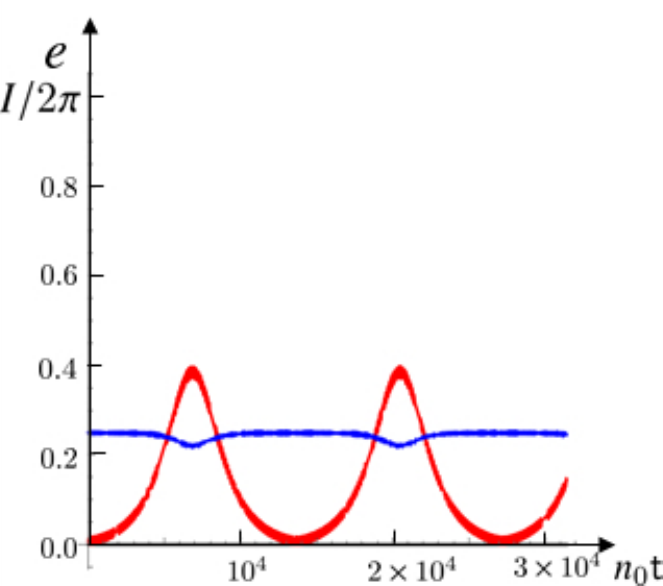}\\
(c) $I_0=45^\circ$
\caption{
Time evolution of the eccentricity $e$ (red curves) and inclination angle $I$ (blue curves) for the case of a circular orbit ($e_{\rm out}=0$) around a Schwarzschild SMBH ($a=0$).
We choose the orbital parameters as $\mathfrak{a}_{\rm out}=10M$ for the center-of-mass orbit, and $e_0=0.01$, $\mathfrak{a}_0=0.005M$, $\omega_0=0$, and $\Omega_0=0$ for the binary orbit. The initial inclination angles are taken to be $I_0=80^\circ$ in panel (a), $60^\circ$ in panel (b), and $45^\circ$ in panel (c).
}
\label{fig:vZLK_eout=0}
\end{center}
\end{figure}

In order to compare the present results with those for a circular orbit, we first present the case of a circular orbit around a Schwarzschild SMBH ($a=0$) in Fig.~\ref{fig:vZLK_eout=0}.
We choose the orbital parameters as $e_{\rm out}=0$ and $\mathfrak{a}_{\rm out}=10M$ for the center-of-mass orbit, and $e_0=0.01$, $\mathfrak{a}_0=0.005M$, and $I_0=80^\circ$, $60^\circ$, and $45^\circ$ for the binary orbit (The red dot in Fig. \ref{fig:parameter_range}). We also fix the other orbital parameters to $\omega_0=0$ and $\Omega_0=0$ throughout this section.
We find clear vZLK oscillations. The maximum eccentricities reach $0.97$, $0.76$, and $0.39$ for $I_0=80^\circ$, $60^\circ$, and $45^\circ$, respectively, while the corresponding vZLK oscillation periods are approximately $1200,P_0$, $1400,P_0$, and $2200,P_0$, where $P_0=2\pi/n_0$ is the binary orbital period.
These results confirm the expected tendency: as the initial inclination $I_0$ decreases, the oscillation period becomes longer and the maximum eccentricity becomes smaller\cite{Maeda:2023tao,Maeda:2023uyx}.

\begin{figure}[htbp]
\begin{center}
\includegraphics[width=6cm]{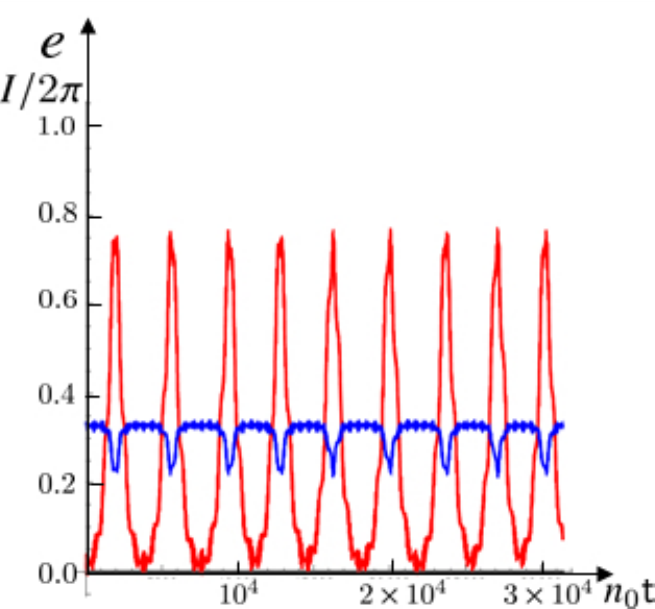}\\
(a)
\\
\includegraphics[width=5.5cm]{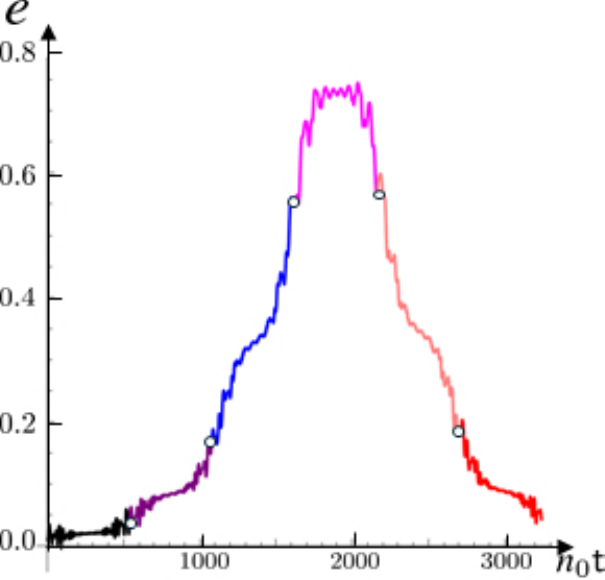}
\\
(b)
\caption{(a) Time evolution of the eccentricity $e$ (red curve) and inclination angle $I$ (blue curve) 
for the case of an eccentric orbit ($e_{\rm out}=0.5$). The other orbital parameters are the same as those in Fig. \ref{fig:vZLK_eout=0}(b).\\
(b) Evolution of the eccentricity $e$ during a single vZLK cycle. Each period of the center-of-mass orbit is indicated by a different color (black, purple, blue, pink, orange, and red), and the small circles denote the periapsis passages. }
\label{fig:vZLK_eout=0.5}
\end{center}
\end{figure}

In Fig.~\ref{fig:vZLK_eout=0.5}(a), we show the case with $e_{\rm out} = 0.5$. The other orbital parameters are the same as those in Fig.~\ref{fig:vZLK_eout=0}(b): $\mathfrak{a}_{\rm out} = 10M$ for the center-of-mass orbit, and $e_0 = 0.01$, $\mathfrak{a}_0 = 0.005M$, and $I_0 = 60^\circ$ for the inner binary. We observe vZLK oscillations with a period of $570\,P_0$, which is approximately half that of the circular orbit case ($e_{\rm out} = 0$). The maximum eccentricity is $0.75$, nearly the same as in the circular case. The oscillation profile appears to be slightly irregular.

To examine the oscillations in more detail, we show a single vZLK cycle in Fig.~\ref{fig:vZLK_eout=0.5}(b). Each period of the center-of-mass orbit is indicated by a different color (black, purple, blue, pink, orange, and red), and the small circles denote the periapsis passages. We find that the eccentricity increases when the center-of-mass approaches periapsis. This behavior can be understood in terms of binary scattering: when the binary is far from periapsis, the enhancement is weak, whereas as it approaches periapsis, the eccentricity is strongly amplified, similar to the binary scattering discussed in the previous section. As a result, the eccentricity evolves in a step-like manner.

\begin{figure}[htbp]
\begin{center}
\includegraphics[width=6.cm]{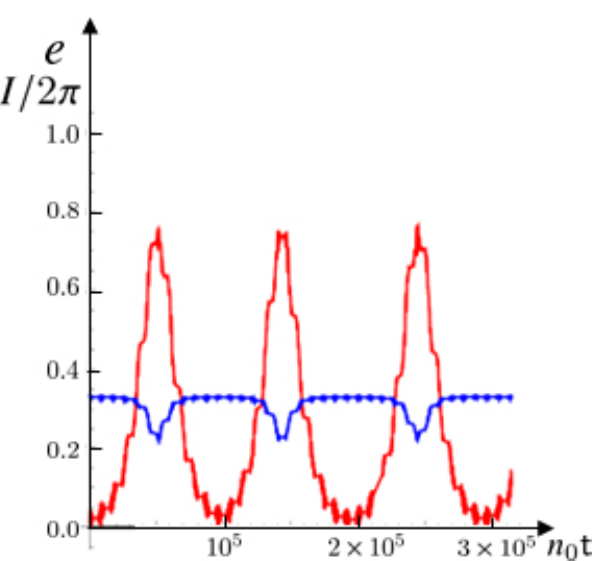}\\
(a)
\\
\includegraphics[width=5.5cm]{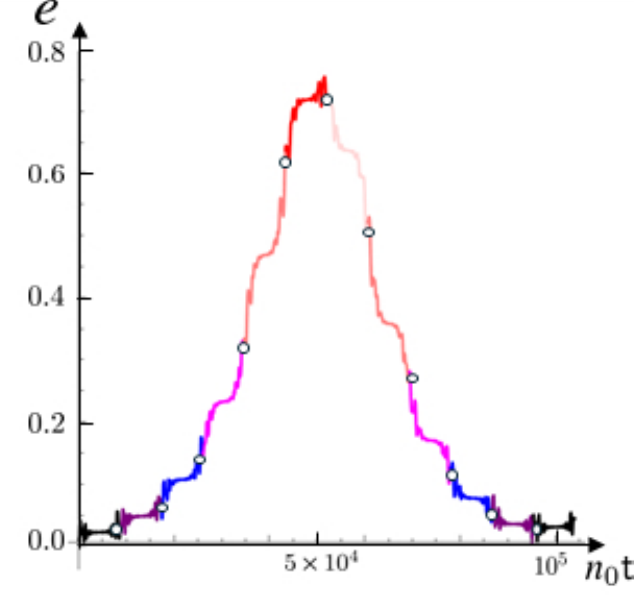}\\
(b)
\caption{(a) Time evolution of the eccentricity $e$ (red curve) and inclination angle $I$ (blue curve) 
for the case of an eccentric orbit ($e_{\rm out}=0.9$). The other orbital parameters are the same as those in the previous model.
(b) Evolution of the eccentricity $e$ during a single vZLK cycle. The other details are the same as those in Fig. \ref{fig:vZLK_eout=0.5}(b).}
 \label{fig:vZLK_eout=0.9}
\end{center}
\end{figure}

The enhancement near the periapsis becomes more 
pronounced for a highly eccentric orbit.
As an example, we show the case with $e_{\rm out}=0.9$ in Fig.~\ref{fig:vZLK_eout=0.9}.
To realize such a highly eccentric orbit, we choose the semi-major axis of the center-of-mass orbit as $\mathfrak{a}_{\rm out}=50M$.
The other orbital parameters are taken to be the same as in the previous model, namely,
$\mathfrak{a}_0=0.005M$, $e_0=0.01$, and $I_0=60^\circ$.
We again find clear vZLK oscillations.
The oscillation timescale is about $15{,}000\,P_0$, which is much longer because of the larger semi-major axis.
The maximum eccentricity reaches approximately $0.75$, which is nearly the same as in the previous cases.

We also show the time evolution of the eccentricity during 
a single vZLK cycle in Fig.~\ref{fig:vZLK_eout=0.9}(b). 
The eccentricity increases or decreases clearly when the center-of-mass approaches periapsis. 
When the binary is far from periapsis, the enhancement is weak, whereas as it approaches periapsis, the eccentricity is strongly amplified or reduced, similar to the binary scattering. 
The eccentricity evolves with multiple steps.

\begin{figure}[htbp]
\begin{center}
\includegraphics[width=6cm]{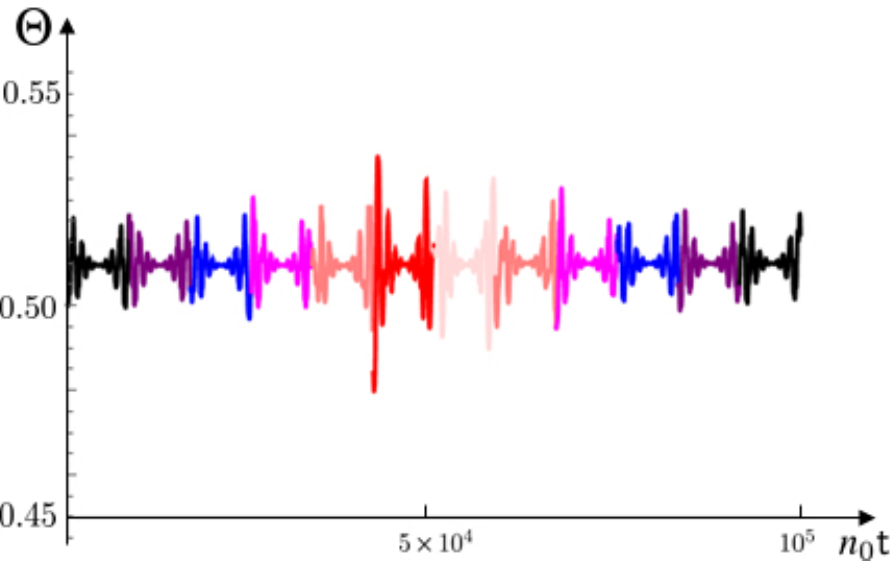}
\caption{Single vZLK cycle evolution of $\Theta\equiv \sqrt{1-e^2}\cos I$. 
Each period of the center-of-mass orbit is indicated by a different color.
We adopt the same  parameters as Fig. \ref{fig:vZLK_eout=0.9}.}
\label{fig:Lz_e=0.9_I=60}
\end{center}
\end{figure}

In Fig.~\ref{fig:Lz_e=0.9_I=60}, we also show the ``conserved'' quantity
$\Theta \equiv \sqrt{1-e^2}\cos I$ during the vZLK oscillation, which is proportional to the $z$-component of the angular momentum in the non-rotating local inertial frame. Although its value deviates slightly (by at most $\sim 10\%$) near periapsis, its mean value remains nearly constant, ensuring the exchange between the eccentricity and the mutual inclination.

Although the oscillation profile differs from that of the standard vZLK oscillation, it still exhibits the essential feature of vZLK dynamics, namely, the exchange between the eccentricity and the mutual inclination while approximately conserving the $z$-component of the angular momentum. Hence, this oscillation can be regarded as a new type of vZLK oscillation. We may refer to it as a {\it scattering vZLK mechanism}.

As shown in \cite{Maeda:2023tao,Maeda:2023uyx}, one of the notable features of vZLK oscillations in the circular orbit is that a chaotic aspect appears when the binary is soft.
A similar chaotic behavior is found for the eccentric orbit, even in cases where regular vZLK oscillations appear in the circular case.

\begin{figure}[htbp]
\begin{center}
\includegraphics[width=6.cm]{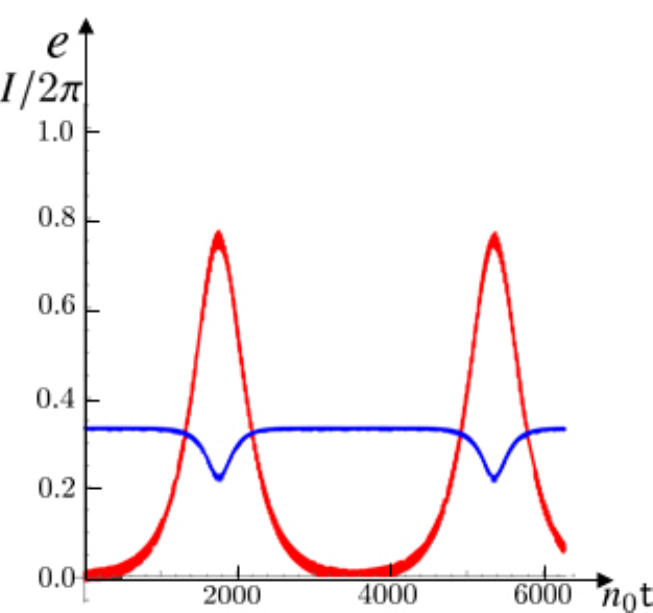}
\\
(a) $e_{\rm out}=0$
\\
\includegraphics[width=6.cm]{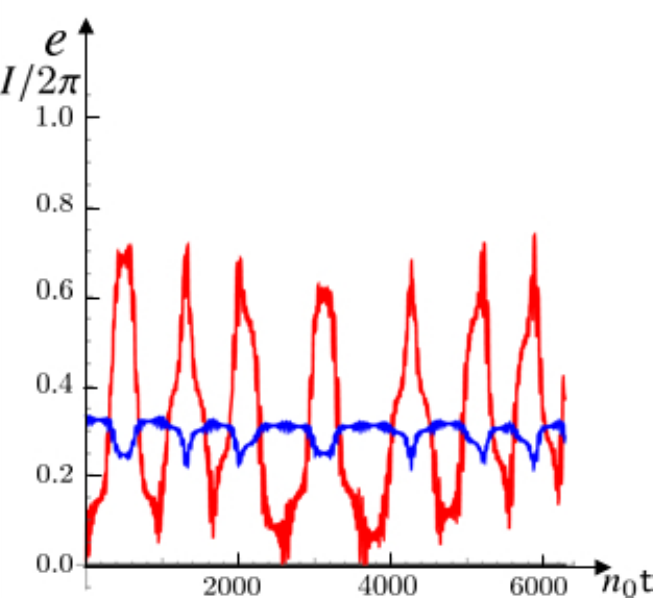}
\\
(b) $e_{\rm out}=0.5$

\caption{Time evolution of the eccentricity $e$ (red curve) and inclination angle $I$ (blue curve) 
for the case of $\mathfrak{a}_0=0.007M$. Panel (a) shows the case with $e_{\rm out}=0$, while panel (b) corresponds to $e_{\rm out}=0.5$.  The other orbital parameters are the same as those in the previous model.}
\label{fig:vZLK_I=60_a=0.007}
\end{center}
\end{figure}

We consider a model with parameters $\mathfrak{a}_{\rm out}=10M$ for the center-of-mass orbit, and $e_0=0.01$, $\mathfrak{a}_0=0.007M$, and $I_0=60^\circ$ for the inner binary.
For the circular orbit ($e_{\rm out}=0$), we still find regular vZLK oscillations, as shown in Fig.~\ref{fig:vZLK_I=60_a=0.007}(a).
The oscillations for the circular orbit are quite regular.
The maximum eccentricity is $e_{\rm max}\approx 0.759$, and the vZLK oscillation period is $P_{\rm vZLK}\approx 135\,P_{\rm out}\approx 551\,P_{\rm in}$.

\begin{figure}[htbp]

\begin{center}
%\includegraphics[width=5cm]{vZLK_e=0_I=60_a=0.007_2.pdf}
%\\
%(a) $e_{\rm out}=0$
%\\
\includegraphics[width=5cm]{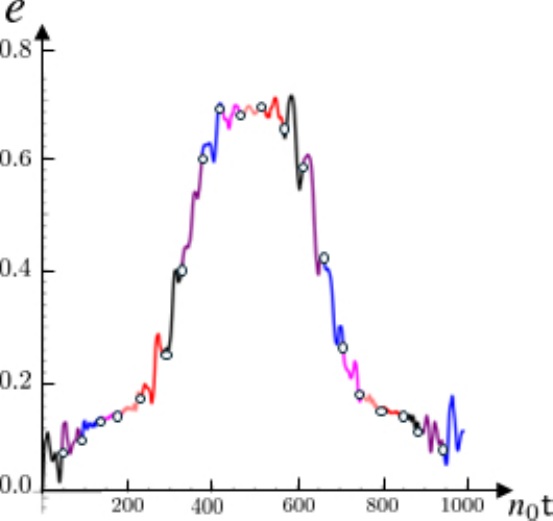}
\\
(a) first vZLK cycle
\\
\includegraphics[width=5cm]{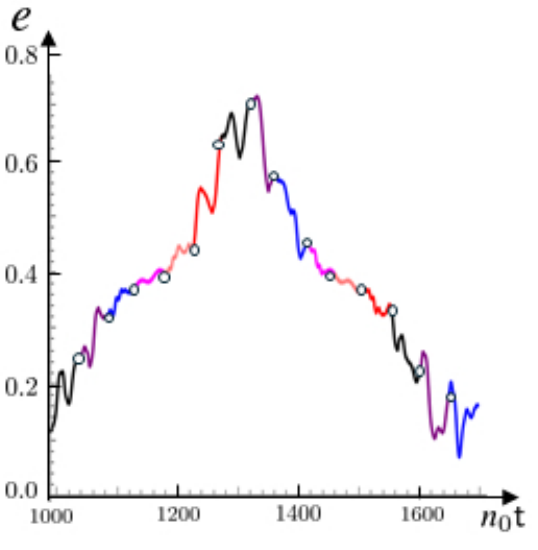}
\\
(b) second vZLK cycle

\caption{Evolution of the eccentricity $e$ during the first (a) and second (b) vZLK cycles for the model with $\mathfrak{a}_0=0.007M$ and $e_{\rm out}=0.5$. 
The other details are the same as those in Fig. \ref{fig:vZLK_eout=0.5}(b).}
\label{fig:vZLK_I=60_a=0.007_2}
\end{center}
\end{figure}

While for the eccentric orbit ($e_{\rm out}=0.5$), the oscillations become chaotic, as seen in Fig.~\ref{fig:vZLK_I=60_a=0.007}(b).
We also show the evolution of the eccentricity $e$ during a single vZLK cycle in Fig.~\ref{fig:vZLK_I=60_a=0.007_2}.
The behavior for the eccentric orbit looks similar to Figs.~\ref{fig:vZLK_eout=0.5}(b) and \ref{fig:vZLK_eout=0.9}(b), but the profile of the first cycle (Fig.~\ref{fig:vZLK_I=60_a=0.007_2}(a)) differs from that of the second one (Fig.~\ref{fig:vZLK_I=60_a=0.007_2}(b)).
The maximum eccentricity and the vZLK oscillation period are $e_{\rm max}\approx 0.706$ and $P_{\rm vZLK}\approx 21\,P_{\rm out}\approx 158\,P_{\rm in}$ for the first cycle, and $e_{\rm max}\approx 0.712$ and $P_{\rm vZLK}\approx 15\,P_{\rm out}\approx 113\,P_{\rm in}$ for the second cycle, respectively.

\begin{figure}[htbp]

\begin{center}
\includegraphics[width=6cm]{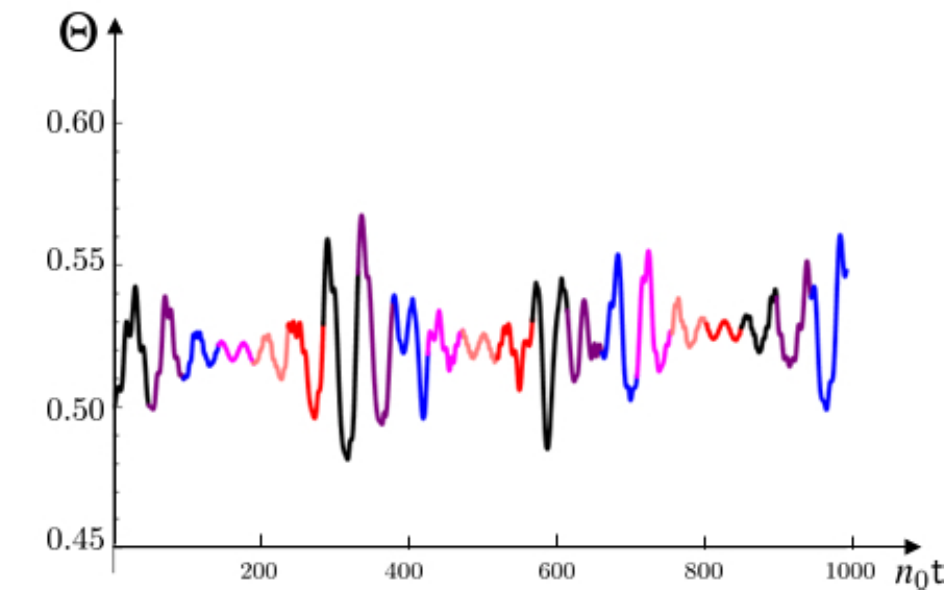}

\caption{Evolution of the ``conserved'' quantity $\Theta$ during the 
first vZLK cycle. 
Each period of the center-of-mass orbit is indicated by a different color.
The adopted parameters are the same as those in Fig. \ref{fig:vZLK_I=60_a=0.007}(b).}
\label{fig:Lz_e=0.5_I=60_a=0.007}
\end{center}
\end{figure}

The evolution of the ``conserved'' quantity  $\Theta$ is shown in Fig.~\ref{fig:Lz_e=0.5_I=60_a=0.007}.
Although its averaged value is nearly conserved, the deviations are relatively large.
We may refer to this behavior as a {\it chaotic scattering vZLK oscillation}.

%%%%%%%%%%%%%%%%%%%%%%%%%%%%%%%%%%%%%%%%%%%%%%%%%
%%%%%%%%%%%%%%%%%%%%%%%%%%%%%%%%%%%%%%%%%%%%%%%%%
\subsection{Kerr SMBH}
%%%%%%%%%%%%%%%%%%%%%%%%%%%%%%%%%%%%%%%%%%%%%%%%%
%%%%%%%%%%%%%%%%%%%%%%%%%%%%%%%%%%%%%%%%%%%%%%%%%
In this subsection, we discuss the case of a Kerr SMBH.
To examine the effect of the rotation parameter $a$, we fix the semi-major axis 
$\mathfrak{a}_{\rm out}$ and 
the eccentricity $e_{\rm out}$.
There exists an upper bound on the eccentricity, denoted by $e_{\rm out}^{\rm (max)}$,
 which is obtained numerically.
This bound also vanishes at the ISCO radius.

\begin{figure}[htbp]

\begin{center}
\includegraphics[width=7.5cm]{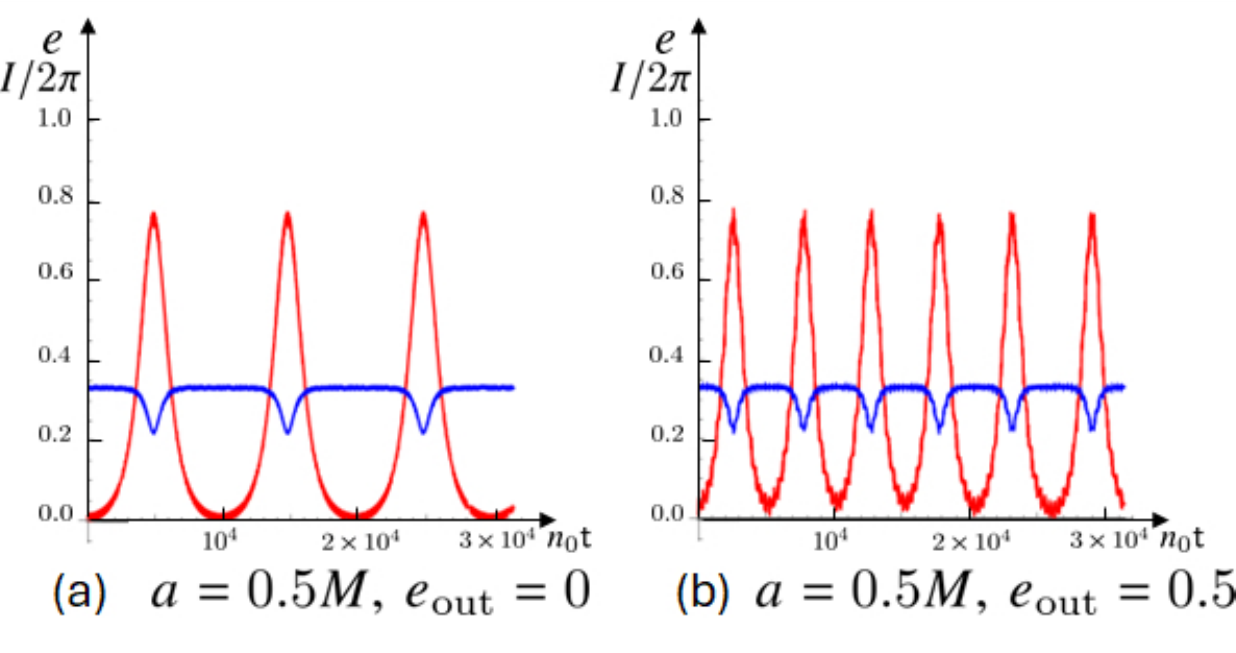}
\includegraphics[width=7.5cm]{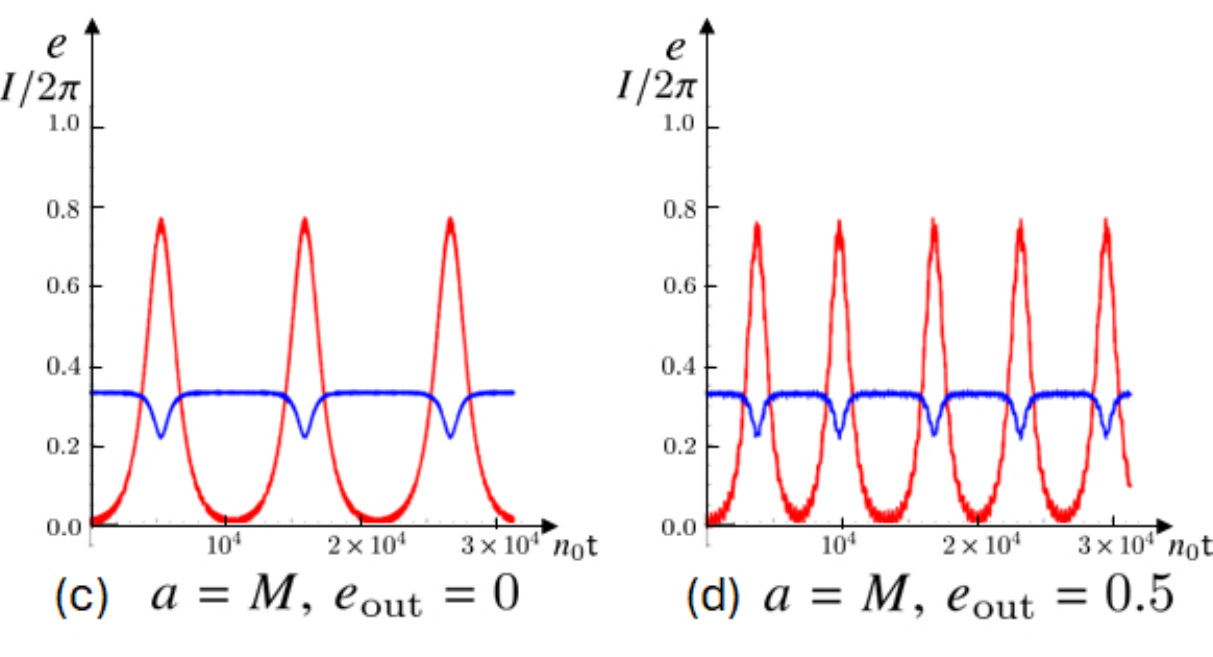}

\caption{Time evolution of the eccentricity $e$ (red curve) and inclination angle $I$ (blue curve) for the cases of
(a) $a=0.5M$, $e_{\rm out}=0$,
(b) $a=0.5M$, $e_{\rm out}=0.5$,
(c) $a=M$, $e_{\rm out}=0$, and
(d) $a=M$, $e_{\rm out}=0.5$.
The orbital parameters are chosen as $\mathfrak{a}_{\rm out}=10M$ for the center-of-mass orbit, and $\mathfrak{a}_0=0.005M$, $e_0=0.01$, and $I_0=60^\circ$ for the inner binary.}
\label{fig:vZLK_I=60}
\end{center}
\end{figure}

We choose the same parameters as in the Schwarzschild case, i.e.,
$\mathfrak{a}_{\rm out}=10M\,, e_{\rm out}=0.5$ 
for the center-of-mass orbit and $\mathfrak{a}_0=0.005M\,,e_0=0.01\,,I_0=60^\circ\,,\omega_0=0\,,\Omega_0=0$ for the inner binary orbit.
In Fig. \ref{fig:vZLK_I=60}, we show the results for $a=0.5M$ and $a=M$.
The vZLK oscillation period increases both as the outer-orbit eccentricity $e_{\rm out}$ decreases and as the SMBH spin parameter $a$ increases.

\begin{figure}[htbp]
\begin{center}
\includegraphics[width=7.5cm]{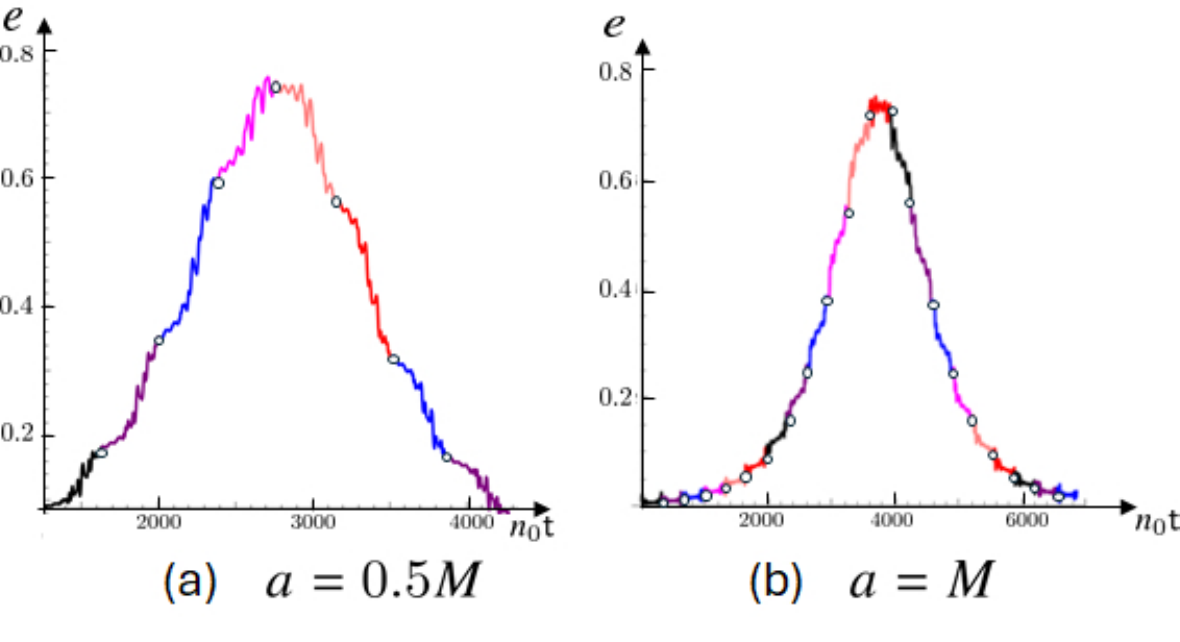}

\caption{Evolution of the eccentricity $e$ during a single vZLK cycle for the models shown in Fig.~\ref{fig:vZLK_I=60}(b) and (d), respectively.}
\label{fig:vZLK_eout=0.5_I=60}
\end{center}
\end{figure}

Apart from the period, the profiles in these four cases
($a=0.5M,\, M$ and $e_{\rm out}=0,\,0.5$) 
 appear similar, but their details differ.
For circular orbits ($e_{\rm out}=0$), the vZLK oscillations are quite regular, 
whereas for eccentric orbits($e_{\rm out}=0.5$), the profiles resemble the scattering-type vZLK oscillations in the Schwarzschild case, as shown in Fig. \ref{fig:vZLK_eout=0.5_I=60}.

\begin{figure}[htbp]
\begin{center}
\includegraphics[width=7.5cm]{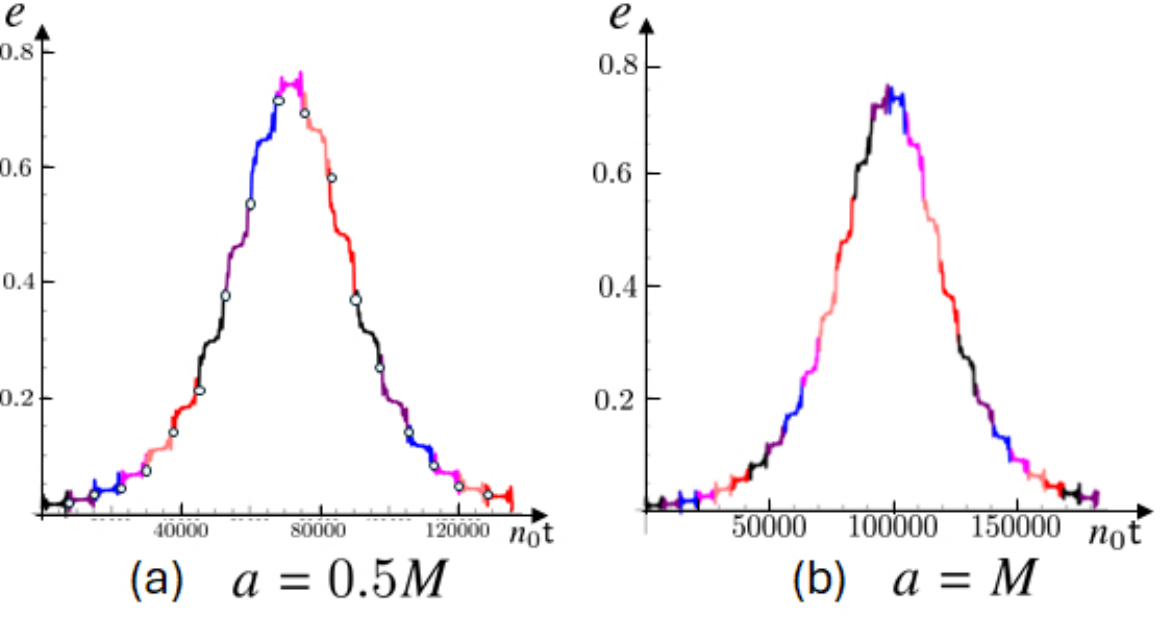}

\caption{Evolution of the eccentricity $e$ during a single vZLK cycle for the models with $e_{\rm out}=0.9$ and $\mathfrak{a}_{\rm out}=50M$ for the center-of-mass orbit, and $\mathfrak{a}_0=0.005M$, $e_0=0.01$, and $I_0=60^\circ$ for the inner binary.
Panel (a) shows the case with $a=0.5M$, while panel (b) corresponds to $a=M$. }
\label{fig:vZLK_eout=0.9_I=60_2}
\end{center}
\end{figure}

Since the step-like evolution of the eccentricity in Fig.~\ref{fig:vZLK_eout=0.5_I=60} is not very clear, we also analyze a more eccentric center-of-mass orbit with $e_{\rm out}=0.9$.
One vZLK cycle for the cases $a=0.5M$ and $a=M$ is shown in Fig.~\ref{fig:vZLK_eout=0.9_I=60_2}.

For $a=0.5M$, we find a vZLK oscillation period and maximum eccentricity of
$P_{\rm vZLK}\approx 18\,P_{\rm out}\approx 21{,}417\,P_{\rm in}$ and
$e_{\rm max}\approx 0.759$, respectively.
For $a=M$, the corresponding values are
$P_{\rm vZLK}\approx 27\,P_{\rm out}\approx 30{,}117\,P_{\rm in}$ and
$e_{\rm max}\approx 0.757$.

The step-like profiles are clearly observed in both cases.
We therefore confirm the existence of scattering-type vZLK oscillations also for a Kerr SMBH. Although we do not present the corresponding figures here, we expect chaotic scattering vZLK oscillations to occur for softer binary systems.

%\begin{figure}[htbp]
%\begin{center}
%\includegraphics[width=7cm]{vZLK_eout=0.9_I=60.pdf}

%\caption{$e_{\rm out}=0.9 ,\, \mathfrak{a}_{\rm out}=50M,\, 
%\mathfrak{a}_0=0.005M\,, e_0=0.01\,,I_0=60^\circ$}
%\label{fig:vZLK_eout=0.9_I=60}
%\end{center}
%\end{figure}

\section{Summary and Discussion}

In this paper, we have investigated the dynamics of a binary system orbiting a SMBH, focusing on two complementary phenomena: binary scattering in unbound orbits and eccentric vZLK oscillations in bound orbits. Our analysis is based on the equations of motion derived in a local inertial frame in Kerr spacetime, where the tidal effects of the SMBH are incorporated through the Riemann curvature tensor.

For unbound orbits, we classified the outcomes of binary–SMBH encounters into four categories: adiabatic (A), tidally affected (T), chaotic (C), and disruptive (D) scattering. We find that the binary behavior depends strongly on its initial semi-major axis and the argument of periapsis. As the binary becomes softer, the tidal interaction becomes stronger, leading to larger changes in the orbital parameters and eventually to chaotic scattering or tidal disruption. In the chaotic regime, the final orbital parameters are highly sensitive to the initial conditions, indicating the presence of strong nonlinear effects.

We also examined the effects of SMBH rotation. For nearly circular binaries, we find that eccentricity excitation is enhanced for prograde orbits and suppressed for retrograde orbits, reflecting the influence of frame dragging. However, this difference becomes less significant for initially eccentric binaries, where the tidal interaction dominates over rotational effects.

For bound orbits, we studied vZLK oscillations when the center-of-mass orbit is eccentric. In contrast to the standard vZLK mechanism, the oscillation profile changes on a dynamical timescale and exhibits qualitatively different features. In particular, we find that the eccentricity evolves in a step-like manner, with significant changes occurring near the periapsis of the center-of-mass orbit. This behavior can be understood as a sequence of scattering-like interactions between the binary and the SMBH during successive periapsis passages.

We have shown that these oscillations still preserve the essential property of the vZLK mechanism, namely the approximate conservation of the quantity
$\Theta\equiv \sqrt{1-e^2}\cos I$,
which corresponds to the $z$-component of the angular momentum in the local inertial frame. Although this quantity exhibits deviations near periapsis, its average value remains nearly constant, indicating that the oscillation is driven by the exchange between eccentricity and inclination. We refer to this behavior as the scattering vZLK mechanism, since it combines features of both vZLK dynamics and binary scattering.

Furthermore, we find that eccentric vZLK oscillations can exhibit chaotic behavior, especially for softer binaries. Even in cases where the corresponding circular-orbit configuration shows regular oscillations, introducing eccentricity into the center-of-mass orbit can lead to irregular, cycle-dependent behavior. We refer to this regime as chaotic scattering vZLK oscillations.

Although the vZLK oscillation period increases as the eccentricity of the outer orbit $e_{\rm out}$ decreases and as the SMBH spin parameter $a$ increases, the inclusion of black hole rotation does not qualitatively change the above features. This is because the Riemann curvature components of Kerr spacetime on the equatorial plane are identical to those of Schwarzschild spacetime.

Our results suggest that binaries in galactic nuclei may experience a rich variety of dynamical behaviors due to repeated strong tidal interactions with the central SMBH. The excitation of high eccentricities through scattering or vZLK oscillations could play an important role in enhancing the merger rate of compact binaries and in shaping their gravitational-wave signals. These effects—particularly the enhancement of eccentricity and the shortening of the oscillation timescale—may also have important implications for gravitational-wave astronomy and for the dynamical evolution of galactic nuclei.

Several important extensions remain for future work. A systematic exploration of the parameter space, including different mass ratios and inclined orbits in Kerr spacetime, would be valuable. The inclusion of dissipative effects such as gravitational-radiation reaction is also essential for assessing the long-term evolution and merger outcomes.

To compute gravitational waves near the SMBH horizon, black hole perturbation theory may provide a useful framework, since the binary motion has already been determined in our previous works \cite{Maeda:2023tao, Maeda:2023uyx, Maeda:2025row} as well as in the present study. However, because of the complexity of the motion, this approach remains technically challenging\cite{Santos:2025bEMRI}. Alternatively, the quadrupole formula may provide a reasonable approximation when the orbital radius exceeds approximately 10$M$ \cite{Shibata:1994qd}. An analysis along these lines is currently in progress and will help clarify the gravitational-wave signatures of binaries evolving in the strong-field environment of supermassive black holes.

\begin{acknowledgments}
 This work was supported in part by the JSPS KAKENHI
Grant Number JP24K07058(K.M.), and by JP23K03222 (H.O.). 
K.M. would like to acknowledge the Yukawa Institute for Theoretical
Physics at Kyoto University, where the present work  was partially performed during
the YITP long-term workshop, 
Multi-Messenger Astrophysics in the Dynamic Universe 2026.

\end{acknowledgments}

%%%%%%%%%%%%%%%%%%%%%%%%%%%%%%%%%%%%
%%%%%%%%%%%%%%%%%%%%%%%%%%%%%%%%%%%%

\begin{appendices}
\renewcommand{\theequation}{\Alph{section}.\arabic{equation}}

%%%%%%%%%%%%%%%%%%%%%%%%%%%%%%%%%%%%
%%%%%%%%%%%%%%%%%%%%%%%%%%%%%%%%%%%% 
%%%%%%%%%%%%%%%%%%%%%%%%%%%%%%%%%%%%%%%
%%%%%%%%%%%%%%%%%%%%%%%%%%%%%%%%%%%%%%%

%%%%%%%%%%%%%%%%%%%%%%%%%%%%%%%%%%%%%%%
%%%%%%%%%%%%%%%%%%%%%%%%%%%%%%%%%%%%%%%
\section{Test Particle  in Kerr Spacetime}
\label{test_particle_Kerr}
%%%%%%%%%%%%%%%%%%%%%%%%%%%%%%%%%%%%%%%%%%%%%%%%%%%%%%%%%%%%%%%%%%%%%%%%%%%%%%
%%%%%%%%%%%%%%%%%%%%%%%%%%
%%%%%%%%%%%%%%%%%%%%%%%%%%
%\subsection{General Orbits in Kerr spacetime}
%\label{general_orbit_Kerr}
%%%%%%%%%%%%%%%%%%%%%%%%%%
%%%%%%%%%%%%%%%%%%%%%%%%%%
We consider a test particle with a unit mass in the Kerr spacetime, which metric is given by Eq. (\ref{Kerr_metric}).
There are two Killing vectors, which guarantee two conserved quantities
in a test particle motion; the proper energy $E$ and the $z$-component of the proper 
angular momentum $L$.
In addition, there exists one more conserved quantity, 
so-called Carter constant ${\cal C}$ or $K$.
These constants are  defined by the Killing tensor $K_{\mu\nu}$ as
\beann
K&=&K_{\mu\nu} p^\mu p^\nu
\\
{\cal C}&=&K-\ell^2
\enann
where 
$p_\mu\equiv u_\mu$ is the proper 4-momentum of a test particle and $\ell\equiv L-aE$.

The Carter constant ${\cal C}$ in BL coordinates is given by
\beann
{\cal C}\equiv p_\theta^2+\cos^2\theta\left[a^2(1-E^2)+{L^2\over \sin^2\theta}\right]
\,,
\enann
where  $p_\theta$ is the $\theta$-component of the proper 4-momentum.
Note that ${\cal C}=0$ when the orbit is on the equatorial plane ($\theta=\pi/2$). While $K$ is non-negative definite.

In order to decouple the equations of motion (EOM),
 we shall introduce the Mino time $\mathsf{t}$ \cite{Mino:2003yg},
which is defined by
\beann
d\mathsf{t}\equiv {1 \over \Sigma} d\tau
\,,
\enann
where $\tau$ is the proper time of a test particle.

We then find two decoupled equations
 for $\mathfrak{r}$ and $\zeta$ as
\bea
\dot{\mathfrak{r}}^2 &=&\left[Er^2-a\ell \right]^2-\Delta \left[ r^2+\ell^2+{\cal C}\right]
\label{eq_r}
\\
\dot \zeta^2 &=&-\zeta^2\left[a^2\left(1-E^2\right)\left(1-\zeta^2\right)+L^2\right]
\nonumber \\
&&
+{\cal C}\left(1-\zeta^2\right)
\label{eq_th}
\,,
\ena
where a dot denotes the derivative with respect to the Mino time $\mathsf{t}$.
The other two EOM for $t$ and $\phi$ are given by 
\beann
\dot{t} &=&{r^2+a^2\over \Delta}\left[Er^2-a\ell\right]-a^2 E\left(1-\zeta^2\right)+aL
\label{eq_t}
\\
\dot \phi&=&{a\over \Delta}\left[E\mathfrak{r}^2-a\ell\right]+{L\over 1-\zeta^2}-aE
\,,
\label{eq_phi}
\enann
which can be integrated once we find the solutions of $\mathfrak{r}(\mathsf{t})$ and $\zeta(\mathsf{t})$.

The above equations are analytically  integrated by use of the elliptic functions as follows.
Here we summarize eccentric orbits of test particles on the equatorial plane ($\zeta=0$), 
which we use in the text.
The eccentric orbits are classified into three types: elliptic, parabolic, and hypabolic orbits, which energies 
are given by $E<1$, $E=1$, and $E>1$, respectively.
The Carter constant ${\cal C}$ vanishes.

%%%%%%%%%%%%%%%%%%%%%%%%%%%%%%%%%%%%%%%
\subsubsection{\rm {\bf elliptic orbit} ($E<1$)}
%%%%%%%%%%%%%%%%%%%%%%%%%%%%%%%%%%%%%%%

Eq. (\ref{eq_r}) is  rewritten as
\bea
\dot{\mathfrak{r}}^2 &=&(1-E^2)\mathfrak{r}
(\mathfrak{r}_a-\mathfrak{r})(\mathfrak{r}-\mathfrak{r}_p)(\mathfrak{r}-\mathfrak{r}_3)~~~
\label{eq_r_el}
\ena
where $\mathfrak{r}_a$ and $\mathfrak{r}_p$ are two turning points; apoapsis and periapsis 
($\mathfrak{r}_a\geq \mathfrak{r}_p$)  in the radial direction. If the orbit is elliptic, these two values are given by 
the semi-major axis $\mathfrak{r}_0$ and the eccentricity $e_{\rm out}$ as
\bea
\mathfrak{r}_a=\mathfrak{r}_0(1+e_{\rm out})\,,~~{\rm and}~~~\mathfrak{r}_p=\mathfrak{r}_0(1-e_{\rm out})
\,.
\ena

Comparing the coefficients in Eqs. (\ref{eq_r}) and (\ref {eq_r_el}), 
we find
\bea
&&
\mathfrak{r}_a+\mathfrak{r}_p+\mathfrak{r}_3={2M\over 1-E^2}
\label{eq1}
\\
&&\mathfrak{r}_a\mathfrak{r}_p+\mathfrak{r}_3(\mathfrak{r}_a+\mathfrak{r}_p)
\nonumber \\
&&={1\over 1-E^2}\left[(\ell+aE)^2+a^2(1-E^2)\right]
~\label{eq2}
\\
&&
\mathfrak{r}_a\mathfrak{r}_p\mathfrak{r}_3
={2M\over 1-E^2}\ell^2
\label{eq3}
\,.
\ena
%\end{widetext}
For given values of $\mathfrak{r}_a, \mathfrak{r}_p$, 
Eqs (\ref{eq1})-(\ref{eq3}) determine $E, \ell$, 
and  $\mathfrak{r}_3$ as follows. 
From Eqs. (\ref{eq1}) and (\ref{eq3}), we find
\beann
\mathfrak{r}_a\mathfrak{r}_p\left[{2M\over 1-E^2}-(\mathfrak{r}_a+\mathfrak{r}_p)
\right]={2M\over 1-E^2}\ell^2
\,,
\enann
which gives 
\beann
\ell=\ell_\pm(E) \equiv 
\pm \sqrt{\mathfrak{r}_a\mathfrak{r}_p
\left[1-{1-E^2\over 2M}(\mathfrak{r}_a+\mathfrak{r}_p)\right]}
\,,
\enann
where $\pm$ correspond to the prograde and retrograde orbits, respectively.
Inserting this equation into Eq. (\ref{eq2}), we find the equation for $E$  as
\beann
&&
\mathfrak{r}_a\mathfrak{r}_p+(\mathfrak{r}_a+\mathfrak{r}_p)\left[{2M\over 1-E^2}-(\mathfrak{r}_a+\mathfrak{r}_p)
\right]
\\
&&
={1\over 1-E^2}\left[(\ell_\pm(E)+aE)^2+a^2(1-E^2)\right]
\,,
\enann
which solves $E=E_\pm$ for the prograde and retrograde orbits, respectively.
We then find 
\beann
\mathfrak{r}_3={2M\over 1-E_\pm^2}-(\mathfrak{r}_a+\mathfrak{r}_p)
\,.
\enann

Since $0<\mathfrak{r}_3<\mathfrak{r}_p<\mathfrak{r}_a$, 
 integrating Eq. (\ref{eq_r_el}), choosing the initial 
 position as $\mathfrak{r}=\mathfrak{r}_p$  at $\mathsf{t}=0$, 
we find 
the solution for $\mathfrak{r}$  as
\beann
\mathfrak{r}&=&\mathfrak{r}_p{\displaystyle{1-{\mathfrak{r}_3(\mathfrak{r}_a-\mathfrak{r}_p)\over \mathfrak{r}_p(\mathfrak{r}_a-\mathfrak{r}_3)}}{\rm sn}^2\left(\mathfrak{w}_r\mathsf{t}/2 ; k_r^2\right)\over 
1-\displaystyle{{(\mathfrak{r}_a-\mathfrak{r}_p)\over (\mathfrak{r}_a-\mathfrak{r}_3)}}
{\rm sn}^2\left(\mathfrak{w}_r\mathsf{t}/2 ; k_r^2\right) }
\,,
\enann
where
${\rm sn}(x;m)$ is the Jacobi elliptic sine function with 
the elliptic parameter $m\equiv k^2$ (the modulus $k$), and 
\beann
\mathfrak{w}_r&\equiv &\sqrt{(1-E_\pm^2)(\mathfrak{r}_a-\mathfrak{r}_3)\mathfrak{r}_p}
\\
k_r^2&\equiv&{\mathfrak{r}_3(\mathfrak{r}_a-\mathfrak{r}_p)\over \mathfrak{r}_p(\mathfrak{r}_a-\mathfrak{r}_3)}
 \,.
\enann
 
The frequency of oscillations in radial directions
is given by
 \beann
 \Upsilon_r&\equiv&{\pi\mathfrak{w}_r\over 2\mathsf{K}(k_r)}={\pi\over 2\mathsf{K}(k_r)}\sqrt{(1-E^2)\mathfrak{r}_p(\mathfrak{r}_a-\mathfrak{r}_3)}
 \,,
 \enann
 where
 $\mathsf{K}(k_r)$ is the complete elliptic integral of the first kind 
 with the modulus $k_r$.
 
 \begin{widetext}
 The solutions of $t$ and $\phi$ are given by
 \bea
 t(\mathsf{t})&=&\Upsilon_t\mathsf{t}+t_r(\mathsf{t})
 \label{t_el}
 \\
 \phi(\mathsf{t})&=&\Upsilon_\phi\mathsf{t}+\phi_r(\mathsf{t})
 \,,
  \label{phi_el}
 \ena
 where
 \beann
\Upsilon_t&=&4E+E\Big{\{}{1\over 2}\Big{[}(4+\mathfrak{r}_a+\mathfrak{r}_p+\mathfrak{r}_3)\mathfrak{r}_3-\mathfrak{r}_a\mathfrak{r}_p+\mathfrak{r}_p(\mathfrak{r}_a-\mathfrak{r}_3){\mathsf{E}(k_r)\over \mathsf{K}(k_r)}
\\
&&
+(4+\mathfrak{r}_a+\mathfrak{r}_p+\mathfrak{r}_3)(\mathfrak{r}_p-\mathfrak{r}_3){\mathsf{\Pi}(h_r;k_r)\over \mathsf{K}(k_r)}
\Big{]}
\\
&&
+{2\over \mathfrak{r}_+-\mathfrak{r}_-}
\left[{(4-aL/E)\mathfrak{r}_+-2a^2\over \mathfrak{r}_3-\mathfrak{r}_+}\left(1-{ \mathfrak{r}_p-\mathfrak{r}_3\over  \mathfrak{r}_p-\mathfrak{r}_+}
{\mathsf{\Pi}(h_+;k_r)\over \mathsf{K}(k_r)}
\right)-(+\leftrightarrow-)\right]
\Big{\}}
\\
t_r(\mathsf{t})&=&\tilde t_r\left(\mathsf{am}(\mathfrak{w}_r\mathsf{t}/2;k_r)\right)-{\tilde t_r(\pi)\mathfrak{w}_r\mathsf{t}\over 4\mathsf{K}(k_r)}
\\
\Upsilon_\phi
&=&
{a\over \mathfrak{r}_+-\mathfrak{r}_-}\left[{2E\mathfrak{r}_+-aL\over \mathfrak{r}_3-\mathfrak{r}_+}\left(1-{\mathfrak{r}_p-\mathfrak{r}_3\over \mathfrak{r}_p-\mathfrak{r}_+}{\mathsf{\Pi}(h_+;k_r)\over \mathsf{K}(k_r)}\right)-(+\leftrightarrow-)\right]+L
\\
\phi_r(\mathsf{t})&=&\tilde \phi_r\left(\mathsf{am}(\mathfrak{w}_r\mathsf{t}/2;k_r)\right)-{\tilde \phi_r(\pi)\mathfrak{w}_r\mathsf{t}\over 4\mathsf{K}(k_r)}
 \,.
 \enann
  Here the functions $\tilde t_r$ and $\tilde \phi_r$
 are defined by
\beann
\tilde t_r(\xi)
&=&{E(\mathfrak{r}_p-\mathfrak{r}_3)\over \sqrt{(1-E^2)\mathfrak{r}_p(\mathfrak{r}_a-\mathfrak{r}_3)}}\Big{\{}
(4+\mathfrak{r}_a+\mathfrak{r}_p+\mathfrak{r}_3)\mathsf{\Pi}(h_r,\xi;k_r)
\\
&&
-{4\over \mathfrak{r}_+-\mathfrak{r}_-}\left[{\mathfrak{r}_+(4-aL/E)-2a^2\over 
(\mathfrak{r}_p-\mathfrak{r}_+)(\mathfrak{r}_3-\mathfrak{r}_+)}\mathsf{\Pi}(h_+,\xi;k_r)-(+\leftrightarrow-)\right]
\\
&&
+{\mathfrak{r}_p(\mathfrak{r}_a-\mathfrak{r}_3)\over (\mathfrak{r}_p-\mathfrak{r}_3)}\left[\mathsf{E}(\xi;k_r)-
h_r{\sin \xi \cos \xi\sqrt{1-k_r^2\sin^2 \xi}\over 1-h_r\sin^2\xi}\right]
\Big{\}}
\\
\tilde \phi_r(\xi)
&=&-{2aE(\mathfrak{r}_p-\mathfrak{r}_3)\over (\mathfrak{r}_+-\mathfrak{r}_-)\sqrt{(1-E^2)\mathfrak{r}_p(\mathfrak{r}_a-\mathfrak{r}_3)}}\left[{2\mathfrak{r}_+-aL/E\over 
(\mathfrak{r}_p-\mathfrak{r}_+)(\mathfrak{r}_3-\mathfrak{r}_+)}\mathsf{\Pi}(h_+,\xi;k_r)-(+\leftrightarrow-)\right]
\enann
with
\beann
h_r={\mathfrak{r}_a-\mathfrak{r}_p\over \mathfrak{r}_a-\mathfrak{r}_3}
\,,~
{\rm and}~~
h_\pm=h_r{\mathfrak{r}_3-\mathfrak{r}_\pm \over \mathfrak{r}_p-\mathfrak{r}_\pm}
\enann

 The proper time $\tau$ is given by the integration
 \beann
 \tau=\int_0^\mathsf{t} d\mathsf{t}\, \Sigma=\int_0^\mathsf{t} d\mathsf{t} \, \mathfrak{r}^2(\mathsf{t})
 \enann
 \end{widetext}
 
%%%%%%%%%%%%%%%%%%%%%%%%%%%%%%%%%%%%%%%
\subsubsection{\rm {\bf parabolic orbit} ($E=1$)}
%%%%%%%%%%%%%%%%%%%%%%%%%%%%%%%%%%%%%%%
Eq. (\ref{eq_r}) is  rewritten as
\bea
\dot{\mathfrak{r}}^2 &=&2M\mathfrak{r}\left(\mathfrak{r}^2-{(\ell+a)^2\over 2M}\mathfrak{r}+\ell^2\right)
\nonumber \\
&=&2M\mathfrak{r}(\mathfrak{r}-\mathfrak{r}_p)(\mathfrak{r}-\mathfrak{r}_3)
\label{eq_r_para}
\,,
\ena
where $\mathfrak{r}_p$ is a turning point
in the radial direction (periapsis ). 

Comparing the coefficients in Eq. (\ref{eq_r_para}), 
we find
\beann
\mathfrak{r}_p+\mathfrak{r}_3&=&{(\ell+a)^2\over 2M}
\label{eq1_para}
\\
\mathfrak{r}_p\mathfrak{r}_3&=&\ell^2
\label{eq2_para}
\,.
\enann
For given value of $\mathfrak{r}_p$, eliminating $\mathfrak{r}_3$, we find the equation for $\ell$ as
\beann
(\mathfrak{r}_p-2M)\ell^2+2\mathfrak{r}_p a\ell -\mathfrak{r}_p(2M\mathfrak{r}_p-a^2)=0
\,,
\enann
which gives
\beann
\ell=\ell_\pm(\mathfrak{r}_p)\equiv {1\over (\mathfrak{r}_p-2M)}\left[-a\mathfrak{r}_p\pm \sqrt{2M\mathfrak{r}_p
\Delta(\mathfrak{r}_p)}\right]
\,.
\enann
We then find  $\mathfrak{r}_3$ as
\beann
 \mathfrak{r}_3={\ell^2_\pm(\mathfrak{r}_p)\over \mathfrak{r}_p}
 \,.
\enann

Integrating Eq.(\ref{eq_r_para}), we find
\beann
\mathfrak{r}&=&\mathfrak{r}_p{\displaystyle{1-{\mathfrak{r}_3\over \mathfrak{r}_p}}{\rm sn}^2\left(\mathfrak{w}_r\mathsf{t}/2 ; k_r^2\right)\over 
1-
{\rm sn}^2\left(\mathfrak{w}_r\mathsf{t}/2 ; k_r^2\right) }
\,,
\enann
where
\beann
\mathfrak{w}_r&=&\sqrt{2M\mathfrak{r}_p}
\\
k_r^2&=&{\mathfrak{r}_3\over \mathfrak{r}_p}
\enann

For an unbound orbit, $ \mathfrak{r}_3< \mathfrak{r}_p$, which gives a constraint for $ \mathfrak{r}_p$. We obtain 
\beann
\mathfrak{r}_p>\mathfrak{r}_{p{\rm (min)}}^{\rm (p)}\equiv 2M-a+2\sqrt{M(M-a)}
\enann
for a prograde orbit, while 
\beann
\mathfrak{r}_p>\mathfrak{r}_{p{\rm (min)}}^{\rm (r)}\equiv 2M+a+2\sqrt{M(M+a)}
\enann
for a retrograde orbit. 
In the case of Schwarzschild BH ($a=0$),  $\mathfrak{r}_{p{\rm (min)}}=4M$, while for extreme Kerr BH ($a=M$), we find
$\mathfrak{r}_{p{\rm (min)}}^{\rm (p)}=M$ and $\mathfrak{r}_{p{\rm (min)}}^{\rm (r)}=(3+2\sqrt{2})M$.
$\mathfrak{r}_{p{\rm (min)}}^{\rm (p)}$ and $\mathfrak{r}_{p{\rm (min)}}^{\rm (r)}$ are the minima of the periastra of parabolic orbits, which are shown by the solid curves in Fig. \ref{fig:minimum_periastra}.
The blue and red solid curves denote the minimum of the periastra 
for prograde and retrograde orbits, respectively.

\begin{widetext}

\begin{figure}[htbp]
\begin{center}
\includegraphics[width=6.0cm]{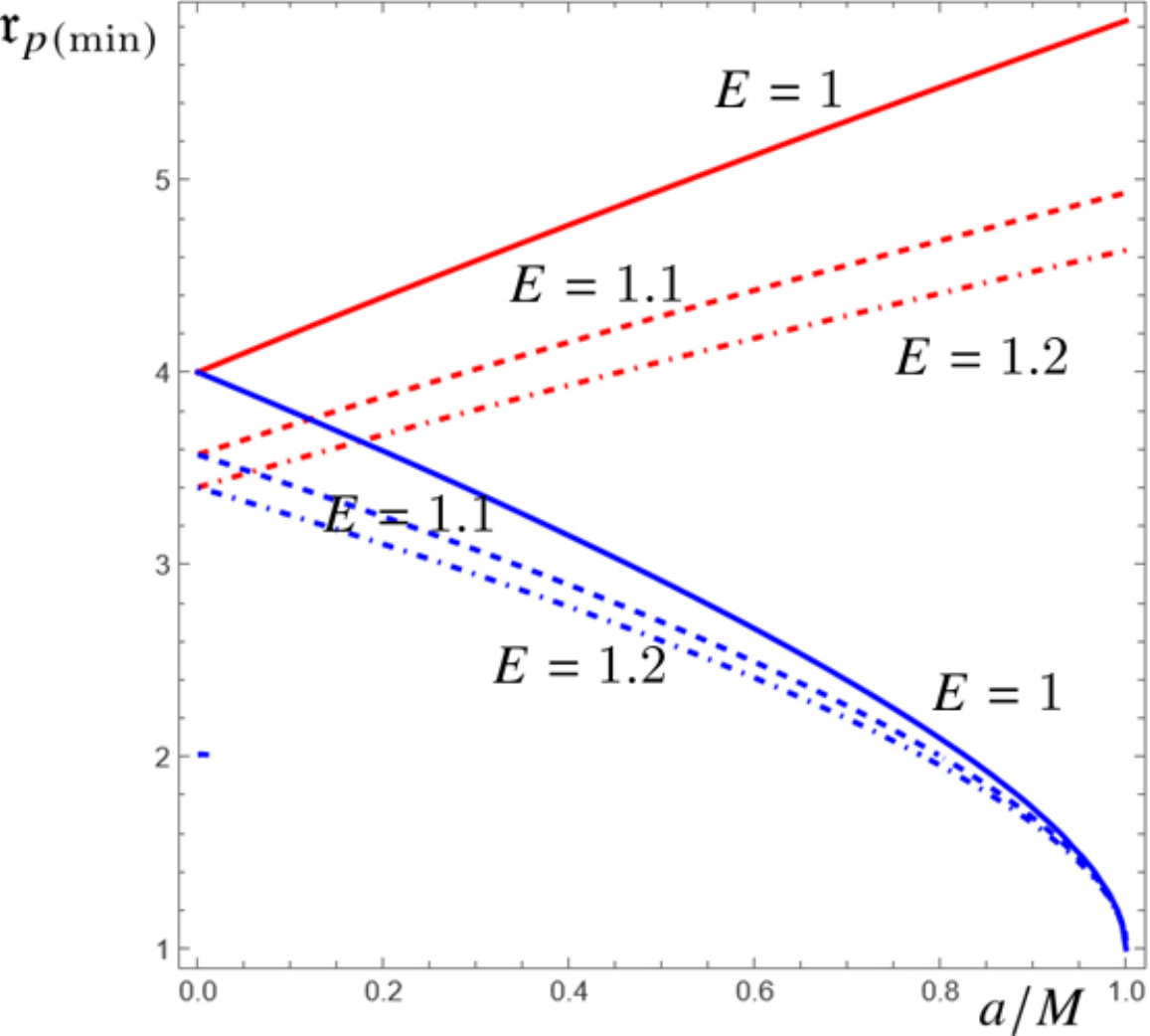}
\caption{The minima of the periastra, $\mathfrak{r}_{p{\rm (min)}}^{\rm (p)}$ and $\mathfrak{r}_{p{\rm (min)}}^{\rm (r)}$, are shown in terms of the Kerr parameter $a$. The blue and red curves denote those
for prograde and retrograde orbits.
The solid, dashed, and dot-dashed curves denote those for $E=1, 1.1$, and 1.2, respectively.}
\label{fig:minimum_periastra}
\end{center}
\end{figure}

 The solutions of $t$ and $\phi$ are given by
 \beann
 t(\mathsf{t})&=&\mathsf{t}\left[E\left({2\mathfrak{r}_p^2+\mathfrak{r}_3^2\over 3}+\mathfrak{r}_p(\mathfrak{r}_++\mathfrak{r}_-)+\mathfrak{r}_+^2+\mathfrak{r}_-^2+\mathfrak{r}_+\mathfrak{r}_-+a^2\right)+{(\mathfrak{r}_+^2+a^2)(2ME\mathfrak{r}_+-aL)\over (\mathfrak{r}_+-\mathfrak{r}_-)(\mathfrak{r}_3-\mathfrak{r}_+)}-(+\leftrightarrow -)\right]
 \\&&+{2E\over \mathfrak{w}_r}\left({2\mathfrak{r}_p^2+\mathfrak{r}_3^2\over 3}+\mathfrak{r}_p(\mathfrak{r}_++\mathfrak{r}_-)\right) \left(\sqrt{\mathfrak{r}(\mathsf{t})(\mathfrak{r}(\mathsf{t})-\mathfrak{r}_p)\over \mathfrak{r}_p(\mathfrak{r}(\mathsf{t})-\mathfrak{r}_3)}-\mathsf{E}(\mathsf{am}(\mathfrak{w}_r\mathsf{t}/2;k_r),k_r)\right)
 \\
 &&
 -{2(\mathfrak{r}_p-\mathfrak{r}_3)\over \mathfrak{w}_r(\mathfrak{r}_+-\mathfrak{r}_-)}
 \left[{(\mathfrak{r}_+^2+a^2)(2ME\mathfrak{r}_+-aL)\over (\mathfrak{r}_3-\mathfrak{r}_+)(\mathfrak{r}_p-\mathfrak{r}_+)}
\mathsf{\Pi}(h_+;\mathsf{am}(\mathfrak{w}_r\mathsf{t}/2;k_r),k_r)-(+\leftrightarrow -)\right]
 \\
 \phi(\mathsf{t})&=&\mathsf{t}\left[L+{a\over (\mathfrak{r}_+-\mathfrak{r}_-)}
\left({(2ME\mathfrak{r}_+-aL)\over (\mathfrak{r}_3-\mathfrak{r}_+)}-{(2ME\mathfrak{r}_--aL)\over (\mathfrak{r}_3-\mathfrak{r}_-)}\right)\right]
  \\
 &&
 -{2a (\mathfrak{r}_p-\mathfrak{r}_3)\over \mathfrak{w}_r(\mathfrak{r}_+-\mathfrak{r}_-)}
 \left[{(2ME\mathfrak{r}_+-aL)\over (\mathfrak{r}_3-\mathfrak{r}_+)(\mathfrak{r}_p-\mathfrak{r}_+)}
\mathsf{\Pi}(h_+;\mathsf{am}(\mathfrak{w}_r\mathsf{t}/2;k_r),k_r)-(+\leftrightarrow -)\right]
 \,.
 \enann
\end{widetext}
%%%%%%%%%%%%%%%%%%%%%%%%%%%%%%%%%%%%%%%
\subsubsection{\rm {\bf hyperbolic orbit} ($E>1$)}
%%%%%%%%%%%%%%%%%%%%%%%%%%%%%%%%%%%%%%%
This case is almost the same as the elliptic case.
However there is no apoapsis. Then we rewrite
Eq. (\ref{eq_r})  as
\bea
\dot{\mathfrak{r}}^2 &=&(E^2-1)\mathfrak{r}
(\mathfrak{r}-\mathfrak{r}_p)(\mathfrak{r}-\mathfrak{r}_2)(\mathfrak{r}-\mathfrak{r}_3)
\label{eq_r_hy}
\ena
where  $\mathfrak{r}_p$ is a turning point in the radial direction (periapsis). 
$\mathfrak{r}_2$ corresponds to $\mathfrak{r}_a$ in an elliptic orbit, but it becomes negative.
We find the same equations as the elliptic case if we replace $\mathfrak{r}_a$ by $\mathfrak{r}_2$.
The radial orbit is given by
\beann
\mathfrak{r}&=&\mathfrak{r}_p{\displaystyle{1-{\mathfrak{r}_3(\mathfrak{r}_2-\mathfrak{r}_p)\over \mathfrak{r}_p(\mathfrak{r}_2-\mathfrak{r}_3)}}{\rm sn}^2\left(\mathfrak{w}_r\mathsf{t}/2 ; k_r^2\right)\over 
1-\displaystyle{{(\mathfrak{r}_2-\mathfrak{r}_p)\over (\mathfrak{r}_2-\mathfrak{r}_3)}}
{\rm sn}^2\left(\mathfrak{w}_r\mathsf{t}/2 ; k_r^2\right) }
\,,
\enann
just as the elliptic case.
But since $\mathfrak{r}_2<0<\mathfrak{r}_3<\mathfrak{r}_p$ in hyperbolic orbit, we find  
$\displaystyle{{(\mathfrak{r}_2-\mathfrak{r}_p)\over (\mathfrak{r}_2-\mathfrak{r}_3)}}>1$,
with which  
the denominater vanishes at finite time $\mathsf{t}=\mathsf{t}_\infty (<\infty)$,
 i.e., the radius $\mathfrak{r}$ diverges then.
When we convert the Mino time $\mathsf{t}$ to the propert time $\tau$,
we find 
\beann
\tau_\infty \equiv \int_0^{\mathsf{t}_\infty} d\mathsf{t} \,\mathfrak{r}^2(\mathsf{t} )=\infty
\,.
\enann

Since $\mathfrak{r}_2$ is not apoapsis, we may give the energy $E$ instead of  $\mathfrak{r}_2$.
If we fix $E$ as well as $\mathfrak{r}_p$, the other parameters are determined as follows:
From Eqs. (\ref{eq1}) and (\ref{eq3}), we find
\beann
\mathfrak{r}_2+\mathfrak{r}_3&=&-{2M\over E^2-1}-\mathfrak{r}_p
\\
\mathfrak{r}_2\mathfrak{r}_3&=&-{2M\over E^2-1}{\ell^2\over \mathfrak{r}_p}
\enann

\begin{widetext}
Inserting these two relations into Eq. (\ref{eq2}),
we find
\beann
(\mathfrak{r}_p-2M)\ell^2+2a\mathfrak{r}_p E\ell+\mathfrak{r}_p(\Delta(\mathfrak{r}_p)-\mathfrak{r}_p^2E^2)=0
\,,
\enann
which gives 
\beann
\ell=\ell_\pm(E)\equiv  {1\over (\mathfrak{r}_p-2M)}\left[-a\mathfrak{r}_p E\pm \sqrt{a^2\mathfrak{r}_p^2E^2
-\mathfrak{r}_p(\mathfrak{r}_p-2M)
(\Delta(\mathfrak{r}_p)-\mathfrak{r}_p^2E^2})\right]
\enann
Since $\mathfrak{r}_2<0<\mathfrak{r}_3$, we then find 
\beann
\mathfrak{r}_2&=&{1\over 2}\left[-{2M\over E^2-1}-\mathfrak{r}_p
-\sqrt{\left({2M\over E^2-1}+\mathfrak{r}_p\right)^2+{8M\ell_\pm^2\over \mathfrak{r}_p(E^2-1)}}\right]
\\
\mathfrak{r}_3&=&{1\over 2}\left[-{2M\over E^2-1}-\mathfrak{r}_p
+\sqrt{\left({2M\over E^2-1}+\mathfrak{r}_p\right)^2+{8M\ell_\pm^2\over \mathfrak{r}_p(E^2-1)}}\right]
\,.
\enann
The condition such that $\mathfrak{r}_3<\mathfrak{r}_p$ gives 
the minimum of the periastra of the hyperbolic orbits, which depends on the energy $E$.
We show some examples for the case of $E=1.1$ (the dashed curves) and $1.2$ (the dot-dashed curves) in 
Fig. \ref{fig:minimum_periastra}, respectively.

We find the solutions of $t$ and $\phi$, which are given by 
 the same equations as Eqs. (\ref{t_el}) and (\ref{phi_el}) 
 in the elliptic case by replacing 
$\mathfrak{r}_a$ by $\mathfrak{r}_2$.

\section{Binary system in a curved spacetime}
\label{binary_in_curved_ST}
%%%%%%%%%%%%%%%%%%%%%%%%%%%%%%%%%%%%%%%
%%%%%%%%%%%%%%%%%%%%%%%%%%%%%%%%%%%%%%%
In this Appendix, we summarize how to discuss a binary system in a fixed curved background, which was discussed in details in \cite{Maeda:2023tao,Maeda:2023uyx}.
A binary consists of two point particles with the masses $m_1$ and $m_2$.
The Lagrangian up to 0.5 PN order is given by
\beann
{\cal L}_{\rm binary}={\cal L}_{\rm N}+{\cal L}_{1/2},
\label{Lagrangian_binary}
\enann
where
\bea
{\cal L}_{\rm N}&\equiv& {1\over 2} \left[ m_1 \left({d\vect{x}_1\over d\tau}\right)^2+m_2 \left({d\vect{x}_2\over d\tau}\right)^2\right]
+  {G m_1m_2\over |\vect{x}_1-\vect{x}_2|}
+{\cal L}_{a}+{\cal L}_{\omega}+{\cal L}_{\bar{\cal R}}
\label{Lagrangian_N},
\ena
with
\beann
{\cal L}_{a}
&=&
-\sum_{I=1}^2 m_I a_{\hat k}x_I ^{\hat k},
\\
{\cal L}_{\omega}
&=&
-\sum_{I=1}^2 m_I \left[\epsilon_{\hat j\hat k\hat \ell}\omega^{\hat \ell}
x_I^{\hat k}{dx_I^{\hat j}\over d\tau}-
{1\over 2} \left(\vect{\omega}^2 \vect{x}_I^2 -(\vect{\omega} \cdot \vect{x}_I)^2
\right)\right],
\\
{\cal L}_{\bar{\cal R}}
&=&
-{1\over 2}  \sum_{I=1}^2 m_I \bar{\cal R}_{\hat 0\hat k\hat 0\hat \ell}x_I^{\hat k} x_I^{\hat \ell}
\,,
\enann
and 
\beann
{\cal L}_{1/2}&\equiv &-{2\over 3}\sum_{I=1}^2
m_I c^2\bar{\cal R}_{\hat 0\hat k \hat j  \hat \ell }
x_I^{\hat k}x_I^{\hat \ell}\, {v_I^{\hat j}\over c}
\,.
\label{Lagrangian_0.5PN}
\enann

Introducing the center of mass coordinates and the relative coordinates by
\beann
\vect{R}&=&{m_1\vect{x}_1+m_2\vect{x}_2\over m_1+m_2},
\\
\vect{r}&=& \vect{x}_2-\vect{x}_1
\,,
\enann
we find the Newtonian Lagrangian (Eq.~\eqref{Lagrangian_N}) in terms of  $\vect{R}$  and $\vect{r}$ as
\beann
{\cal L}_{\rm N}={\cal L}_{\rm CM}\left(\vect{R}, {d\vect{R}\over d\tau}\right)+
{\cal L}_{\rm rel}\left(\vect{r}, {d\vect{r}\over d\tau}\right)
\label{Lagrangian_N_Rr}
\,,
\enann
where
\beann
{\cal L}_{\rm CM}\left(\vect{R}, {d\vect{R}\over d\tau}\right)
&=&{1\over 2} (m_1+m_2) \left({d\vect{R}\over d\tau}\right)^2
+{\cal L}_{{\rm CM}\mathchar`-a}\left(\vect{R}, {d\vect{R}\over d\tau}\right)
+{\cal L}_{{\rm CM}\mathchar`-\omega}\left(\vect{R}, {d\vect{R}\over d\tau}\right)
+{\cal L}_{{\rm CM}\mathchar`-\bar{\cal R}}\left(\vect{R}, {d\vect{R}\over d\tau}\right),
\enann
with 
\beann
{\cal L}_{{\rm CM}\mathchar`-a}
&=&-(m_1+m_2)\vect{a}\cdot\vect{R}
\\
{\cal L}_{{\rm CM}\mathchar`-\omega}
&=&-(m_1+m_2)\left[
\epsilon_{\hat j\hat k\hat \ell}\omega^{\hat \ell}
R^{\hat k}{dR^{\hat j}\over d\tau}
-{1\over 2}\left(\vect{\omega}^2 \vect{R}^2
-\left(\vect{\omega}\cdot \vect{R}\right)^2\right)\right],
\\{\cal L}_{{\rm CM}\mathchar`-\bar{\cal R}},
&=&
-{1\over 2}(m_1+m_2)
\bar{\cal R}_{\hat 0\hat k \hat 0 \hat \ell}R^{\hat k}R^{\hat \ell},
\enann
and
 \beann
{\cal L}_{\rm rel}\left(\vect{r}, {d\vect{r}\over d\tau}\right)&=&
{1\over 2}\mu \left({d\vect{r}\over d\tau}\right)^2+  {G m_1m_2\over r}
+{\cal L}_{{\rm rel}\mathchar`-\omega}\left(\vect{r}, {d\vect{r}\over d\tau}\right)
+{\cal L}_{{\rm rel}\mathchar`- \bar{\cal R}}\left(\vect{r}, {d\vect{r}\over d\tau}\right),
\enann
with 
\beann
{\cal L}_{{\rm rel}\mathchar`-\omega}
&=&-\mu \left[
\epsilon_{\hat j\hat k\hat \ell}\omega^{\hat \ell}
r^{\hat k}{dr^{\hat j}\over d\tau}
-{1\over 2}\left(\vect{\omega}^2 \vect{r}^2
-\left(\vect{\omega}\cdot \vect{r}\right)^2\right)\right]\,,
\\
{\cal L}_{{\rm rel}\mathchar`-\bar{\cal R}}
&=&
-{1\over 2}\mu 
\bar{\cal R}_{\hat 0\hat k \hat 0 \hat \ell}r^{\hat k}r^{\hat \ell}
\,.
\enann
Here, $\mu = m_1 m_2/(m_1+m_2)$ is the reduced mass. When we consider only ${\cal L}_{\rm N}$, 
we can separate the variables $\vect{R}$ and $\vect{r}$.
 In particular, when the observer follows the geodesic ($\vect{a}=0$ and $\vect{\omega}=0$),
the orbit of $\vect{R}=0$ is a solution of the equation for $\vect{R}$.
It means that the center-of-mass follows the observer's geodesic.
We have only the equation for the relative coordinate $\vect{r}$. However, when we include the 0.5 PN term, it is not the case.
The 0.5PN Lagrangian ${\cal L}_{1/2}$  is rewritten by use of  $\vect{R}$  and    $\vect{r}$ as
\beann
{\cal L}_{1/2}={\cal L}_{1/2\mathchar`-{\rm CM}}
\left(\vect{R}, {d\vect{R}\over d\tau}\right)+
{\cal L}_{1/2\mathchar`-{\rm rel}}\left(\vect{r}, {d\vect{r}\over d\tau}\right)
+{\cal L}_{1/2\mathchar`- {\rm int}}\left(\vect{R}, {d\vect{R}\over d\tau}, \vect{r}, {d\vect{r}\over d\tau}\right),
\label{Lagrangian_1/2_Rr}
\enann
where
\beann
{\cal L}_{1/2\mathchar`-{\rm CM}}\left(\vect{R}, {d\vect{R}\over d\tau}\right)&=&-{2\over 3}(m_1+m_2)R_{\hat 0\hat k \hat j \hat \ell}
R^{\hat k}R^{\hat \ell}{dR^{\hat j}\over d\tau},
\nn
{\cal L}_{1/2\mathchar`-{\rm rel}}\left(\vect{r}, {d\vect{r}\over d\tau}\right)&=&- {2\over 3} \mu{(m_1-m_2)\over (m_1+m_2)}R_{\hat 0\hat k \hat j \hat \ell}r^{\hat k}r^{\hat \ell}{dr^{\hat j}\over d\tau},
\nn
{\cal L}_{1/2\mathchar`- {\rm int}}\left(\vect{R}, {d\vect{R}\over d\tau}, \vect{r}, {d\vect{r}\over d\tau}\right)&=&
- {2\over 3} \mu R_{\hat 0\hat k \hat j \hat \ell}
\left[r^{\hat k}r^{\hat \ell}{dR^{\hat j}\over d\tau}
+\left(R^{\hat k}r^{\hat \ell}+r^{\hat k}R^{\hat \ell}
\right){dr^{\hat j}\over d\tau}\right].
\label{interaction_term}
\enann
Due of the last interaction term ${\cal L}_{1/2\mathchar`- {\rm int}}$, the orbit of $\vect{R}=0$ is no longer a solution even if the acceleration vanishes. The motion of  the center-of-mass ($\vect{R}(\tau$)) couples with the relative motion $(\vect{r}(\tau))$. As a result, not only the orbit of a binary but also the motion of the center-of-mass will become complicated even if the observer's orbit is a geodesic.

However, if we introduce an appropriate acceleration $\vect{a}$
in 0.5PN order 
 to cancel the interaction terms, $\vect{R}=0$ will become a solution, i.e., 
 the center-of-mass can follow the observer's motion as follows:
Integrating by parts the interaction term, we find
 \beann
{\cal L}_{1/2\mathchar`- {\rm int}}\left(\vect{R}, {d\vect{R}\over d\tau}, \vect{r}, {d\vect{r}\over d\tau}\right)
&\approx&
2\mu\left[{1\over 3}{d\bar{\cal R}_{\hat 0\hat k \hat j \hat \ell}\over d\tau}
r^{\hat k}r^{\hat \ell} 
+ \bar{\cal R}_{\hat 0\hat k \hat j \hat \ell}
r^{\hat k}{dr^{\hat \ell}\over d\tau} \right]R^{\hat j}~~{\rm (integration~by~part)}
\,,
\enann
where the time derivative of the curvature is evaluated along the observer's orbit.

If we define the acceleration by
 \beann
 a_{\hat j}={2\mu \over m_1+m_2}\left[{1\over 3}{d\bar{\cal R}_{\hat 0\hat k \hat j \hat \ell}\over d\tau}
r^{\hat k}r^{\hat \ell} 
+ \bar{\cal R}_{\hat 0\hat k \hat j \hat \ell}
r^{\hat k}{dr^{\hat \ell}\over d\tau}  \right]
\label{0.5PN_acceleration}
\,,
 \enann
two terms ${\cal L}_{1/2\mathchar`- {\rm int}}$ and ${\cal L}_{{\rm CM}\mathchar`-a}$ cancel each other.
As a result, the Lagrangians for $\vect{R}$ and $\vect{r}$ are decoupled,
and  $\vect{R}=0$ becomes an exact solution of the equation for $\vect{R}$, which is
derived from the Lagrangian
(${\cal L}_{\rm CM}+{\cal L}_{1/2\mathchar`-{\rm CM}}$).
The center-of-mass follows the observer's orbit and therefore, we obtain the decoupled equation for the relative coordinate $\vect{r}$.
 
In order to obtain the proper observer's orbit, 
which is not a geodesic but may be close to the geodesic, 
we have to solve the equation of motion including small acceleration such that
 \bea
 {Du_{\rm CM}^\mu\over d\tau}=a^\mu=e^{\mu\hat j } a_{\hat j}=
{2\mu \over m_1+m_2}e^{\mu\hat j }\left[{1\over 3}{d\bar{\cal R}_{\hat 0\hat k \hat j \hat \ell}\over d\tau}
r^{\hat k}r^{\hat \ell} 
+ \bar{\cal R}_{\hat 0\hat k \hat j \hat \ell}
r^{\hat k}{dr^{\hat \ell}\over d\tau} \right]
\,.
\label{eq_CM}
 \ena

As a result, we first solve the equation for the relative coordinate $\vect{r}$, which is obtained solely from the Lagrangian ${\cal L}_{\rm rel}(\vect{r}) + {\cal L}_{1/2\mathchar`-{\rm rel}}(\vect{r})$.
Note that when $m_1=m_2$, only the Newtonian Lagrangian ${\cal L}_{\rm rel}$ remains because ${\cal L}_{1/2\mathchar`-{\rm rel}}$ vanishes.
After obtaining the solution for $\vect{r}(\tau)$, we determine the motion of the center-of-mass (or the observer) in the background spacetime by solving Eq.~\eqref{eq_CM}.
Using the relative motion $\vect{r}(\tau)$ together with the solution for the center-of-mass motion, $x_{\rm CM}^\mu(\tau)$, we obtain the binary motion in a given curved background spacetime, $(x_1^\mu(\tau), x_2^\mu(\tau))$.
Some concrete examples were given in \cite{Maeda:2023uyx}.
\vskip 1cm

\end{widetext}

\end{appendices}

%\end{widetext}

\newpage
\bibliography{refer}
% Produces the bibliography via BibTeX.

%%%%%%%%%%%%%%%%%%%%%%%%%%
%%%%%%%%%%%%%%%%%%%%%%%%%%
%%%%%%%%%%%%%%%%%%%%%%%%%%
%%%%%%%%%%%%%%%%%%%%%%%%%%
%%%%%%%%%%%%%%%%%%%%%%%%%%
%%%%%%%%%%%%%%%%%%%%%%%%%%
\end{document}